\documentclass[useAMS,usenatbib]{mn2e}
\usepackage{graphicx} 
\usepackage{natbib}
\usepackage{amsmath,amsfonts}
\usepackage{epsfig}
\usepackage{psfrag}
\usepackage{mathptmx}
\usepackage{longtable}
\usepackage{verbatim}
\usepackage{sidecap}

\def\reff@jnl#1{{\rm#1\/}}

\def\aj{\reff@jnl{AJ}}                  
\def\araa{\reff@jnl{ARA\&A}}            
\def\apj{\reff@jnl{ApJ}}                        
\def\apjl{\reff@jnl{ApJ}}               
\def\apjs{\reff@jnl{ApJS}}              
\def\ao{\reff@jnl{Appl.Optics}}         
\def\apss{\reff@jnl{Ap\&SS}}            
\def\aap{\reff@jnl{A\&A}}               
\def\aapr{\reff@jnl{A\&A~Rev.}}         
\def\aaps{\reff@jnl{A\&AS}}             
\def\azh{\reff@jnl{AZh}}                        
\def\baas{\reff@jnl{BAAS}}              
\def\jrasc{\reff@jnl{JRASC}}            
\def\memras{\reff@jnl{MmRAS}}           
\def\mnras{\reff@jnl{MNRAS}}            
\def\pra{\reff@jnl{Phys. Rev. A}}         
\def\prb{\reff@jnl{Phys. Rev. B}}         
\def\prc{\reff@jnl{Phys. Rev. C}}         
\def\prd{\reff@jnl{Phys. Rev. D}}         
\def\prl{\reff@jnl{Phys. Rev. Lett}}      
\def\pasp{\reff@jnl{PASP}}              
\def\pasj{\reff@jnl{PASJ}}              
\def\qjras{\reff@jnl{QJRAS}}            
\def\skytel{\reff@jnl{S\&T}}            
\def\solphys{\reff@jnl{Solar~Phys.}}    
\def\sovast{\reff@jnl{Soviet~Ast.}}     
\def\ssr{\reff@jnl{Space~Sci.Rev.}}     
\def\zap{\reff@jnl{ZAp}}                        
\def\nat{\reff@jnl{Nature}}             

\def\p#1by#2{{\partial{#1} \over \partial{#2}}}
\def\pp#1by#2#3{{\partial^2{#1} \over \partial{#2}\partial{#3}}}
\def\d#1by#2{{{\rm d}{#1} \over {\rm d}{#2}}}
\def\dd#1by#2#3{{{\rm d}^2{#1} \over {\rm d}{#2}{\rm d}{#3}}}

\title[SZ observations of LoCuSS clusters with the AMI: high X-ray luminosity sample]
{Sunyaev-Zel'dovich observations of LoCuSS clusters with the Arcminute Microkelvin Imager: high X-ray luminosity sample \thanks{We request that any reference to this paper cites `AMI Consortium: Shimwell et~al. 2011'}}

\label{firstpage}
\author[Shimwell et~al.]
{AMI Consortium: 
Timothy W. Shimwell$^1$$\thanks{E-mail: tws29@mrao.cam.ac.uk}$, 
Carmen Rodr{\'i}guez-Gonz{\'a}lvez$^1$$\thanks{E-mail: cr384@mrao.cam.ac.uk}$, 
\newauthor
 Matthew L. Davies$^1$, 
 Farhan Feroz$^1$,
 Thomas M. O. Franzen$^1$, 
 Keith J. B. Grainge$^{1,2}$, 
\newauthor
 Michael P. Hobson$^1$,
 Natasha Hurley-Walker$^1$,
 Anthony N. Lasenby$^{1,2}$,
 Malak Olamaie$^1$,
\newauthor
 Guy Pooley$^1$,
 Richard D. E. Saunders$^{1,2}$,
 Anna M. M. Scaife$^3$,
 Michel P. Schammel$^1$,
\newauthor
 Paul F. Scott$^1$, 
 David J. Titterington$^1$,
 Elizabeth M. Waldram$^1$ 
 \vspace{0.03in}\\
$^1$ Astrophysics Group, Cavendish Laboratory, J J Thomson Avenue,
Cambridge CB3 0HE\\
$^2$ Kavli Institute for Cosmology Cambridge, Madingley Road,
Cambridge, CB3 0HA\\
$^{3}$ Dublin Institute for Advanced Studies, 31 Fitzwilliam Place, Dublin 2, Ireland \\
}

\date{Accepted ---; received ---; in original form \today}
\pagerange{\pageref{firstpage}--\pageref{lastpage}}

\begin{document}
\maketitle

\begin{abstract}
We present observations from the Small Array of the Arcminute Microkelvin Imager (AMI) of eight high X-ray luminosity galaxy cluster systems selected from the Local Cluster Substructure Survey (LoCuSS) sample. We detect the  Sunyaev-Zel'dovich (SZ) effect in seven of these clusters. With the assumptions that galaxy clusters are isothermal, have a density profile described by a spherical $\beta$-model and obey the theoretical M-T relation, we are able to derive cluster parameters at $r_{200}$ from our SZ data. With the additional assumption of hydrostatic equilibrium we are able to derive parameters at $r_{500}$. We present posterior probability distributions for cluster parameters such as mass, radius and temperature ($T_{SZ,MT}$). Combining our sample with that of  AMI Consortium: Rodr{\'i}guez-Gonz{\'a}lvez et al. (2011) and using large-radius X-ray temperature estimates ($T_{X}$) from \emph{Chandra} and \emph{Suzaku} observations, we find that there is reasonable correspondence between $T_{X}$ and $T_{SZ,MT}$ values at low  $T_{X}$, but that for clusters with $T_{X}$ above around 6keV the correspondence breaks down with $T_{X}$ exceeding $T_{SZ,MT}$; we stress that this finding is based on just ten clusters.
%
\end{abstract}


\section{Introduction}

The Local Cluster Substructure Survey (LoCuSS see Smith et al. 2003, 2005) 
sample of clusters contains 164 clusters with redshifts between 0.142 and 0.295. The LoCuSS aims to measure the relationship between the structure of galaxy clusters and the evolution of the hot gas and galaxies that inhabit them using gravitational lensing data and other observations spanning the electromagnetic spectrum from the radio to X-ray.
Example LoCuSS analysis papers relevant to this work are \cite{ex_locuss1} and \cite{ex_locuss2}. 

We have imaged a subset of LoCuSS in the Sunyaev-Zel'dovich (SZ; \citealt{SunZel}) effect with the Arcminute Microkelvin Imager (AMI; see e.g. \citealt{2008MNRAS.391.1545Z}) centred at 16~GHz. The SZ signal arises from the inverse Compton scattering of cosmic microwave background (CMB) photons by the hot cluster plasma (see e.g. \citealt{BIRK_SZ_REVIEW} and \citealt{CARL_SZ_REVIEW}), with a surface brightness that is independent of redshift and dependencies on plasma density and temperature that are different to those for e.g. X-ray bremsstrahlung emission.

Hereafter, we assume a concordance $\rm{\Lambda}$CDM cosmology with $\rm{\Omega_{m}}$ = 0.3, $\rm{\Omega_\Lambda}$ = 0.7 and H$_{0}$ = 100km\,s$^{-1}$Mpc$^{-1}$. The dimensionless Hubble parameters are defined as $h_{X}$ = H$_{0}$/(X km\,s$^{-1}$Mpc$^{-1}$). All coordinates are given at equinox J2000.

\section{Cluster sample}\label{sec:CLUSTER_SELECTION}

 In this paper we focus on LoCuSS clusters with a declination greater than 20$^{\circ}$ and an X-ray luminosity ($L_{X}$) greater than 11 $\times 10^{37}$W over the 0.1-2.4 keV band in the cluster rest frame (according to Ebeling et al. 1998, 2000, using $h_{50}=1$). 
Radio source contamination can make it difficult to observe the SZ effect at 16~GHz and we have not studied the clusters with sources brighter than 10~mJy/beam within 10$\arcmin$ of the cluster X-ray centre. Note that our redshifts correspond to those cited in Ebeling et al. (1998, 2000).
We present results from the analysis of eight galaxy cluster systems; Table \ref{tab:cluster_coords} shows the coordinates, redshifts and X-ray luminosities of our selected LoCuSS clusters. A companion paper (\citealt{CARMEN_LOCUS}) presents results from 11 LoCuSS clusters with an X-ray luminosity in the range 7-11 $\times 10^{37}$W ($h_{50}=1$).

\begin{table*}
\caption{Coordinates, redshifts and X-ray luminosities of the observed LoCuSS clusters. Note that Abell~1758B is included even though it is below our luminosity cut; this is because it is within the field of view of our Abell~1758A observations. Redshifts and X-ray luminosities are taken from Ebeling et al. (1998, 2000).}
 \label{tab:cluster_coords}
\begin{tabular}{lccccc}
\hline \hline
Cluster         & Right ascension  & Declination     & Redshift      & X-ray luminosity          & Alternative cluster names \\ 
                &  (J2000)         & (J2000)         &               & in $ 10^{37}$W            & \\ 
                &                  &                 &               & ($h_{50}=1$)              & \\ \hline
Abell 586       & 07:32:12         & +31:37:30       &  0.171        & 11.12  \\ 
Abell 611 	& 08:00:56    	   & +36:03:40       &	0.288 	     & 13.60 	& \\ 
Abell 773       & 09:17:54    	   & +51:42:58       &	0.217        & 13.08 		         & RXJ0917.8+5143  \\ 
Abell 781 	& 09:20:25    	   & +30:31:32       &	0.298 	     & 17.22 	& \\ 
Abell 1413 	& 11:55:18    	   & +23:24:29       &	0.143 	     & 13.28 	& \\ 
Abell 1758B 	& 13:32:29    	   & +50:24:42       &	0.280 	     & 07.25 	& \\ 
Abell 1758A 	& 13:32:45    	   & +50:32:31       &	0.280 	     & 11.68 	& \\ 
Zw1454.8+2233   & 14:57:15    	   & +22:20:34       &	0.258 	     & 13.19 		         & Z7160  \\ 
RXJ1720.1+2638 	& 17:20:10    	   & +26:37:31       &	0.164 	     & 16.12 	& \\ \hline
 \end{tabular}
\end{table*}


\section{Instrument, observations and source finding}\label{sec:OBS_INFO}

\subsection{The Arcminute Microkelvin Imager (AMI)}

AMI consists of a pair of aperture-synthesis interferometric arrays located near Cambridge. The Small Array (SA) is optimised for SZ imaging while the Large Array (LA) is used to measure radio sources that contaminate the SZ-effect in the SA observations. AMI is described in detail in Zwart et al. (2008).

\subsection{Observations}

SA pointed observations centred at the X-ray cluster position (Table \ref{tab:cluster_coords}) for our eight clusters were taken during 2007-2010. The observation lengths were in the range 20-90~hours per cluster before any flagging of the data; the noises on the SA maps reflect the actual observation time used. 

With the SA we observed phase calibrators every hour and used bi-daily observations of 3C48 or 3C286 for amplitude calibration; the assumed flux densities for the calibrator sources are consistent with the \cite{rudy_mars} model of Mars -- see \cite{TMOF_10C}. 

With the LA we typically conduct 61+19pt hexagonal raster observations centred on the cluster X-ray position. The 61 pointing centres are separated by 4$\arcmin$; the inner 19 pointings are observed for approximately six times longer than the outer 42 pointings. We observed phase calibrators every ten minutes. Observations were taken over 2009-2010 and each cluster was observed for 10-25 hours before any flagging of the data. For more details on the rastering techniques and calibration see Franzen et al. (2010).

All our cluster data were passed through \textsc{reduce}, the standard AMI data reduction package which is discussed in detail in e.g. \cite{DAVIES_09}. Thermal noise levels for the SA and the LA maps ($\sigma_{SA}$ and $\sigma_{LA}$ respectively), and phase calibrators that we have taken from the Jodrell Bank VLA Survey (\citealt{JVAS1}, \citealt{JVAS2} and \citealt{JVAS3}) are summarised in Table \ref{hot_observations}. 

\begin{table*}
\caption{Details of AMI observations.} 
 \label{hot_observations}
\begin{tabular}{lcccc}
\hline \hline
Cluster   & $\sigma_{SA}$               & $\sigma_{LA}$   & Number of LA $4\sigma_{LA}$ sources & LA phase calibrator\\ 
          & (mJy)                       & (mJy)           & & \\ \hline  
Abell 586       & 0.172   & 0.09    & 23 & J0741+3112 \\
Abell 611 	& 0.106   & 0.07    & 23 & J0808+408\\ 
Abell 773       & 0.133   & 0.09    & 09  & J0903+468 or J0905+4850 \\ 
Abell 781 	& 0.116   & 0.07    & 24 & J0925+3127 or J0915+2933 \\ 
Abell 1413 	& 0.130   & 0.09    & 17 & J1150+2417 \\ 
Abell 1758A 	& 0.115   & 0.08    & 14 & J1349+536 \\ 
Abell 1758B 	& 0.130   & 0.08    & 14 & J1349+536 \\ 
Zw1454.8+2233   & 0.100   & 0.10    & 16 & J1513+2338 \\ 
RXJ1720.1+2638 	& 0.084   & 0.10    & 17 & J1722+2815\\ \hline
 \end{tabular}
\end{table*}

\subsection{LA mapping and source finding}

Our LA map-making and source-finding procedures follow \cite{SHIMWELL} exactly. We applied standard \textsc{aips} \footnote{http://www.aips.nrao.edu} tasks to image the continuum and individual-channel \textsc{uvfits} data. On the continuum maps we detected sources at $\geq$4$\sigma_{LA}$ and catalogued the source right ascension, $x_{s}$, declination, $y_{s}$ and peak flux, $S_{0}$. To determine whether a source is extended in our LA maps we compared the measured source area with the LA synthesized beam. For extended sources we catalogued the integrated flux rather than the peak flux. We measured the flux of each source in each channel map and assumed a power law relationship between flux and frequency ($S \propto \nu^{-\alpha}$) to determine the source spectral index, $\alpha$. The number of $\geq$4$\sigma_{LA}$ sources detected in our LA observations towards the eight clusters is shown in Table \ref{hot_observations}.

\section{Bayesian Analysis}\label{sec:ANALYSIS}

To analyse the AMI cluster observations we use a Bayesian analysis methodology (\citealt{MARSH_MCADAM} and \citealt{FF_MCADAM}). This Bayesian analysis uses M\textsc{ulti}N\textsc{est} (\citealt{multinest2} and \citealt{2009MNRAS.398.1601F}) to efficiently explore the multidimensional parameter space and to calculate Bayesian evidence. This analysis has been applied to pointed observations of known clusters in AMI Consortium Rodr{\'i}guez-Gonz{\'a}lvez et al. (2010), \cite{7CLUSTERS} \& Rodr{\'i}guez-Gonz{\'a}lvez et al. (2011) and also to detect previously unknown clusters in Shimwell et al. In this paper we use the same model as Shimwell et al., the differences between this model and previous models are explained in \cite{malak_mc} and Rodr{\'i}guez-Gonz{\'a}lvez et al. (2010).

The priors that we use in our Bayesian analysis are shown in Table  \ref{MC_PRIORS}.

\begin{table*}
\caption{Priors used in our Bayesian analysis.}
 \label{MC_PRIORS}
\begin{tabular}{lcc}
\hline \hline
Parameter & Prior \\ \hline
Source position ($\bf x_{s}$) & + or $\times$:  Delta-function using the LA positions. \\ 
                              & $\triangle$: Gaussian centred on the LA postions with $\sigma$=5$''$. \\ 
Source flux densities ($S_{0}/\rm{Jy}$) & $\times$ or $\triangle$: Gaussian centred on the LA continuum value with a $\sigma$ of $0.4S_{0}$. \\ 
                                        & +: Delta-function on the LA value.\\
Source spectral index ($\alpha$) & $\times$ or $\triangle$: Gaussian centred on the value calculated from the LA channel maps with $\sigma$ as the LA error.  \\ 
                                 & +: Delta-function on the LA value. \\
Redshift ($z$) & Delta-function on the X-ray value (Table \ref{tab:cluster_coords}). \\ 
Core radius ($r_{c}/h_{100}^{-1}\rm{kpc}$) & Uniform between 10 and 1000. \\ 
Beta   ($\beta$)  & Uniform between 0.3 and 2.5. \\ 
Mass ($M_{\rm{T},r200}/h_{100}^{-1}M_{\odot}$) & Uniform in log space over, $(0.32-50)$ $\times$ $10^{14}M_{\odot}h_{100}^{-1}$. \\
Gas fraction ($f_{g}$) & Gaussian prior centred on 0.086, $\sigma$=0.02 (\citealt{WMAP_FGAS_VALUE}). \\ 
Cluster position ($\bf x_{c}$) & Gaussian prior on the X-ray position, $\sigma$=60$\arcsec$ (Table \ref{tab:cluster_coords}).  \\\hline
%
 \end{tabular}
\end{table*}

\section{SA source subtraction and mapping}\label{sec:SA_sub}

Source subtraction is performed using the software package \textsc{muesli}. This performs the same function as the \textsc{aips} task \textsc{uvsub} but it is optimised for processing AMI data. For both the LA sources and the sources modelled in our Bayesian analysis we use the source position, mean frequency, spectral index and central flux value to subtract the appropriate flux for each source from each channel of the SA \textsc{uvfits} file. To perform this subtraction from our non-primary-beam-corrected SA \textsc{uvfits} we make use of an accurate SA power primary beam model.


We use \textsc{aips} to produce the SA maps and apply no primary beam correction i.e. the SA thermal noise is constant across the map. All the SA maps presented here have contour levels changing linearly from  2-10$\sigma_{SA}$; positive contours are solid lines and negative contours are dashed lines. The SA synthesized beam FWHM is shown in the bottom left corner of our SA maps. We also present the source-subtracted maps with a $uv$ taper of 600k$\lambda$ (a Gaussian taper of value 1 are low $uv$ falls to 0.3 at 600k$\lambda$) since the shorter SA baselines are more sensitive to the large angular size of the SZ-effect signal. 
Maps before source subtraction have been \textsc{clean}ed with a single box over their total extents, whilst source-subtracted maps have been \textsc{clean}ed with a tight box around the SZ signals.

\section{Results}\label{sec:RESULTS}

For each cluster we present SA maps before and after source subtraction, and posterior probability distributions of the large-scale cluster parameters obtained from running the SA data through our Bayesian analysis software. The derived cluster parameters are given in Table \ref{CLUS_MEAN_VALS}. We also present \emph{Chandra} images taken from the \emph{Chandra} Data Archive. All maps are displayed with right ascension on the x-axis and declination on the y-axis. In Section \ref{sec:TEMP_COMP} we compare our derived cluster temperatures with large-radius X-ray temperatures ($r\approx 500$kpc) taken from the literature.

\begin{table*}
\caption{Derived values for cluster parameters.}
 \label{CLUS_MEAN_VALS}
\begin{tabular}{lcccccccc}
\hline \hline
Cluster name &$M_{T}(r_{200})$ & $M_{T}(r_{500})$ & $M_{g}(r_{200})$&$M_{g}(r_{500})$&$r_{200} $&$r_{500} $& $f_{g}(r_{500})$ & $T_{SZ,MT}(r_{200})$ \\ 
             &$\times10^{14}h_{100}^{-1}M_{\odot}$ & $\times10^{14}h_{100}^{-1}M_{\odot}$ & $\times10^{13}h_{100}^{-2}M_{\odot}$ & $\times10^{13}h_{100}^{-2}M_{\odot}$& $h_{100}^{-1}$Mpc & $\times10^{-1}h_{100}^{-1}$Mpc & $\times10^{-1}h_{100}^{-1}$ & keV \\ \hline

Abell 586 &  $5.1 \pm 2.4$  &  $2.1 \pm 1.1$  &  $4.3 \pm 2.0$  &  $2.6 \pm 0.8$  &  $1.2 \pm 0.2 $  &  $6.6 \pm 1.1$  &  $1.4 \pm 0.4$  &  $5.2 \pm 1.6$  \\

Abell 611   &  $4.0 \pm 0.8$  &  $2.0 \pm 0.5$  &  $3.5 \pm 0.6$  &  $2.8 \pm 0.3$  &  $1.1 \pm 0.1$  &  $6.3 \pm 0.5$  &  $1.5 \pm 0.4$  &  $4.5 \pm 0.6$  \\

Abell 773  &  $3.6 \pm 1.3$  &  $1.7 \pm 0.7$  &  $3.1 \pm 1.1$  &  $2.1 \pm 0.5$  &  $1.1 \pm 0.1 $  &  $6.0 \pm 0.8$  &  $1.4 \pm 0.4$  &  $4.1 \pm 1.0$  \\

Abell 781  &  $4.1 \pm 0.8$  &  $2.0 \pm 0.5$  &  $3.6 \pm 0.6$  &  $2.9 \pm 0.4$  &  $1.1 \pm 0.1$  &  $6.3 \pm 0.5$  &  $1.5 \pm 0.4$  &  $4.5 \pm 0.6$  \\ 

Abell 1413  &  $4.0 \pm 1.0$  &  $1.9 \pm 0.6$  &  $3.5 \pm 0.8$  &  $2.7 \pm 0.4$  &  $1.1 \pm 0.1$  &  $6.6 \pm 0.7$  &  $1.5 \pm 0.4$  &  $4.4 \pm 0.8$  \\

Abell 1758B   &  $4.4 \pm 2.2$  &  $2.2 \pm 1.2$  &  $3.7 \pm 1.8$  &  $2.2 \pm 0.6 $  &  $1.1 \pm 0.2$  &  $6.4 \pm 1.2$  &  $1.2 \pm 0.4$  &  $4.6 \pm 1.5$  \\

Abell 1758A &$4.1 \pm 0.7$ &$2.5 \pm 4.4$ &$3.6 \pm 0.5$  &  $3.4 \pm 0.4$  &  $1.1 \pm 0.1$  &  $6.8 \pm 0.4$  &  $1.4 \pm 0.3$  &  $4.5 \pm 0.5$  \\

RXJ1720.1+2638  &  $2.0 \pm 0.4$  &  $1.2 \pm 0.2$  &  $1.7 \pm 0.3$  &  $1.6 \pm 0.3$  &  $0.9 \pm 0.1$  &  $5.6 \pm 0.4$  &  $1.4 \pm 0.3$  &  $2.8 \pm 0.4$  \\\hline

 \end{tabular}
\end{table*}

\subsection{Abell 586} 

\begin{figure*}
\centerline{Abell~586}
\centerline{\includegraphics[width=7.5cm,height= 7.5cm,clip=,angle=0.]{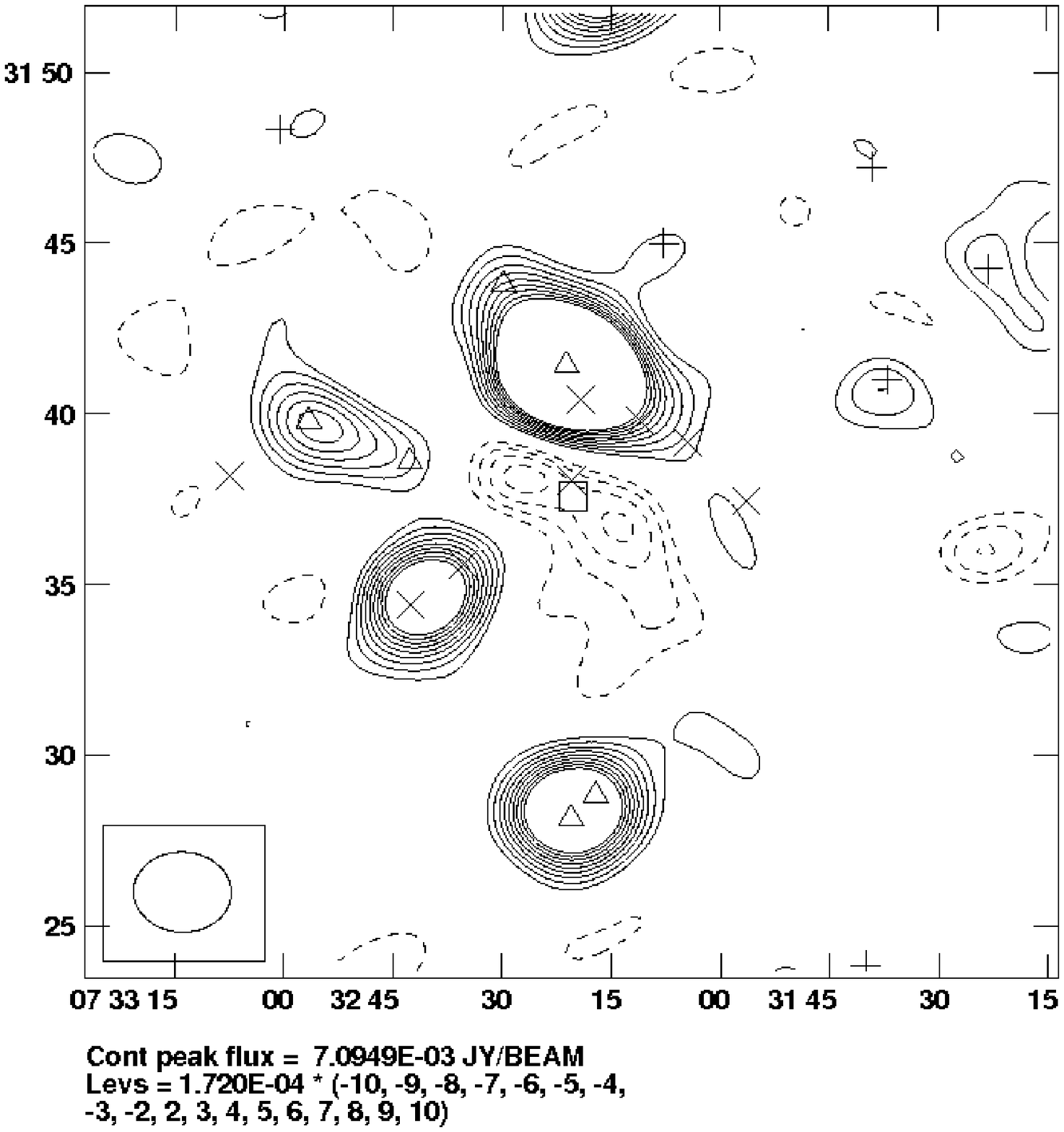}\qquad\includegraphics[width=7.5cm,height= 7.5cm,clip=,angle=0.]{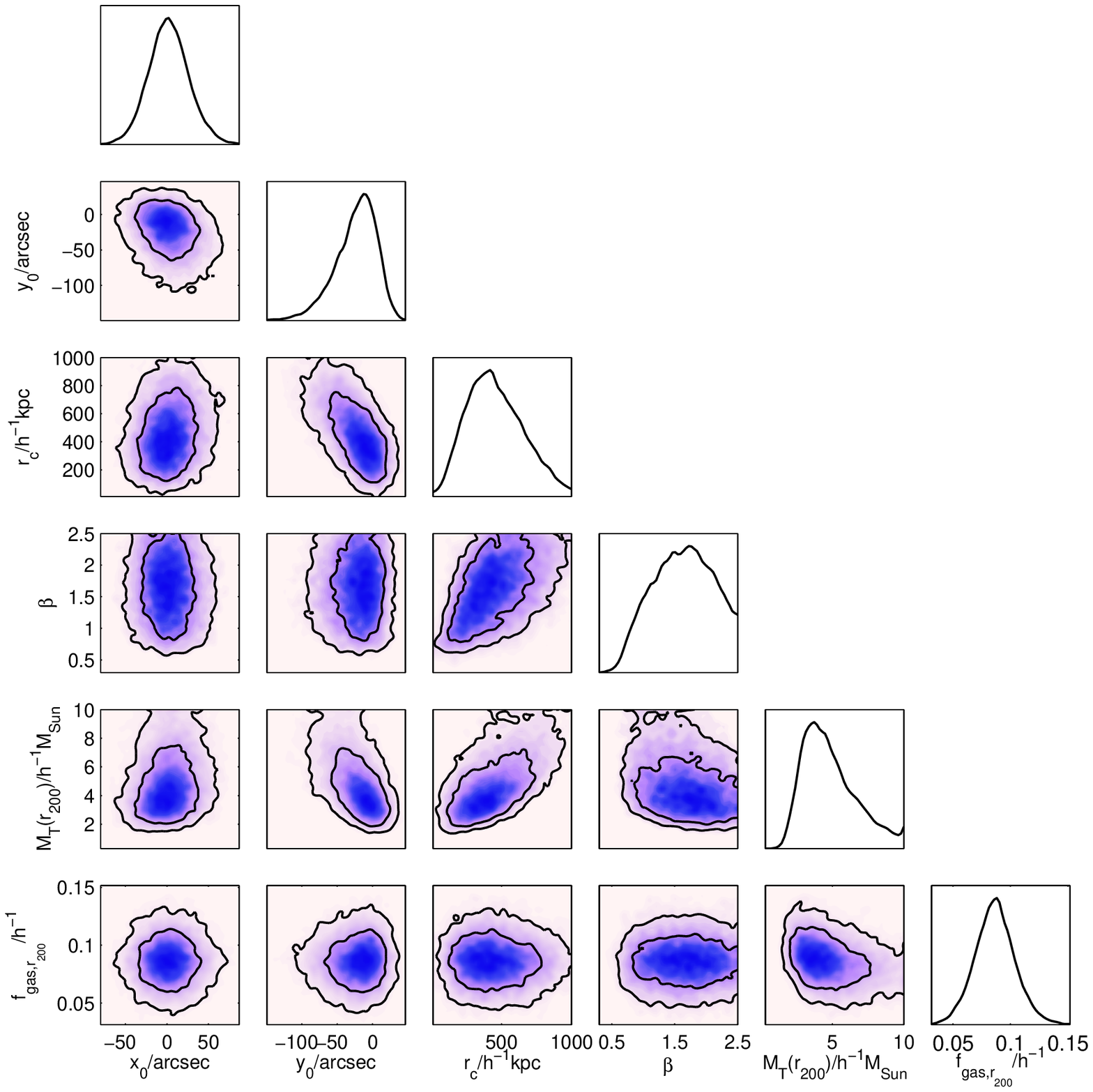}}
 \centerline{\includegraphics[width=7.5cm,height=
     7.5cm,clip=,angle=0.]{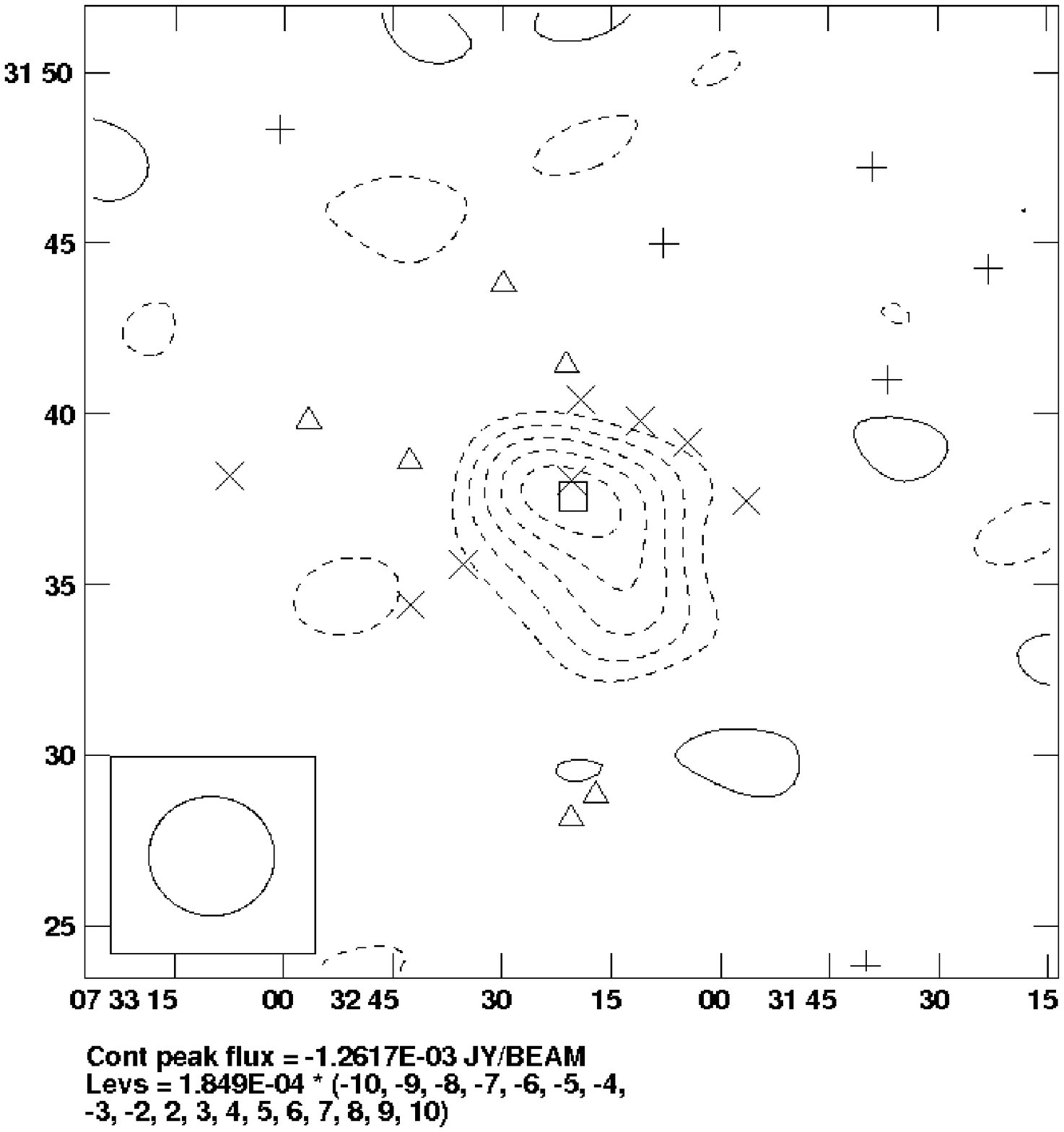}\qquad\includegraphics[width=7.5cm,height= 7.5cm,clip=,angle=0.]{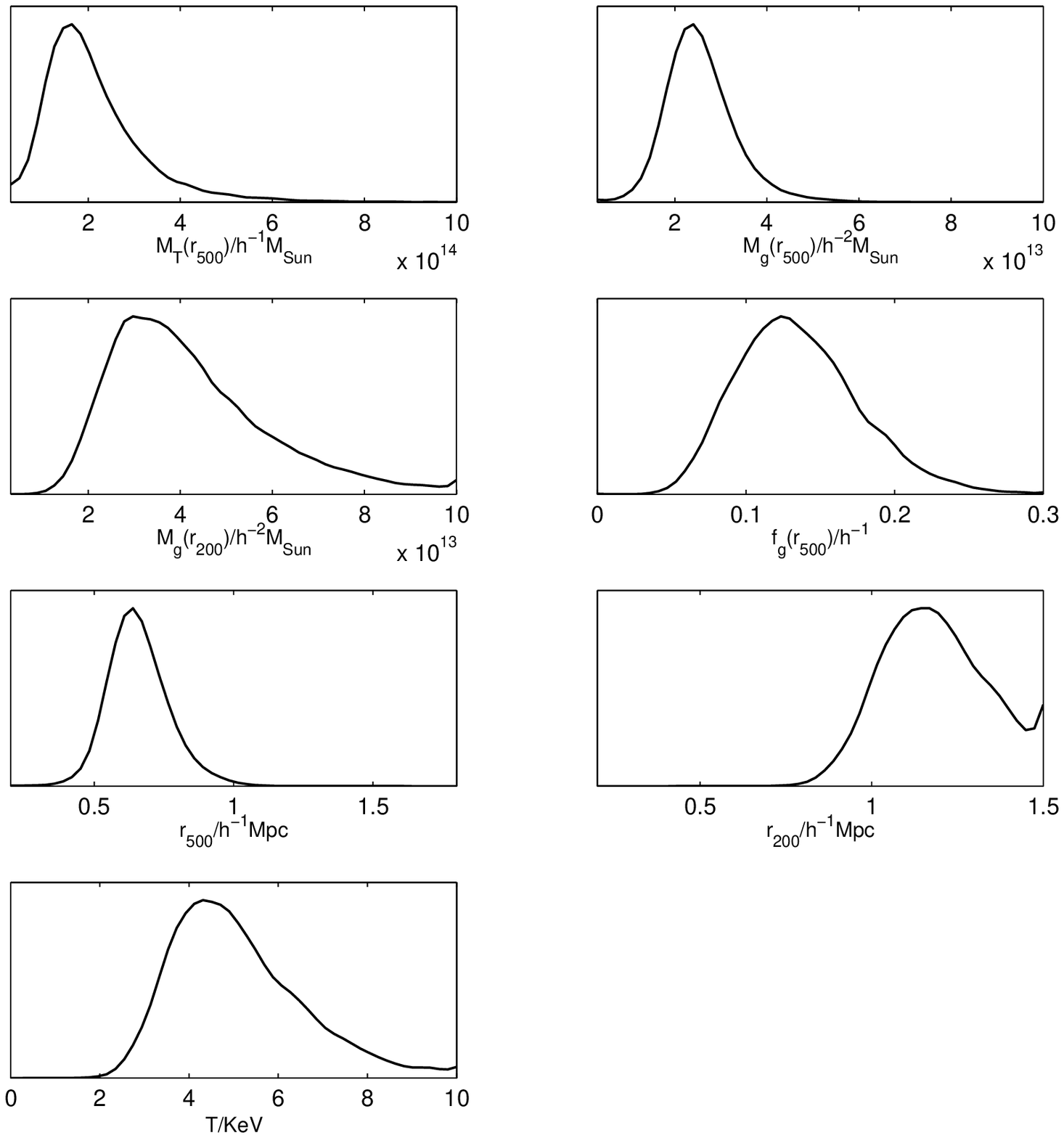}}
   \centerline{
\includegraphics[width=7.5cm,height=6.5cm,clip=,angle=0.]{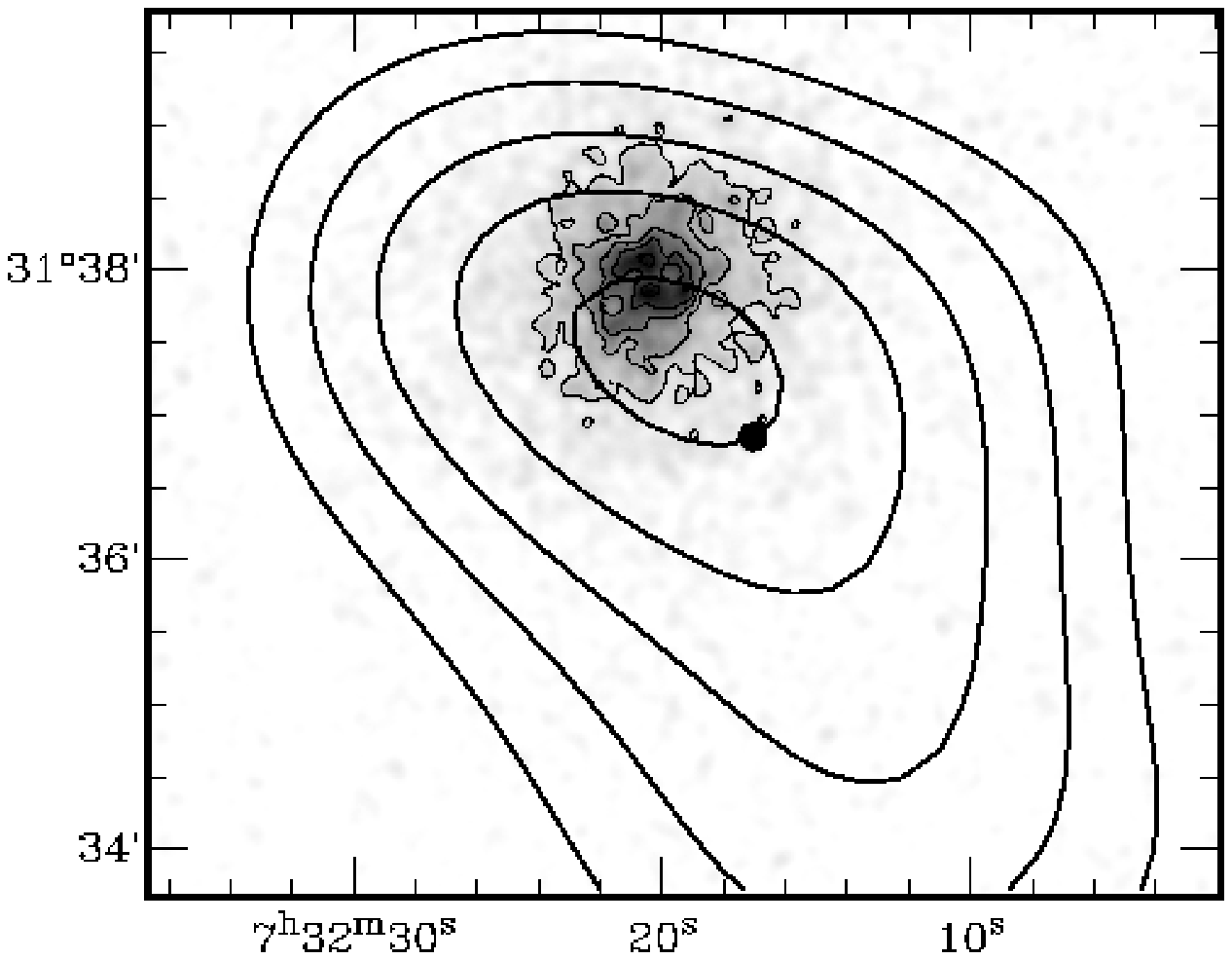}}
\caption{The top left image shows the SA map before subtraction, the map in the middle left has had the sources removed, the top right panel shows the cluster parameters that we sample from in our Bayesian analysis and the middle right plot presents several cluster parameters derived from our sampling parameters. The image at the bottom shows the \emph{Chandra} X-ray map overlayed with SA contours.}
\label{fig:A586}
\end{figure*}

Our AMI SA maps and the parameters we measure are presented in Figure \ref{fig:A586}. We have overlayed our map on an X-ray \emph{Chandra} image; the cluster centroids match and we observe an extension of the cluster towards the south.

The SZ effect from Abell~586 has previously been observed with OVRA/BIMA by \cite{laroque2006} and \cite{bona_chandra}. LaRoque et al. apply an isothermal $\beta$-model to SZ and \emph{Chandra} X-ray observations and find $M_{\rm{g}}(r_{2500})=2.49\pm{0.32}\times10^{13}M_{\odot}$ and $M_{\rm{g}}(r_{2500})=2.26^{+0.13}_{-0.11}\times10^{13}M_{\odot}$ respectively (using $h_{70}=1$). In addition, they determine an X-ray spectroscopic temperature of the cluster gas of $\approx6.35$keV between a radius of 100kpc and $r_{2500}$; $r_{2500}$ is the radius at which the average cluster density falls to 2500 times the critical density at that redshift and is determined from \emph{Chandra} observations by \cite{bona_chandra}. In comparison, \cite{locuss_subaru} use \emph{Subaru} to calculate the cluster mass from weak lensing by applying a Navarro, Frenk \& White (NFW; \citealt{NFW}) profile. They find $M_{\rm{T}}(r_{2500})=2.41^{+0.45}_{-0.41} \times10^{14}M_{\odot}$, whilst at large radii they find  $M_{\rm{T}}(r_{500})=4.74^{+1.40}_{-1.14} \times10^{14}M_{\odot}$ (using $h_{70}=1$).

Abell~586 has been studied extensively in the X-ray e.g. \cite{allen2000} and \cite{white2000}. A recent 
analysis of the temperature profile \citep{cypriano2004} shows how the
temperature falls from $\approx9$ keV at the cluster centre to
$\approx 5.5$keV at a radius $\approx 280\arcsec$. Cypriano et al. have used the Gemini Multi-Object Spectrograph together with X-ray data taken from the \emph{Chandra} archive to measure the properties of Abell~586. They compare mass estimates derived from the velocity distribution and from the X-ray temperature profile and find that both give very similar results, $M_{\rm{g}}\approx 0.46\times10^{14}M_{\odot}$ within 1.3$h_{70}^{-1}$Mpc. They suggest that the cluster is spherical and relaxed with no recent mergers. The elongation of the SZ signal on our map (Figure \ref{fig:A586}) suggests non-sphericity.

\subsection{Abell 611}\label{sec:A611}

\begin{figure*}
\centerline{A611}
\centerline{\includegraphics[width=7.5cm,height= 7.5cm,clip=,angle=0.]{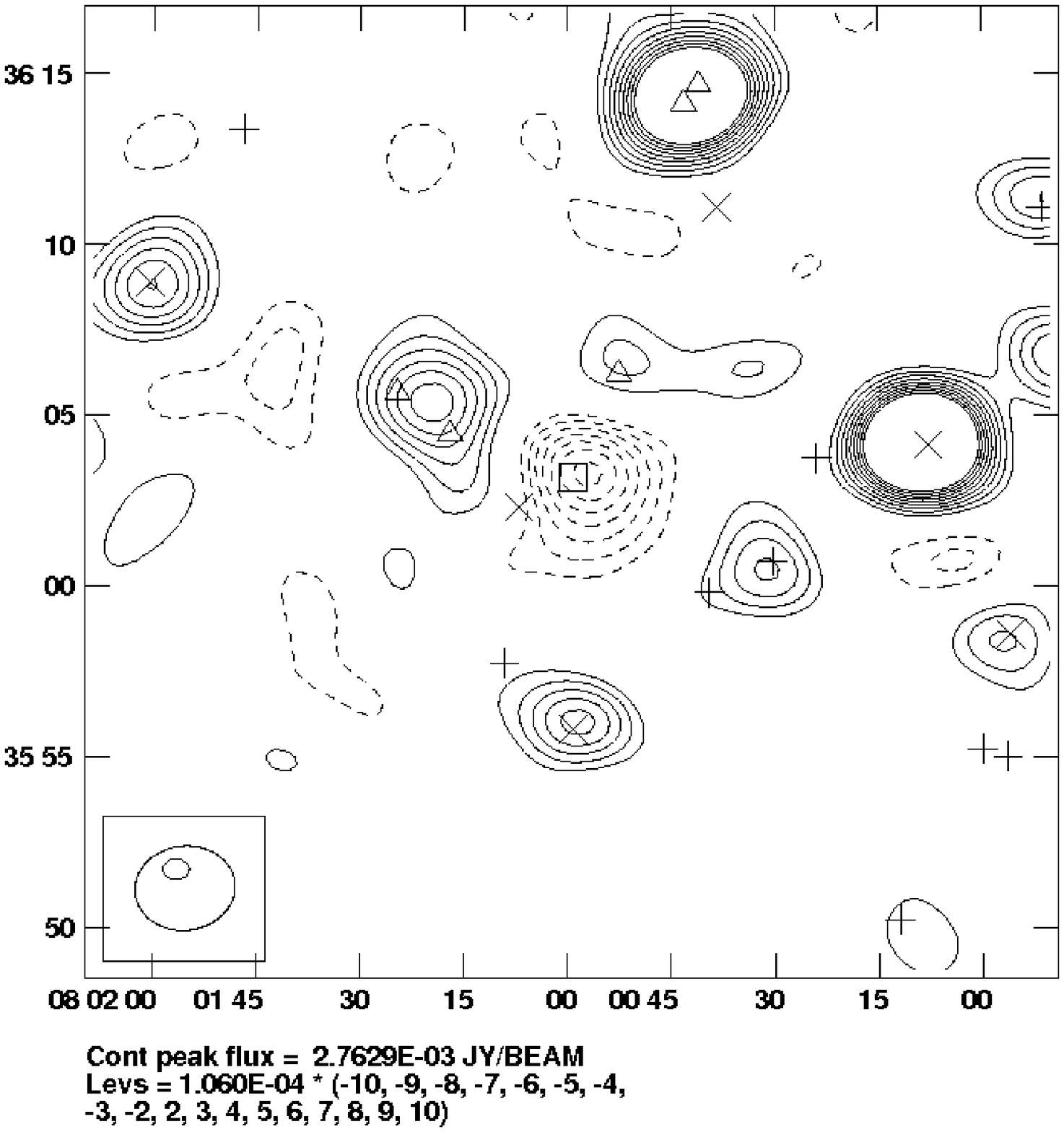}\qquad\includegraphics[width=7.5cm,height= 7.5cm,clip=,angle=0.]{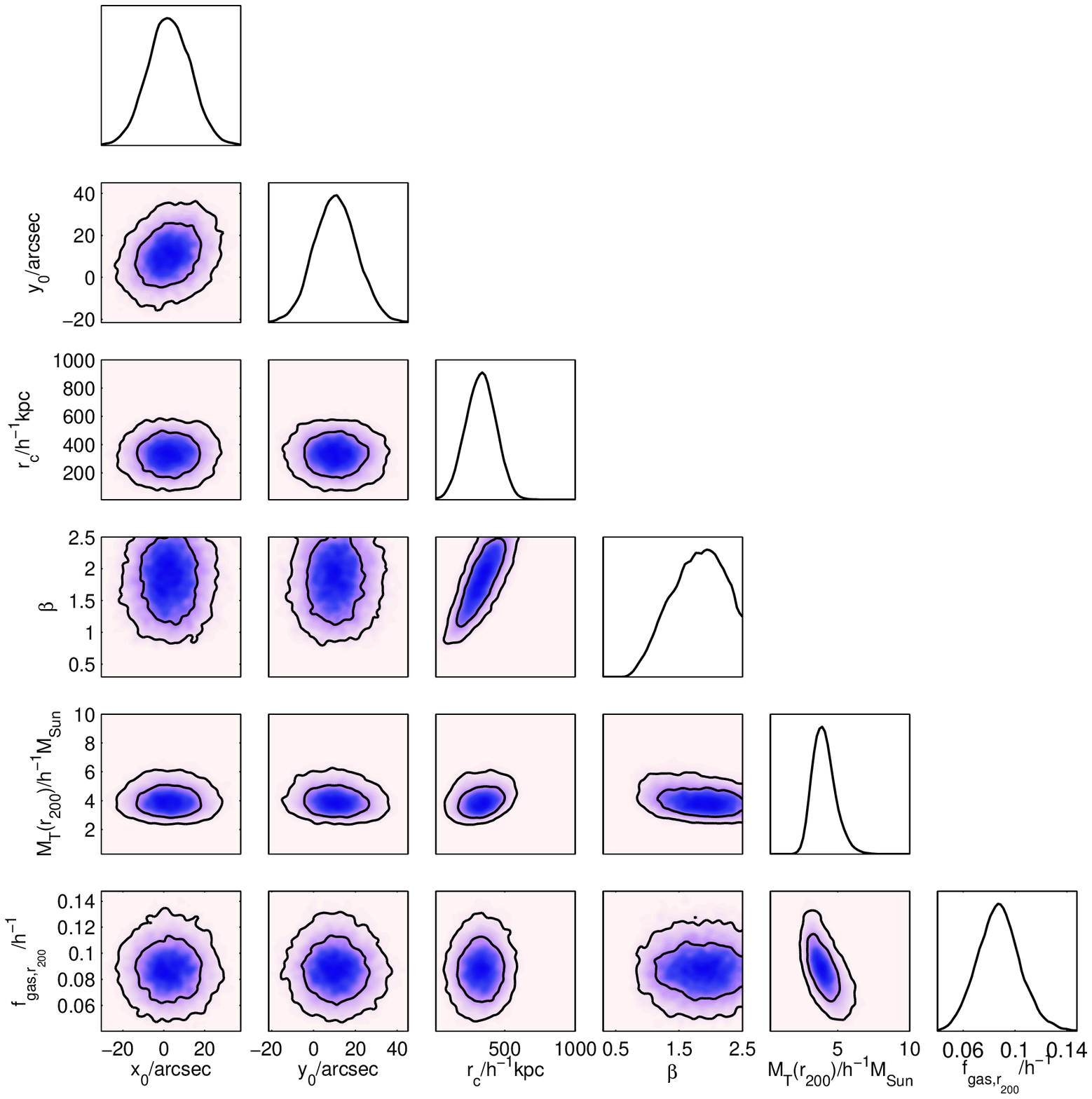}}
 \centerline{\includegraphics[width=7.5cm,height= 7.5cm,clip=,angle=0.]{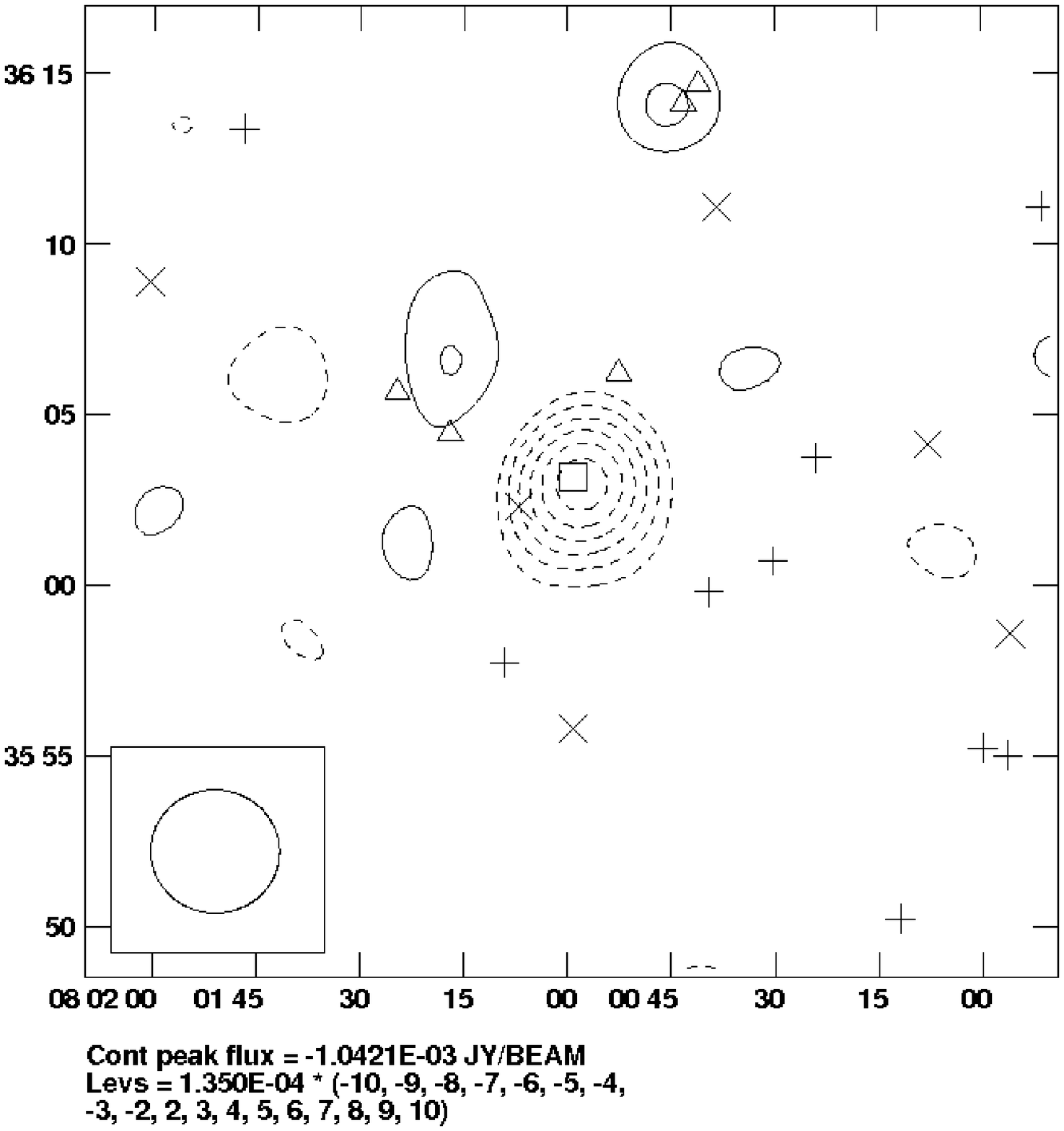}
\qquad\includegraphics[width=7.5cm,height= 7.5cm,clip=,angle=0.]{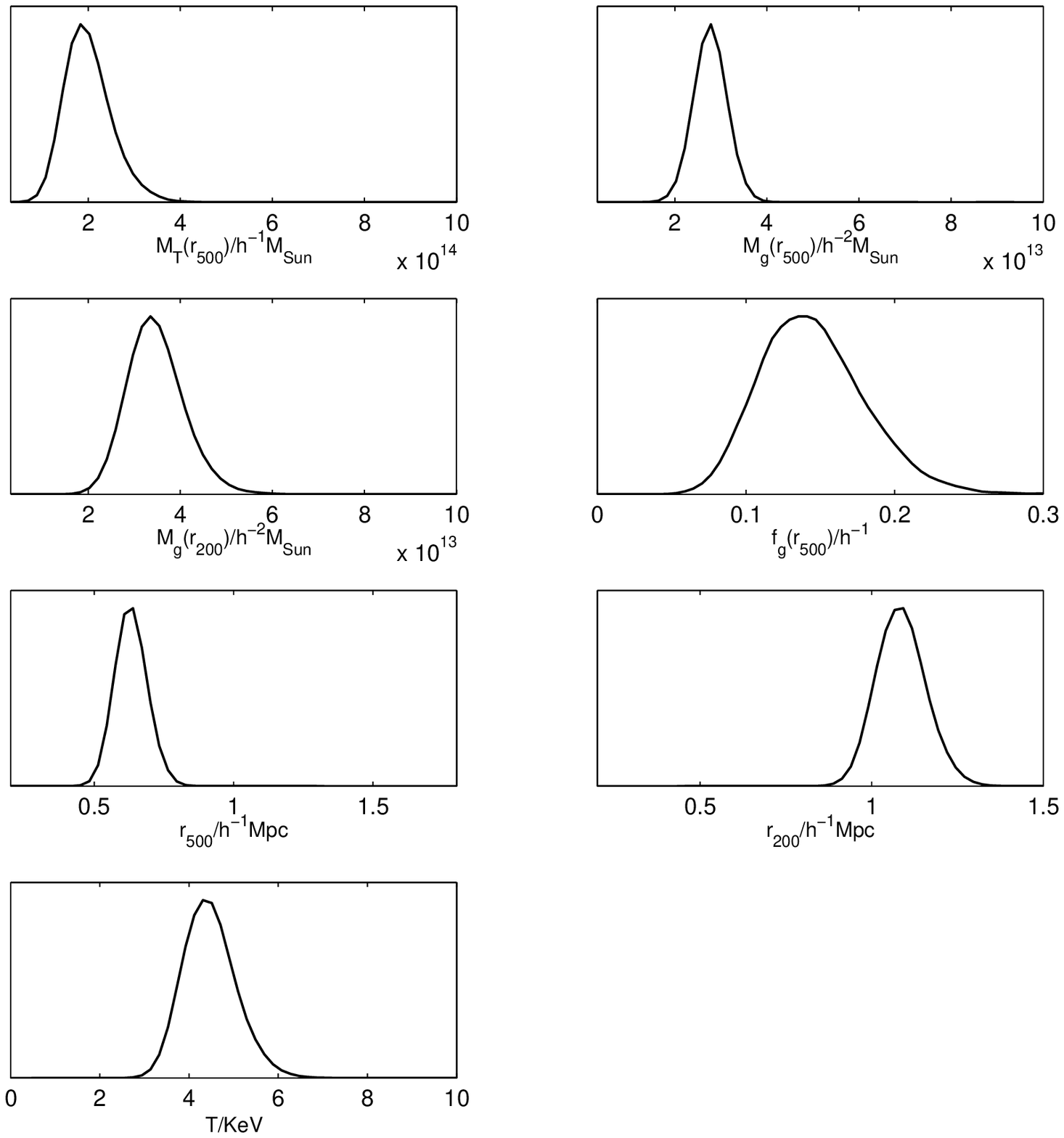}}
 \centerline{
\includegraphics[width=7.5cm,height= 6.5cm,clip=,angle=0.]{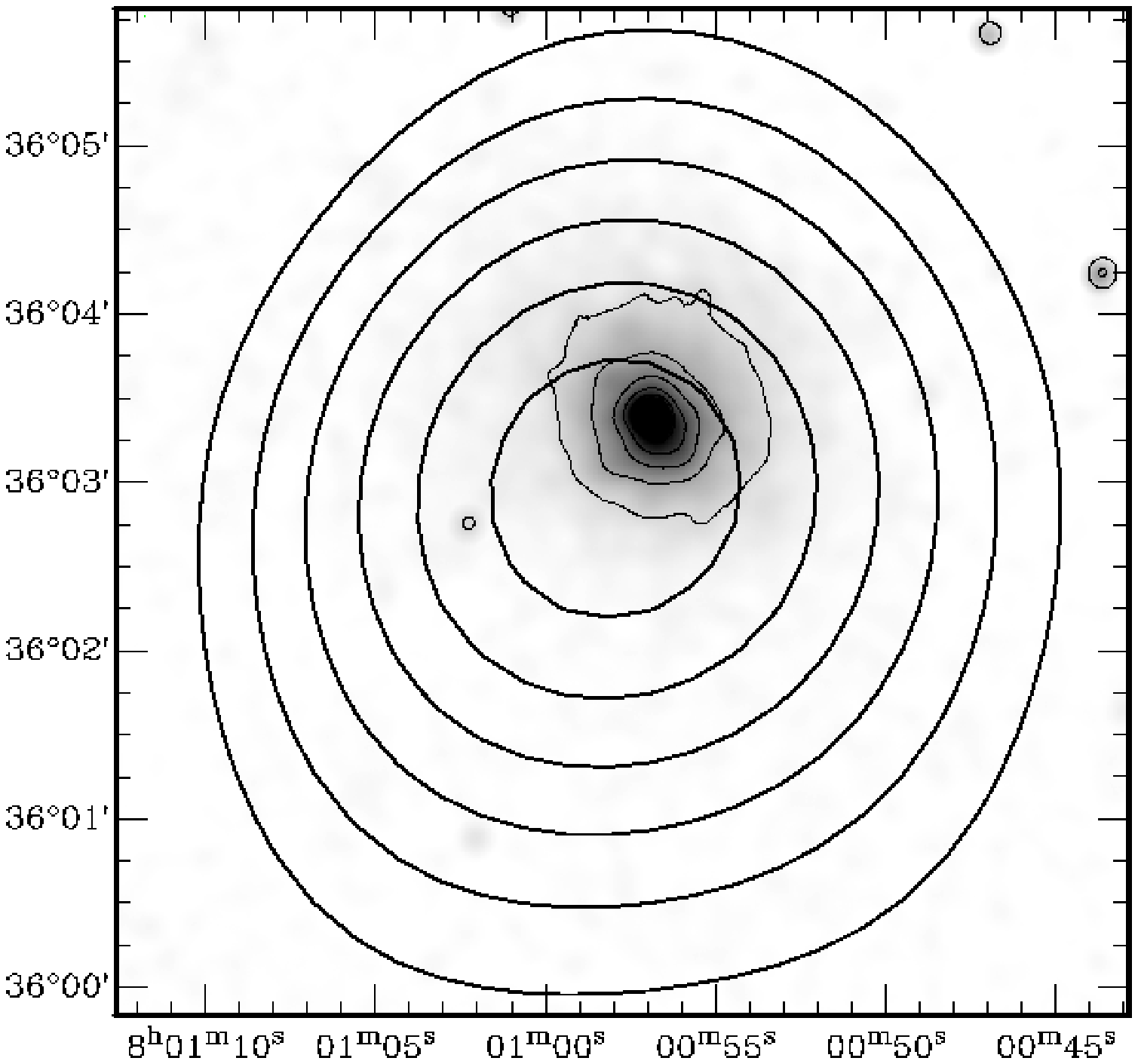}}
\caption{The top left image shows the SA map before subtraction, the map in the middle left has had the sources removed, the top right panel shows the cluster parameters that we sample from in our Bayesian analysis and the middle right plot presents several cluster parameters derived from our sampling parameters. The image at the bottom shows the \emph{Chandra} X-ray map overlayed with SA contours.}
\label{fig:A611}
\end{figure*}

We present AMI SA maps, large-scale cluster parameters and a \emph{Chandra} image of Abell 611 in Figure \ref{fig:A611}. The \cite{Xray_lense_Abell 611_2} analysis of the \emph{Chandra} achive data shows that the X-ray isophotes are quite circular, the surface brightness profile is smooth and the brightest cluster galaxy lies at the centre of the X-ray emission. These results indicate that the cluster is relaxed. From the X-ray data, the cluster mass was estimated using an NFW profile, spherical symmetry and hydrostatic equilibrium to be 9.32$\pm$1.39$\times 10^{14} \rm{M_{\odot}}$ (within a radius of 1.8$\pm$0.5~Mpc). However, the Donnarumma et al. analysis of strong lensing data indicates that the cluster mass could be closer to 4.68$\pm$0.31$\times 10^{14} \rm{M_{\odot}}$ (within a radius of 1.5$\pm$0.2 ~Mpc. Note that the values quoted from Donnarumma et al. are an example of their mass estimates; from fitting different models they find estimated mass varies significantly (between 9.32--11.11$\times 10^{14} \rm{M_{\odot}}$ for the X-ray mass and between 4.01--6.32$\times 10^{14} \rm{M_{\odot}}$ for the lensing mass). Their mass estimates use $h_{70}=1$. Several other analyses of \emph{Chandra} data produce comparable mass estimates (e.g. \citealt{schmidt_A611}, \citealt{MORANDI_1_Abell 611}, \citealt{MORANDI_2_Abell 611} and \citealt{SANDERSON_Abell 611}). 

\cite{romano_a611} perform a weak lensing analysis of Abell~611 using data from the Large Binocular Telescope. With an NFW profile they estimate $M_{\rm{T},r200}$ = 4--7$\times 10^{14} \rm{M_{\odot}}$ and $r_{200}= 1400-1600$kpc, for $h_{70}=1$. These are in agreement with the values obtained from \emph{Subaru} weak lensing observations by Okabe et al.

Using GMRT observations \cite{GMRT_HALO} concluded that Abell 611 has no radio halo at 610MHz.  Abell 611 has also previously been observed in the SZ at 15~GHz by \cite{RT_Abell 611} and Zwart et al. (2010), and at 30~GHz by \cite{OCRO_Abell 611}, \cite{bona_chandra} and LaRoque et al.

From our analysis of the AMI SA observations of Abell 611 presented in this paper we find that $M_{\rm{T},r200}$ = 4.0$^{+0.3}_{-0.4}$$\times 10^{14}\rm{M_{\odot}}$. We note that the mass obtained is significantly smaller than the result given in Zwart et al. (2010); however, their  $M_{\rm{T}}$ estimates are biased high. The bias occurs because they used a low-radius X-ray temperature as a constant temperature throughout the cluster, as is explained by them and in Olamaie et al. (2010). The SZ maps presented in this paper are similar to those in Zwart et al. (2010); both sets of observations indicate that the cluster is extended in the NW direction. However, the analysis presented in this paper differs from that by Zwart et al. (2010) who sample from temperature and $M_{g,r200}$ and derive $M_{T,r200}$ under the additional assumption of hydrostatic equilbrium. Instead we sample from $M_{\rm{T},r200}$ and $f_{g,r200}$ and calculate $T$ using the M-T scaling relation given in Rodr{\'i}guez-Gonz{\'a}lvez et al. (2010). The differences between these two models are described in detail by Olamaie et al. who demonstrate that the mass estimated using the technique in this paper produces a more reliable value and that the Zwart et al. (2010) analysis underestimates the values for $M_{g,r200}$ and $f_{g,r200}$. The values of $\beta$ and $r_{c}/h_{100}^{-1}$kpc presented here agree with those in the Zwart et al. (2010) analysis.

We find no significant contamination from radio sources and detect the cluster with a high signal-to-noise ratio. A comparison of the SZ-effect image and the \emph{Chandra} map shows that the centres of the SZ and X-ray emission are coincident.

\subsection{Abell 773}\label{sec:A773}

\begin{figure*}
\centerline{Abell 773}
\centerline{\includegraphics[width=7.5cm,height= 7.5cm,clip=,angle=0.]{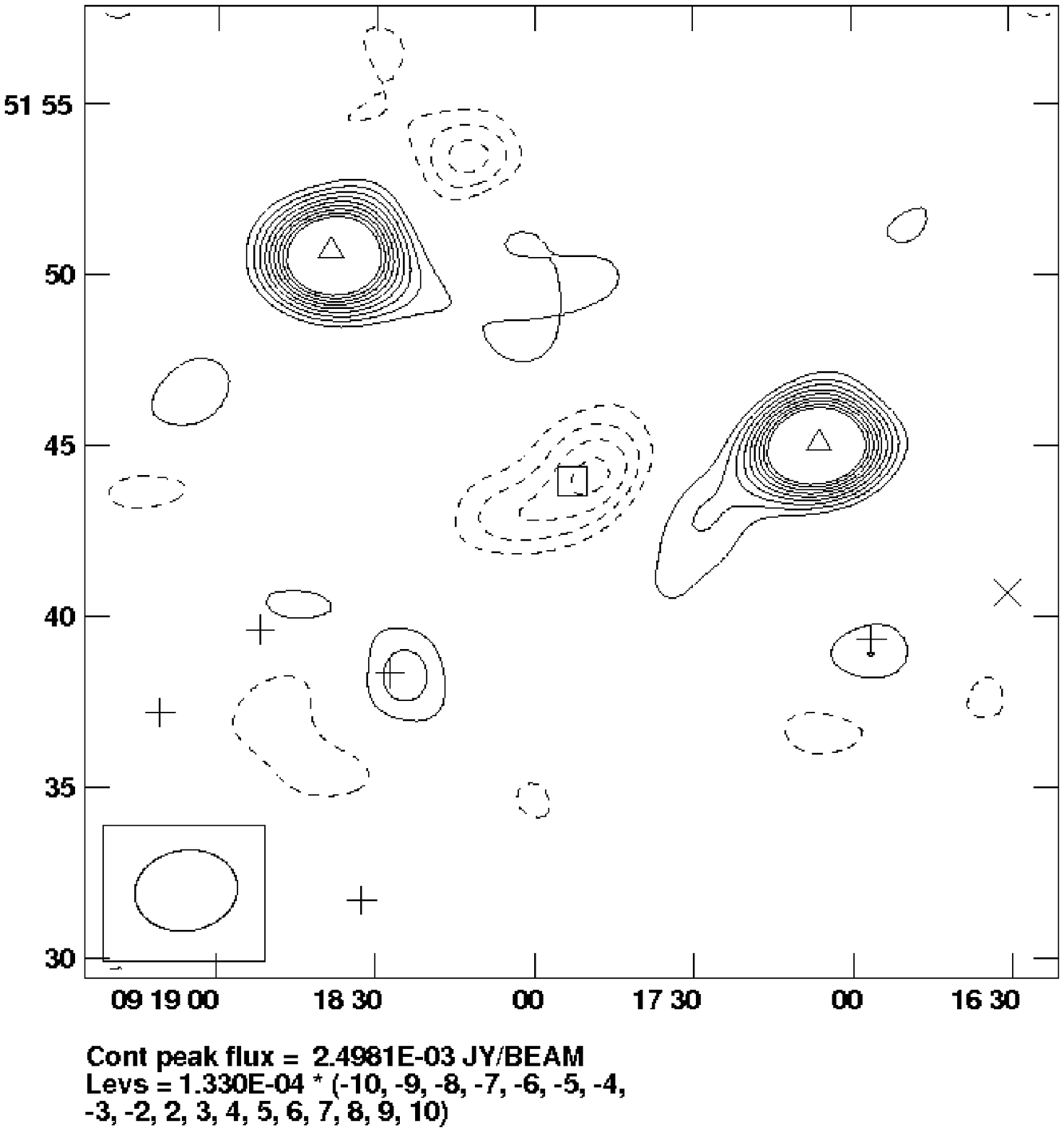}\qquad\includegraphics[width=7.5cm,height= 7.5cm,clip=,angle=0.]{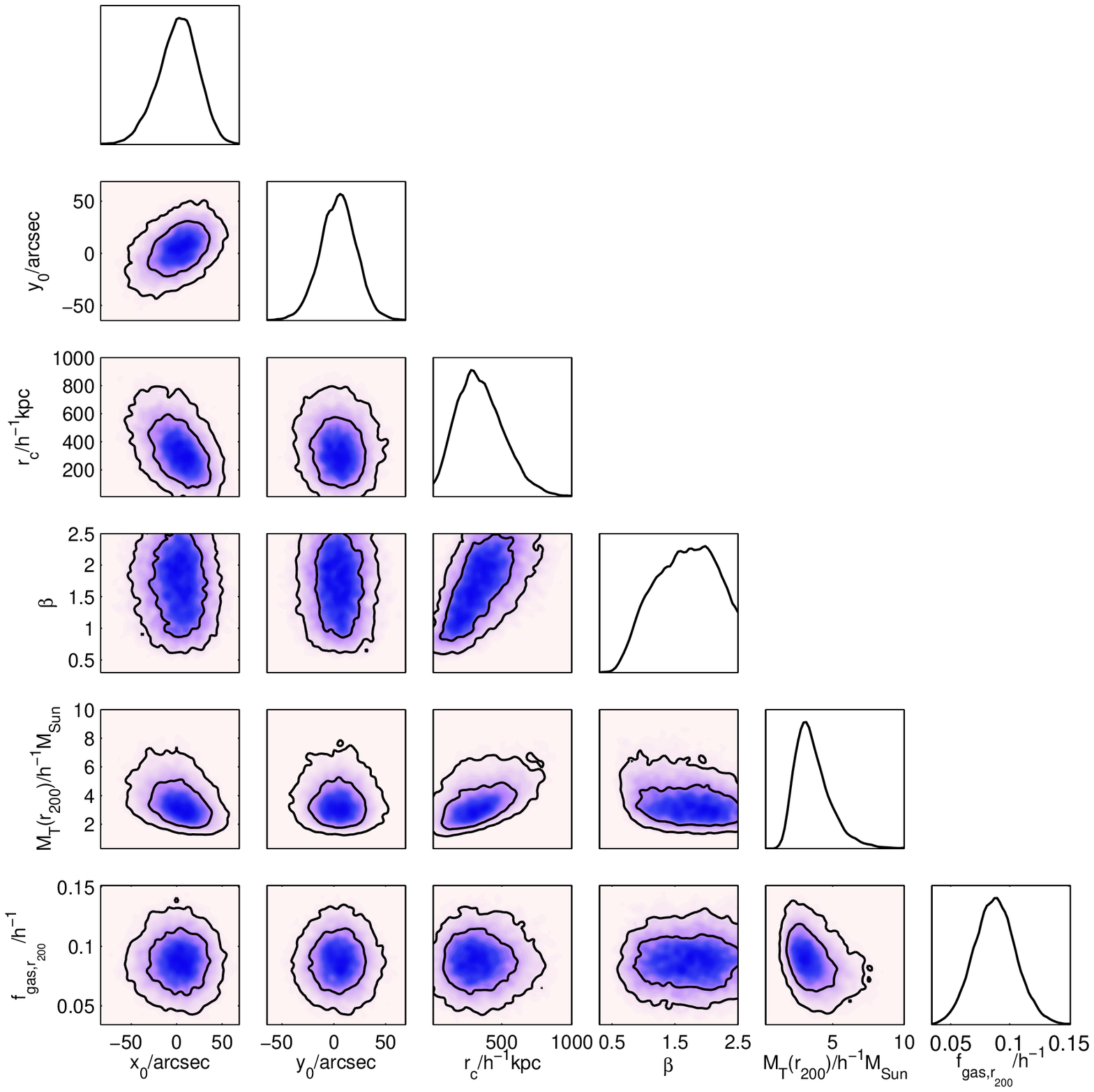}}
 \centerline{\includegraphics[width=7.5cm,height= 7.5cm,clip=,angle=0.]{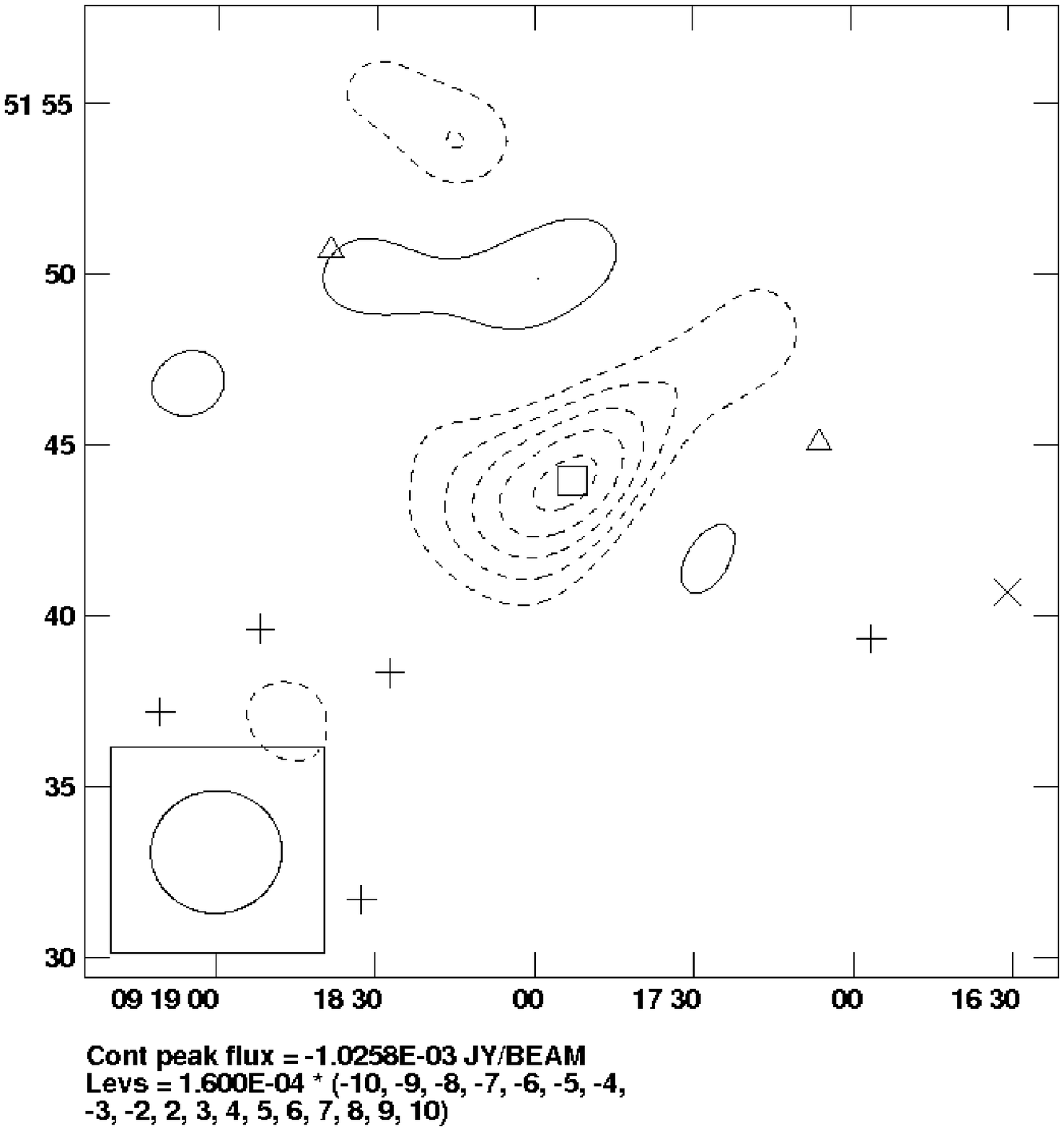}\qquad\includegraphics[width=7.5cm,height= 7.5cm,clip=,angle=0.]{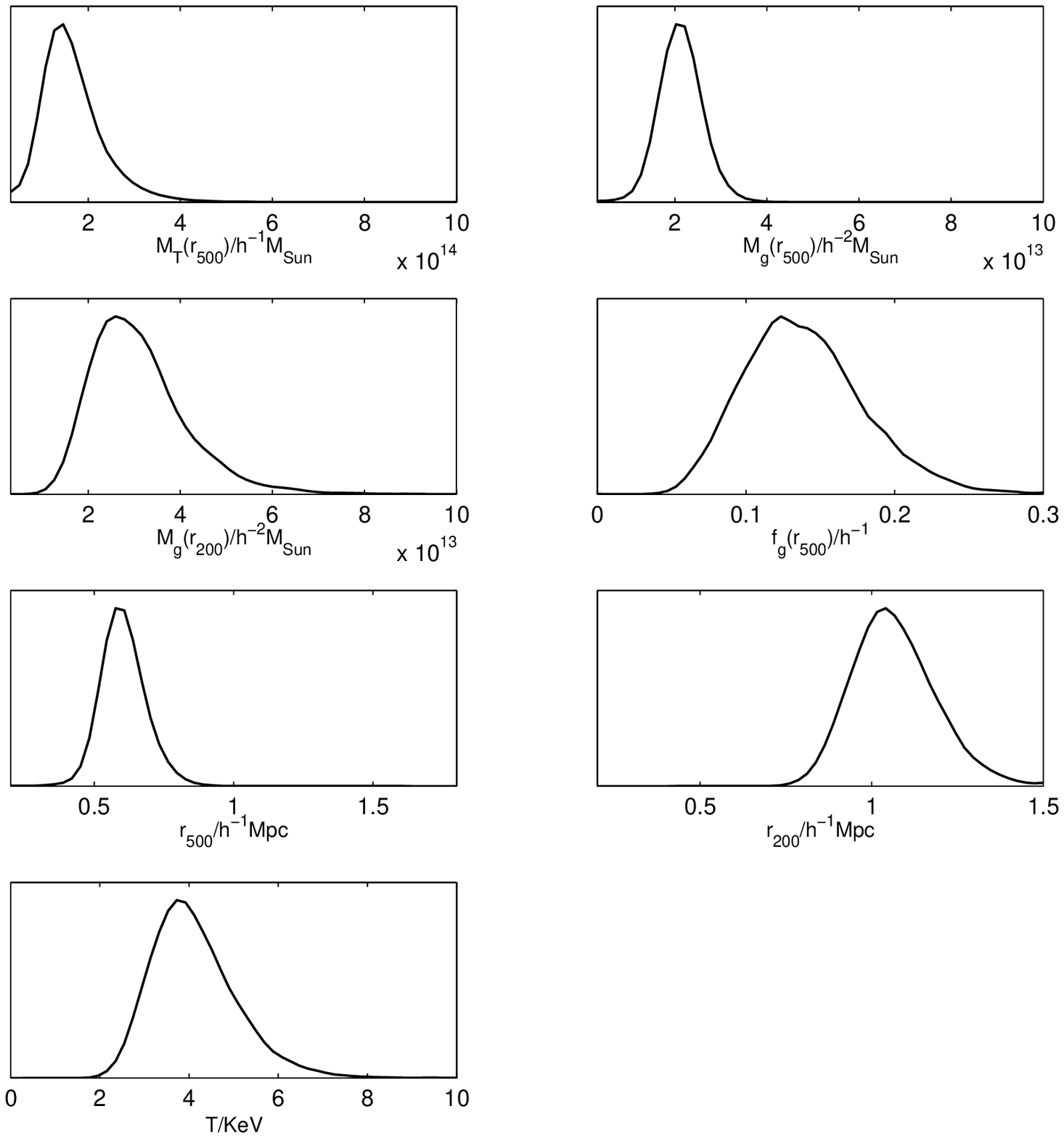}}
 \centerline{
\includegraphics[width=7.5cm,height= 6.5cm,clip=,angle=0.]{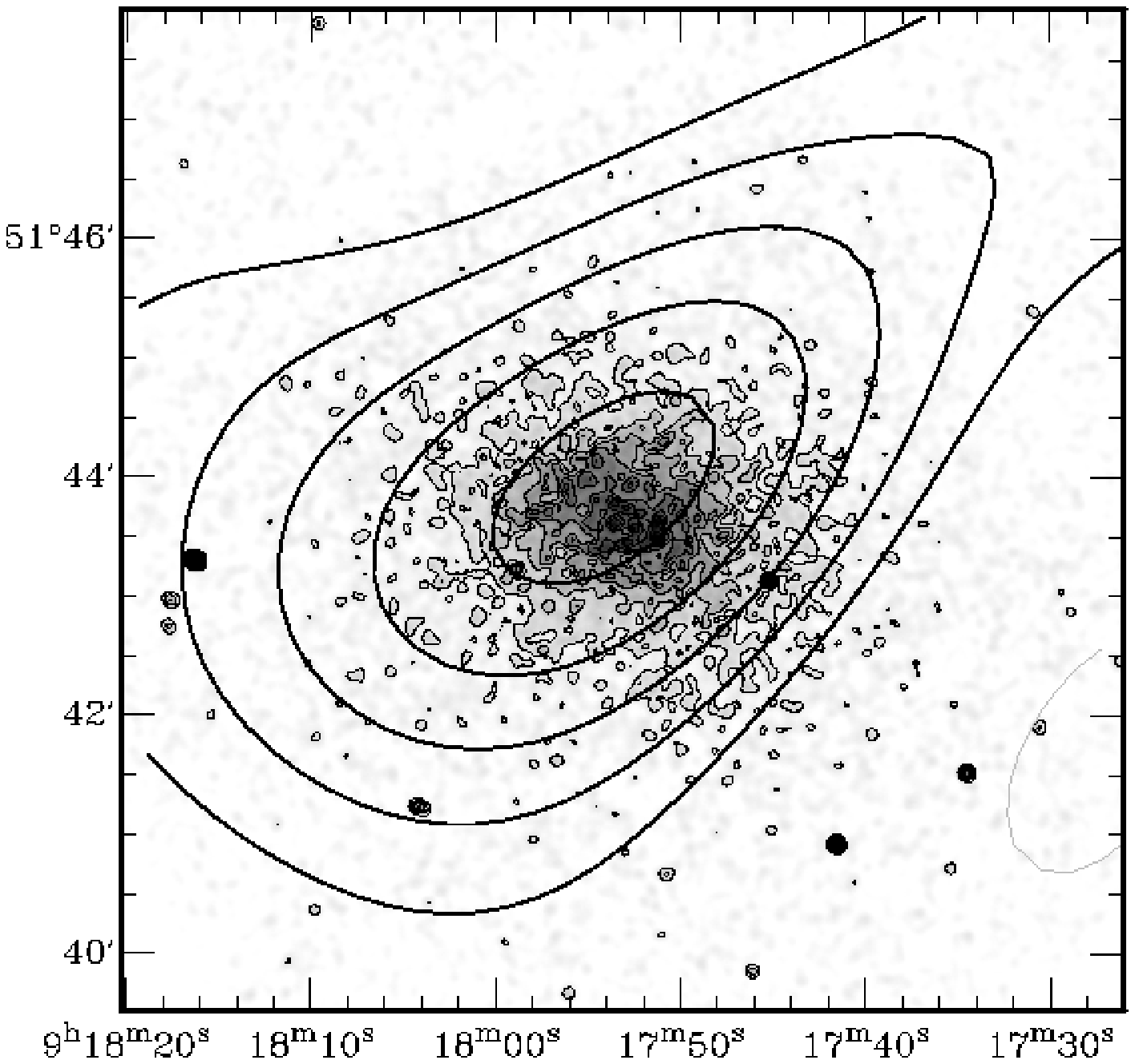}}
\caption{The top left image shows the SA map before subtraction, the map in the middle left has had the sources removed, the top right panel shows the cluster parameters that we sample from in our Bayesian analysis and the middle right plot presents several cluster parameters derived from our sampling parameters. The image at the bottom shows the \emph{Chandra} X-ray map overlayed with SA contours.}
\label{fig:A773}
\end{figure*}

In Figure \ref{fig:A773} we show the AMI SA maps of Abell 773, a \emph{Chandra} X-ray map and the cluster parameters derived from our analysis. The SZ effect associated with Abell 773 has been observed several times (\citealt{1993MNRAS.265L..57G}, \citealt{1996ApJ...456L..75C}, \citealt{bona_chandra}, LaRoque et al. and \citealt{2003MNRAS.341..937S}). Most recently, Zwart et al. (2010) observed the cluster and found a cluster mass of $M_{\rm{T,r200}}$ = 1.9$^{+0.3}_{-0.4}$$\times 10^{15}\rm{M_{\odot}}$ using $h_{70}=1$; however, their  $M_{\rm{T}}$ estimates are biased high -- see Section \ref{sec:A611}.

Inspection of a 10$\arcmin \times 10\arcmin$ region of the Sloan Digital Sky Survey (SDSS%
\footnote{Funding for the SDSS and SDSS-II has been provided by the Alfred P. Sloan Foundation, the Participating Institutions, the National Science Foundation, the U.S. Department of Energy, the National Aeronautics and Space Administration, the Japanese Monbukagakusho, the Max Planck Society, and the Higher Education Funding Council for England. The SDSS Web Site is http://www.sdss.org/.}
) centred on Abell 773 reveals a complex galaxy distribution with some EW extension. Our observations support this extension, but there is no detailed correspondence between the galaxy and gas distributions. The \emph{Chandra} observations appear to show little if any such extension. We find no significant contamination from radio sources and detect the cluster with a high signal to noise ratio. 

For this cluster, \cite{2007A&A...467...37B} present an intensive study of the optical data from the Telescopio Nazionale Galileo (TNG) telescope and X-ray data from the \emph{Chandra} data archive. They find two peaks in the velocity distribution of the cluster members which are separated by 2$\arcmin$ along the E-W direction. Two peaks can also be seen in the X-ray, although these are along the NE-SW direction. Barrena et al. estimate the virial mass of the entire system to be $M_{vir}$ = 1.2-2.7$\times 10^{15}h_{70}^{-1}\rm{M_{\odot}}$. \cite{Gio_A773_halo} reported the existence of a radio halo in Abell~773. This feature, typical of cluster mergers, was confirmed with 1.4~GHz VLA observations by \cite{A773_HALO}. 
\cite{zhang_xmm} used \emph{XMM-Newton} to study Abell~773 and found $M_{500}$ = 8.3 $\pm$2.5 $\times 10^{14}\rm{M_{\odot}}$, where $r_{500}$ = 1.33\,Mpc; they assumed isothermality, spherical symmetry and $h_{70}=1$. \cite{A773_temp} present a \emph{Chandra} temperature map and an X-ray image of Abell~773; they estimate a mean temperature of 7.5$\pm$0.8\,keV within a radius of 800\,kpc ($h_{70}=1$).

\subsection{Abell 781}

\begin{figure*}
\centerline{A781}
\centerline{\includegraphics[width=7.5cm,height= 7.5cm,clip=,angle=0.]{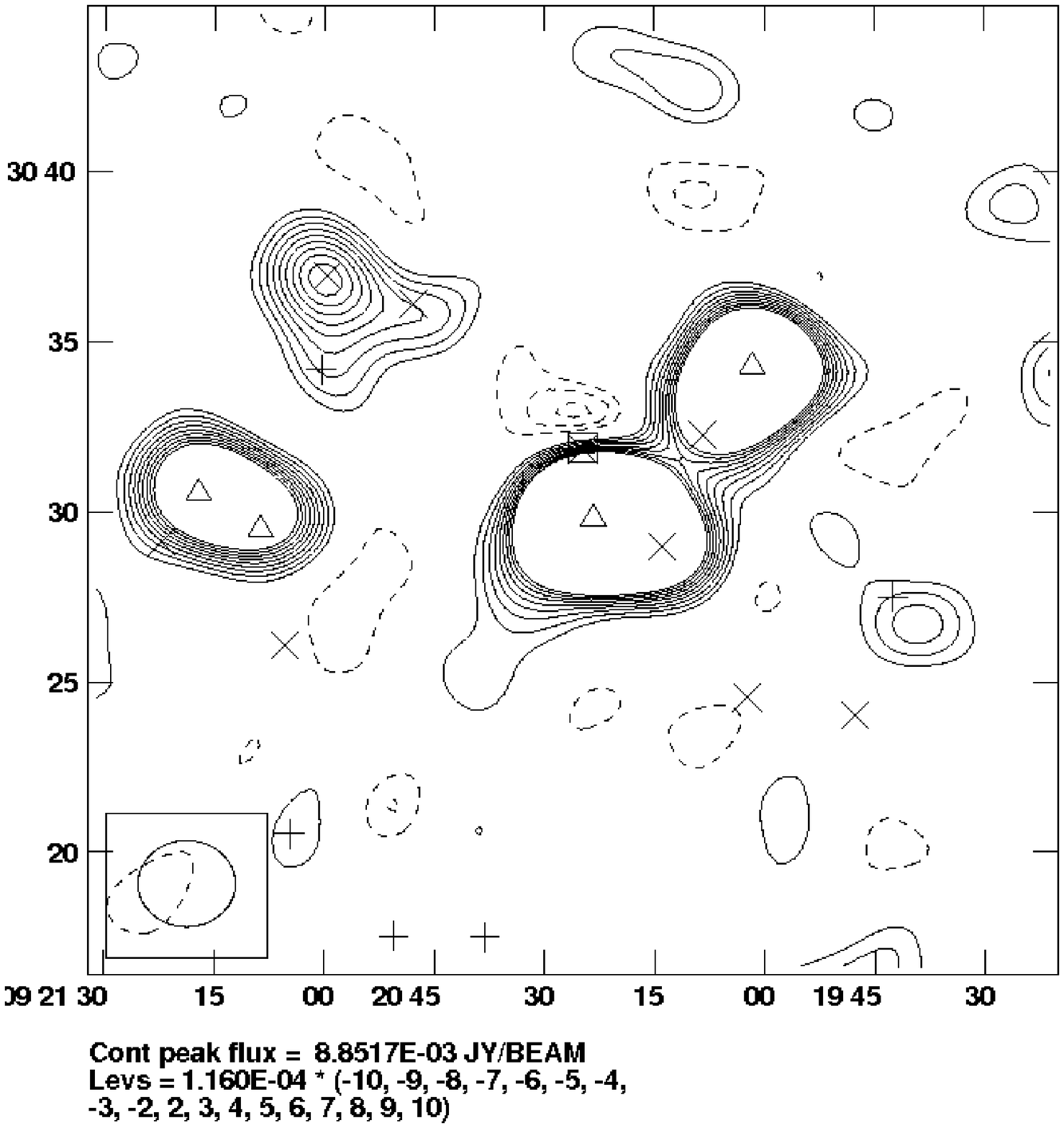}\qquad\includegraphics[width=7.5cm,height= 7.5cm,clip=,angle=0.]{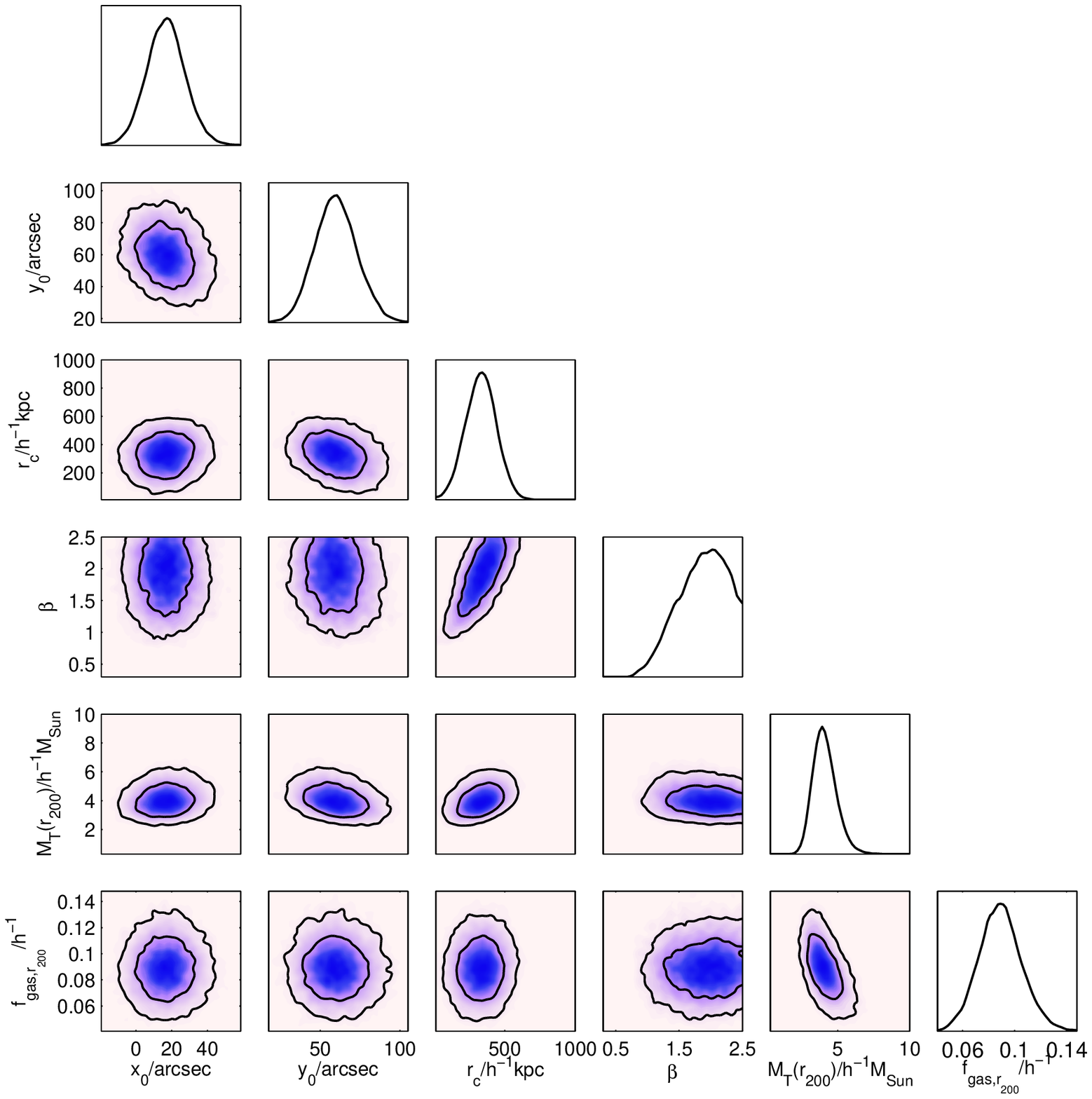}}
 \centerline{\includegraphics[width=7.5cm,height= 7.5cm,clip=,angle=0.]{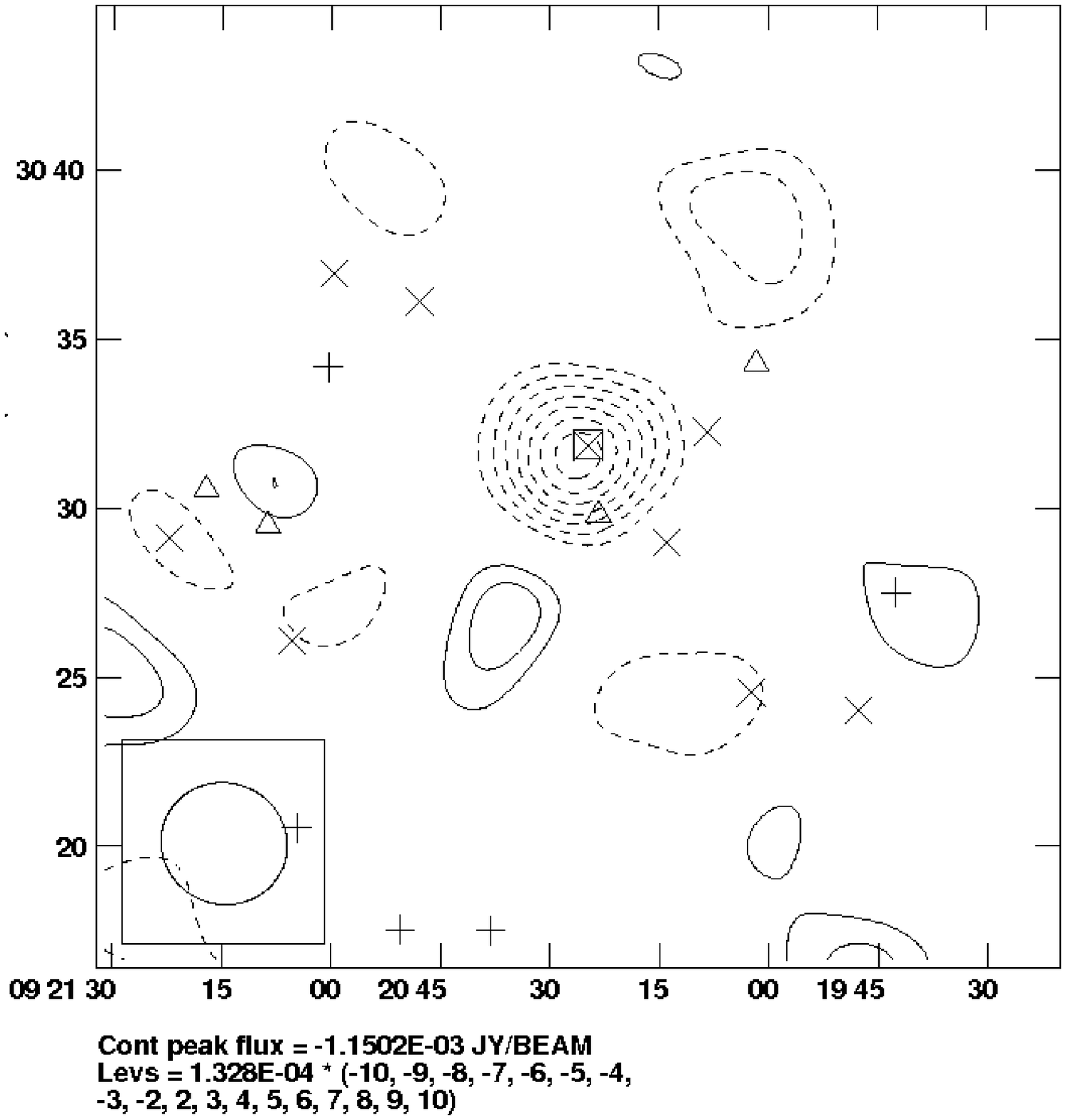}\qquad\includegraphics[width=7.5cm,height= 7.5cm,clip=,angle=0.]{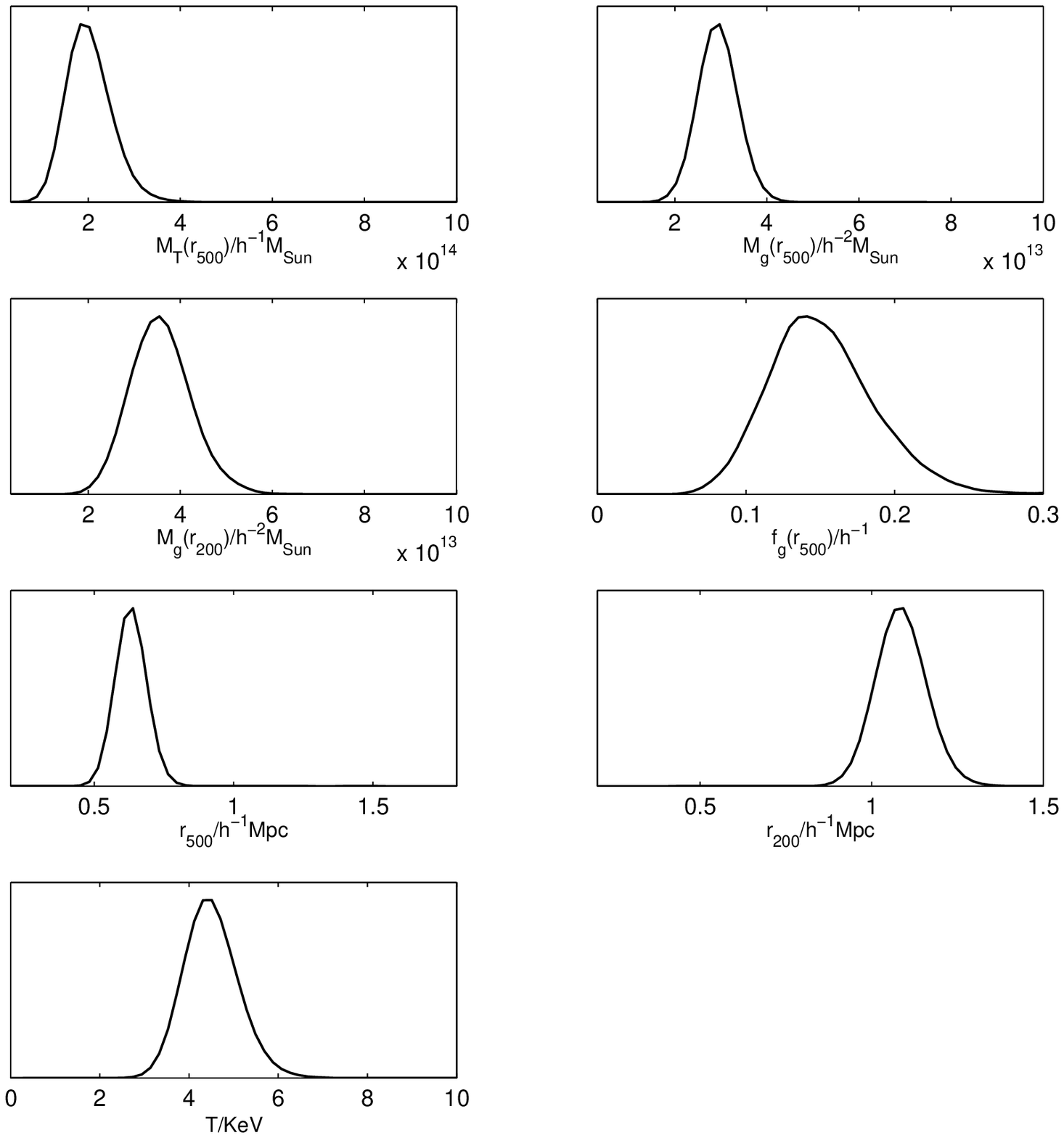}}
 \centerline{
\includegraphics[width=7.5cm,height= 6.5cm,clip=,angle=0.]{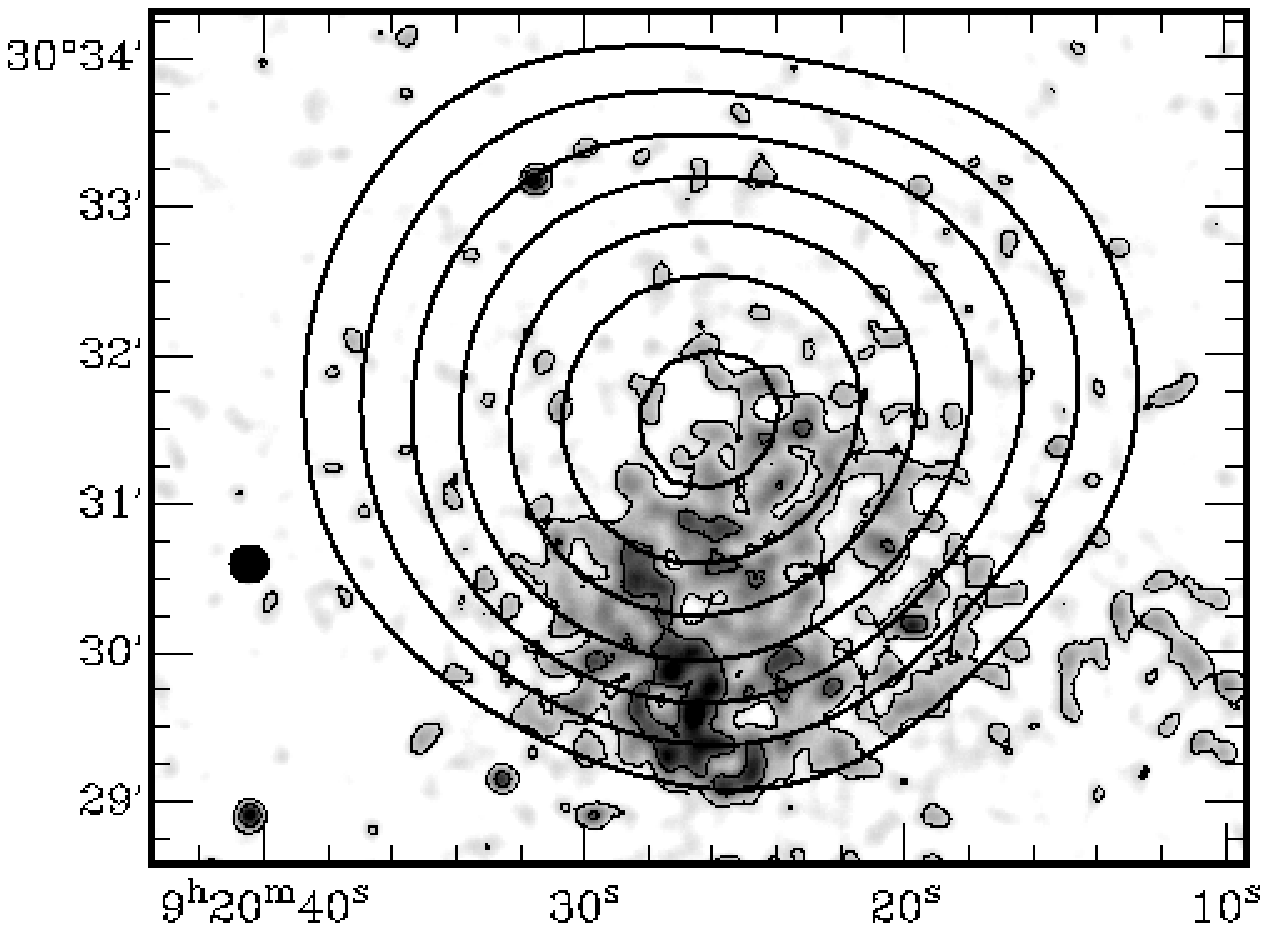}}
\caption{The top left image shows the SA map before subtraction, the map in the middle left has had the sources removed, the top right panel shows the cluster parameters that we sample from in our Bayesian analysis and the middle right plot presents several cluster parameters derived from our sampling parameters. The image at the bottom shows the \emph{Chandra} X-ray map overlayed with SA contours.}
\label{fig:A781}
\end{figure*}

AMI SA maps, derived parameters and a \emph{Chandra} observation of the Abell~781 cluster are presented in Figure \ref{fig:A781}. From X-ray observations with \emph{Chandra} and \emph{XMM-Newton} (\citealt{Abell 781_XMM}) it is apparent that Abell~781 is a complex cluster system. The main cluster is surrounded by three smaller clusters, two to the East of the main cluster and one to the West. They estimate the  mass of the clusters assuming a NFW matter density profile; the results indicate that the cluster mass of Abell~781 within $r_{500}$ is $5.2^{+0.3}_{-0.7}$$\times$$10^{14}\rm{M_{\odot}}$ from \emph{XMM-Newton} and\emph{Chandra} X-ray observations or $2.7^{+1.0}_{-0.9}$$\times$$10^{14}\rm{M_{\odot}}$ from the Kitt Peak Mayall 4-m telescope lensing observations (where $r_{500}$ is  $1.09^{+0.04}_{-0.04}$ and $0.89^{+0.10}_{-0.12}$ respectively). Alternatively, Zhang et al. use XMM Newton observations to estimate $M_{500}$ = 4.5 $\pm$1.3 $\times 10^{14}\rm{M_{\odot}}$, where $r_{500}$ = 1.05Mpc, assuming isothermality and spherical symmetry. Both Zhang et al. and Sehgal et al. use $h_{70}=1$.

The main cluster of Abell 781 is also known to contain a diffuse peripheral source at 610MHz; this was observed with the GMRT by \cite{GMRT_HALO}.

\subsection{Abell 1413}

\begin{figure*}
\centerline{Abell 1413}
\centerline{\includegraphics[width=7.5cm,height= 7.5cm,clip=,angle=0.]{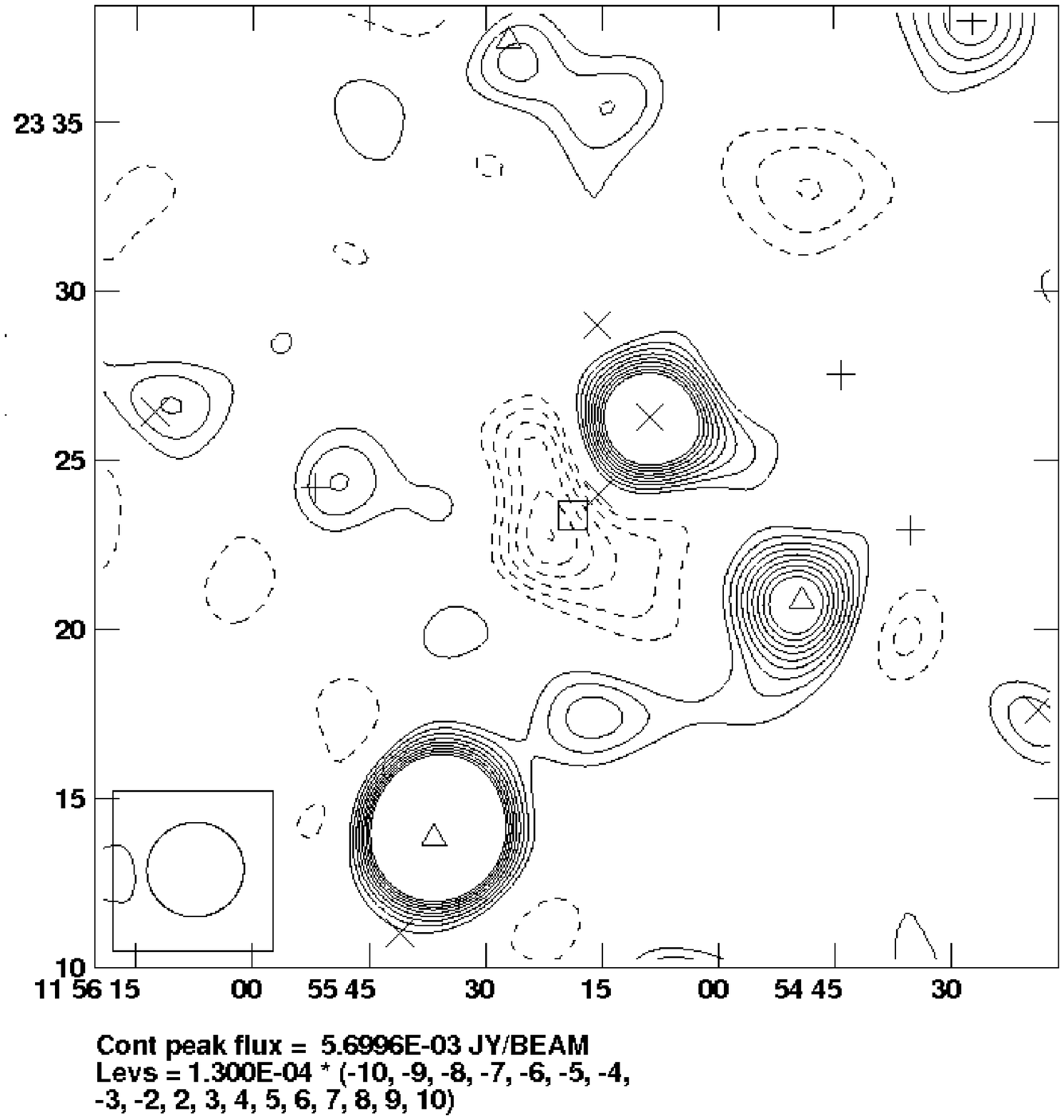}\qquad\includegraphics[width=7.5cm,height= 7.5cm,clip=,angle=0.]{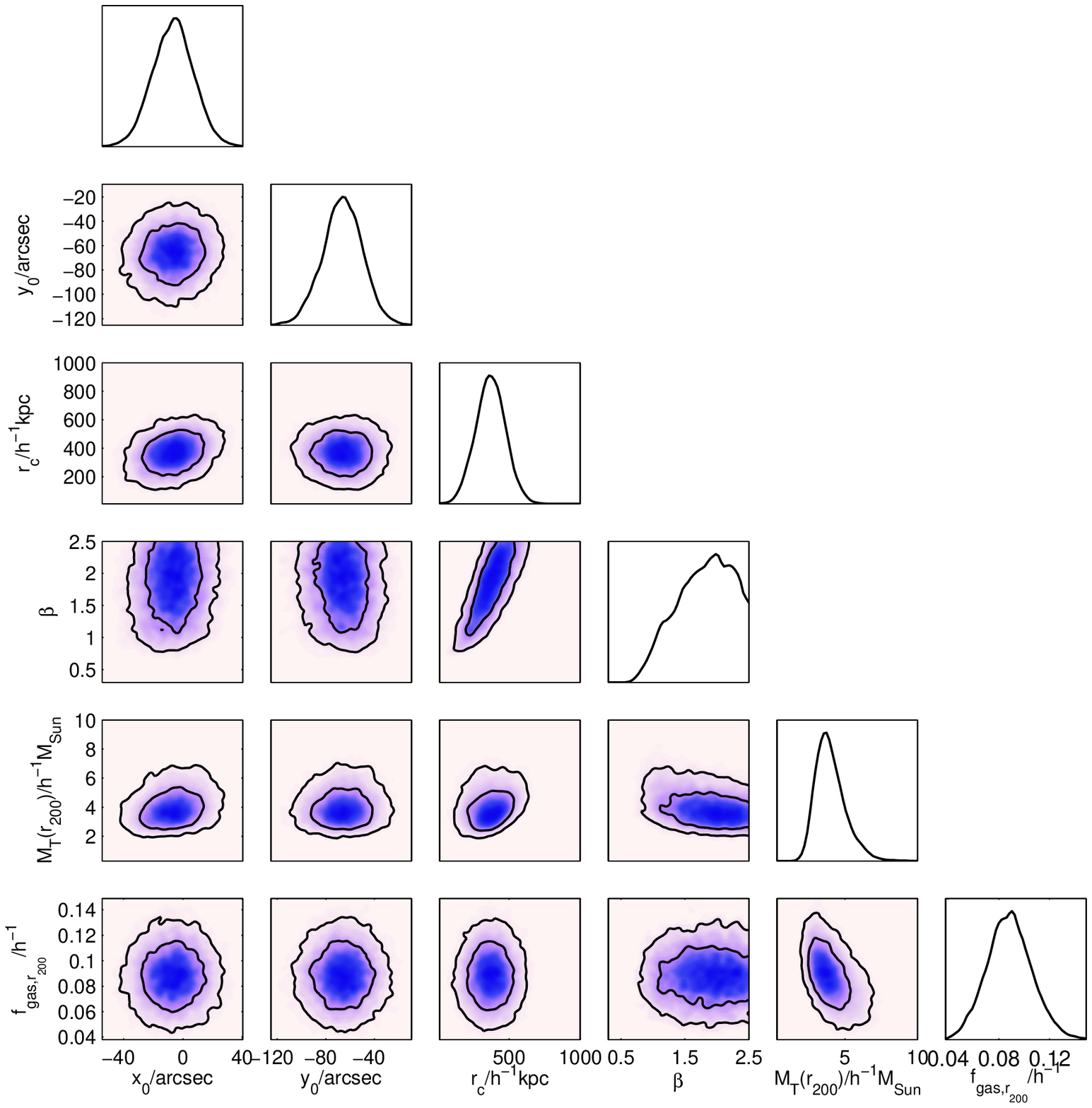}}
 \centerline{\includegraphics[width=7.5cm,height= 7.5cm,clip=,angle=0.]{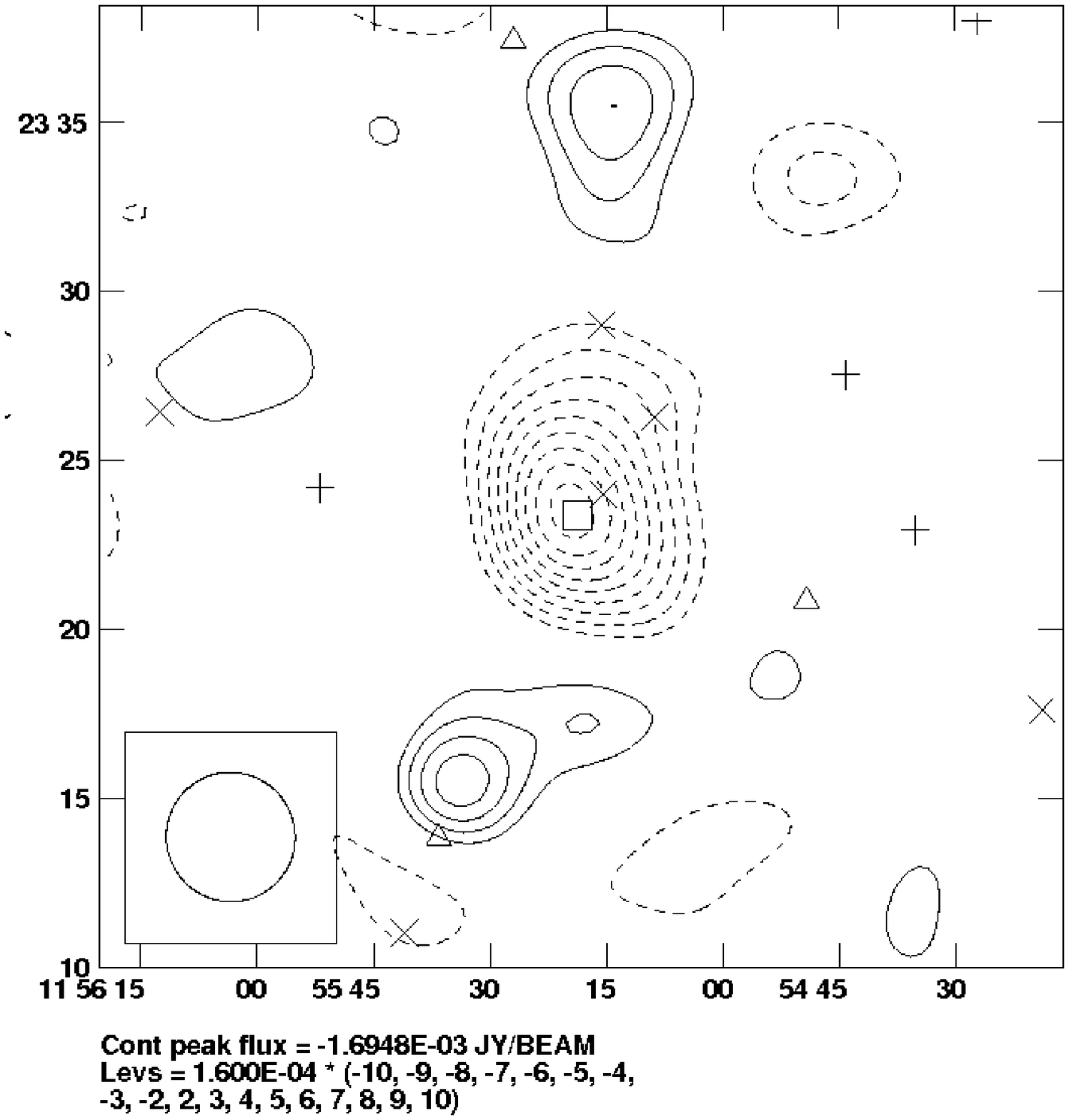}\qquad\includegraphics[width=7.5cm,height= 7.5cm,clip=,angle=0.]{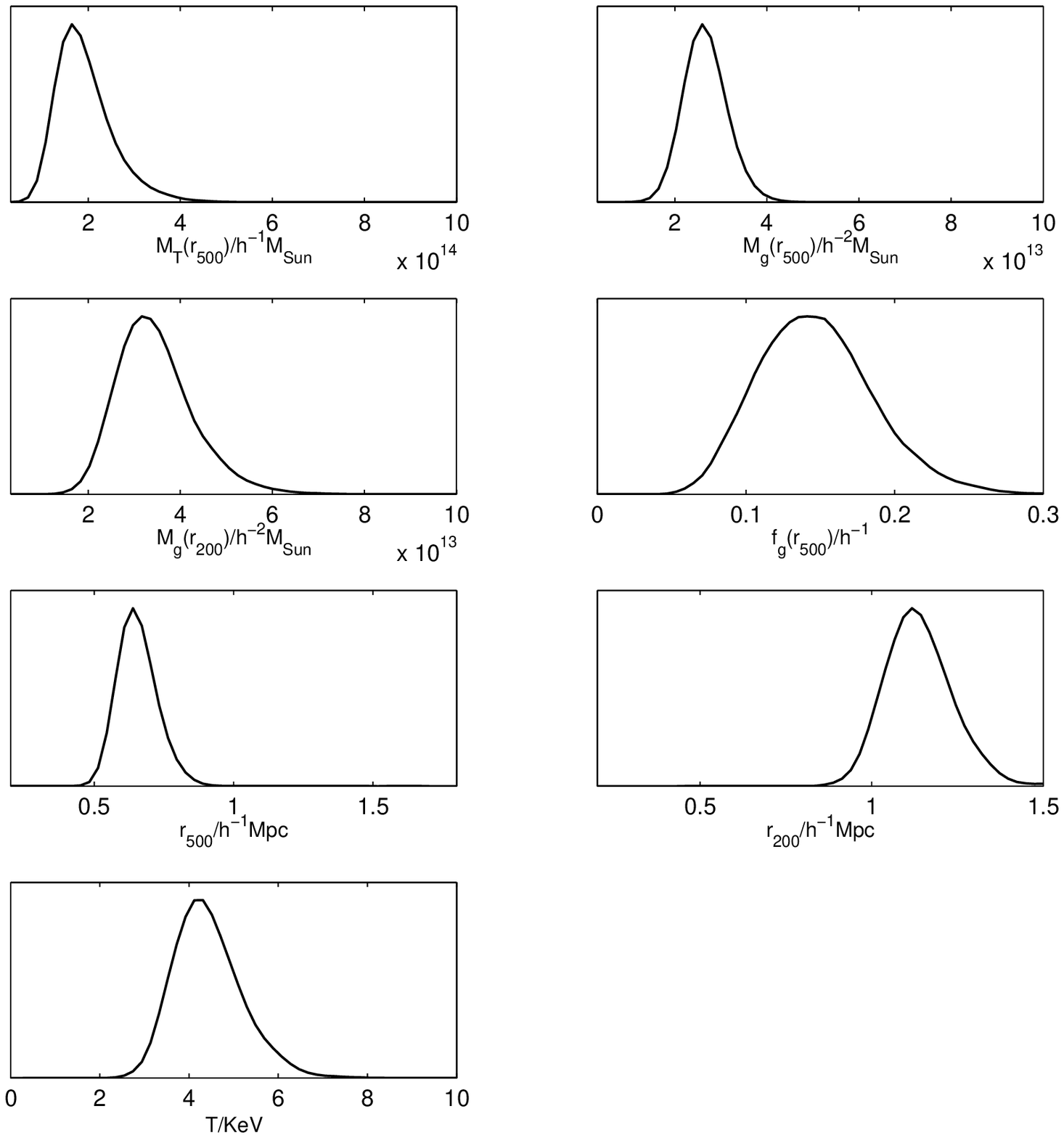}}
 \centerline{
\includegraphics[width=7.5cm,height= 6.5cm,clip=,angle=0.]{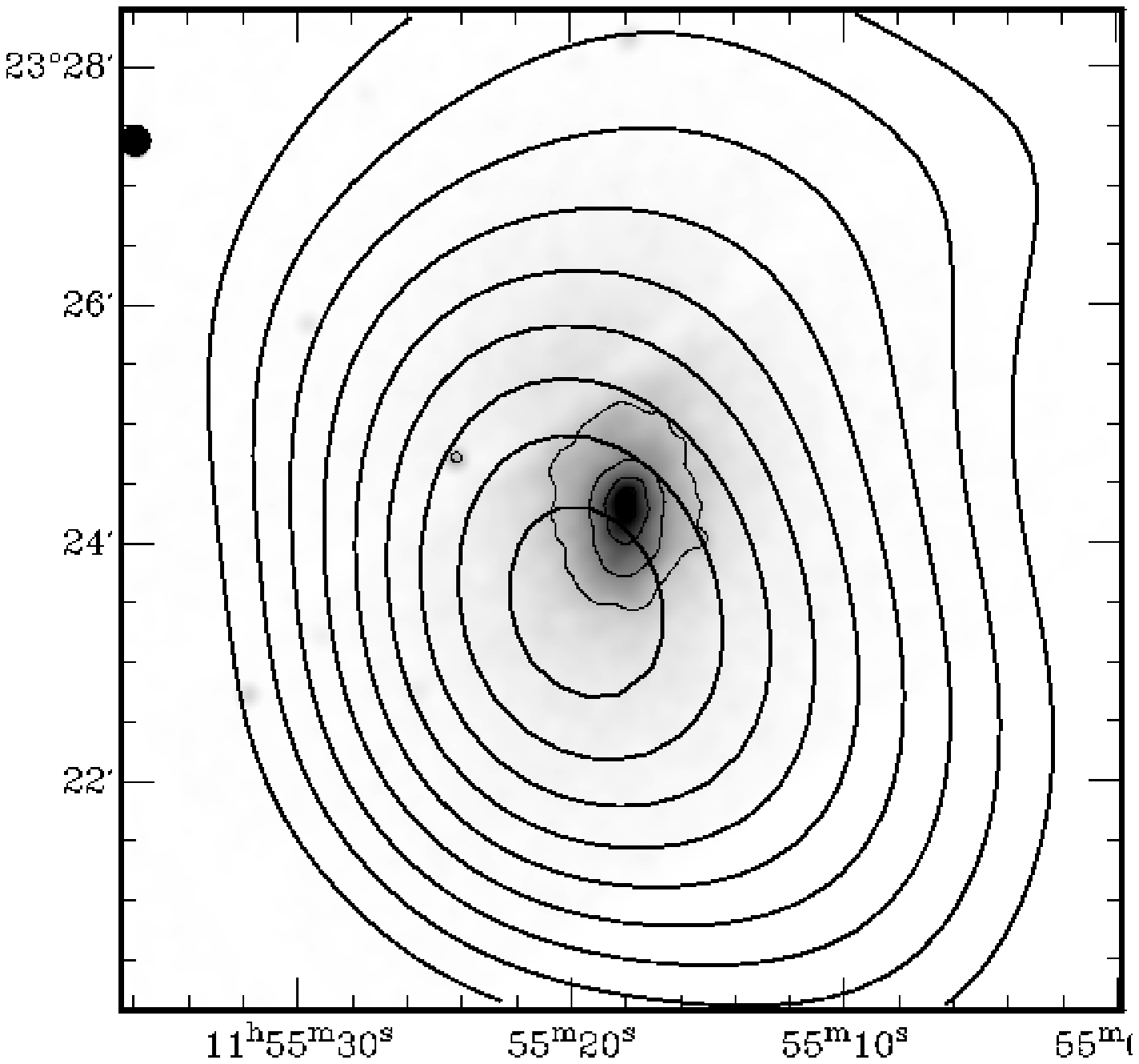}}
\caption{The top left image shows the SA map before subtraction, the map in the middle left has had the sources removed, the top right panel shows the cluster parameters that we sample from in our Bayesian analysis and the middle right plot presents several cluster parameters derived from our sampling parameters. The image at the bottom shows the \emph{Chandra} X-ray map overlayed with SA contours.}
\label{fig:A1413}
\end{figure*}

We present the SA maps before and after source subtraction in Figure \ref{fig:A1413}. We also show the derived cluster parameters and we overlay our SZ image on a \emph{Chandra} X-ray map. Abell 1413 has been observed in the X-ray by \emph{XMM-Newton} (e.g. \citealt{XMM-A1413}), \emph{Chandra} (e.g. \citealt{Chandra-A1413} and \citealt{bona_chandra}) and most recently by the low background \emph{Suzaku} satellite (\citealt{Suzaka_A1413}); SZ images have been made of Abell 1413 with the Ryle Telescope at 15~GHz (\citealt{RT_A1413}) and with OVRO/BIMA at 30~GHz (LaRoque et al. and \citealt{bona_chandra}). These analyses indicate that Abell~1413 is a relaxed cluster with no evidence of recent merging despite its elliptical morphology. Between the X-ray observations there is good agreement in the temperature and density profiles of the cluster out to half the virial radius. Hoshino et al. measure the variation of the electron temperature with radius, finding a temperature of 7.5keV at the centre and 3.5keV at $r_{200}$.  They assume spherical symmetry, an NFW density profile and hydrostatic equilibrium to calculate $M_{\rm{T},r200}$ = 6.6$\pm$2.3$\times 10^{14}h_{70}^{-1}\rm{M_{\odot}}$; where $r_{200}$ = $2.24h^{-1}_{70}$Mpc. Zhang et al. use \emph{XMM-Newton} to study Abell~1413 and find $M_{500}$ = 5.4 $\pm$1.6 $\times 10^{14}\rm{M_{\odot}}$, where $r_{500}$ = 1.18Mpc; they assume isothermality, spherical symmetry and $h_{70}=1$.

Recent VLA observations (\citealt{VLA_A1413}) indicate that there is diffuse 1.4-GHz emission associated with the cluster -- this may be due to a mini-halo around the cluster. 

From the SA observations we detect the SZ decrement at high significance. We determine the cluster mass to be $M_{\rm{T},r200}$ = 4.0$^{+0.3}_{-0.6}$$\times 10^{14}h_{100}^{-1}\rm{M_{\odot}}$ and $r_{200}$ = 1.14$^{+0.4}_{-0.5}$$h_{100}^{-1}$kpc; both these values are comparable with the Hoshino et al. results.

The source environment around the cluster at 16\,GHz is reasonable: the brightest source is 14mJy (11:55:36.63  +23:13:50.1), but this is 700$''$ from the cluster X-ray centre. After this bright source is subtracted from our data we are left with a low level of residual flux density on the map; it is unlikely that this residual flux significantly contaminates our cluster detection or parameters. Both the X-ray map and the SZ image indicate that the cluster is elliptical and extended in the N-S direction. 

\subsection{Abell 1758}
 
\begin{figure*}
\centerline{Abell 1758A}
\centerline{\includegraphics[width=7.5cm,height= 7.5cm,clip=,angle=0.]{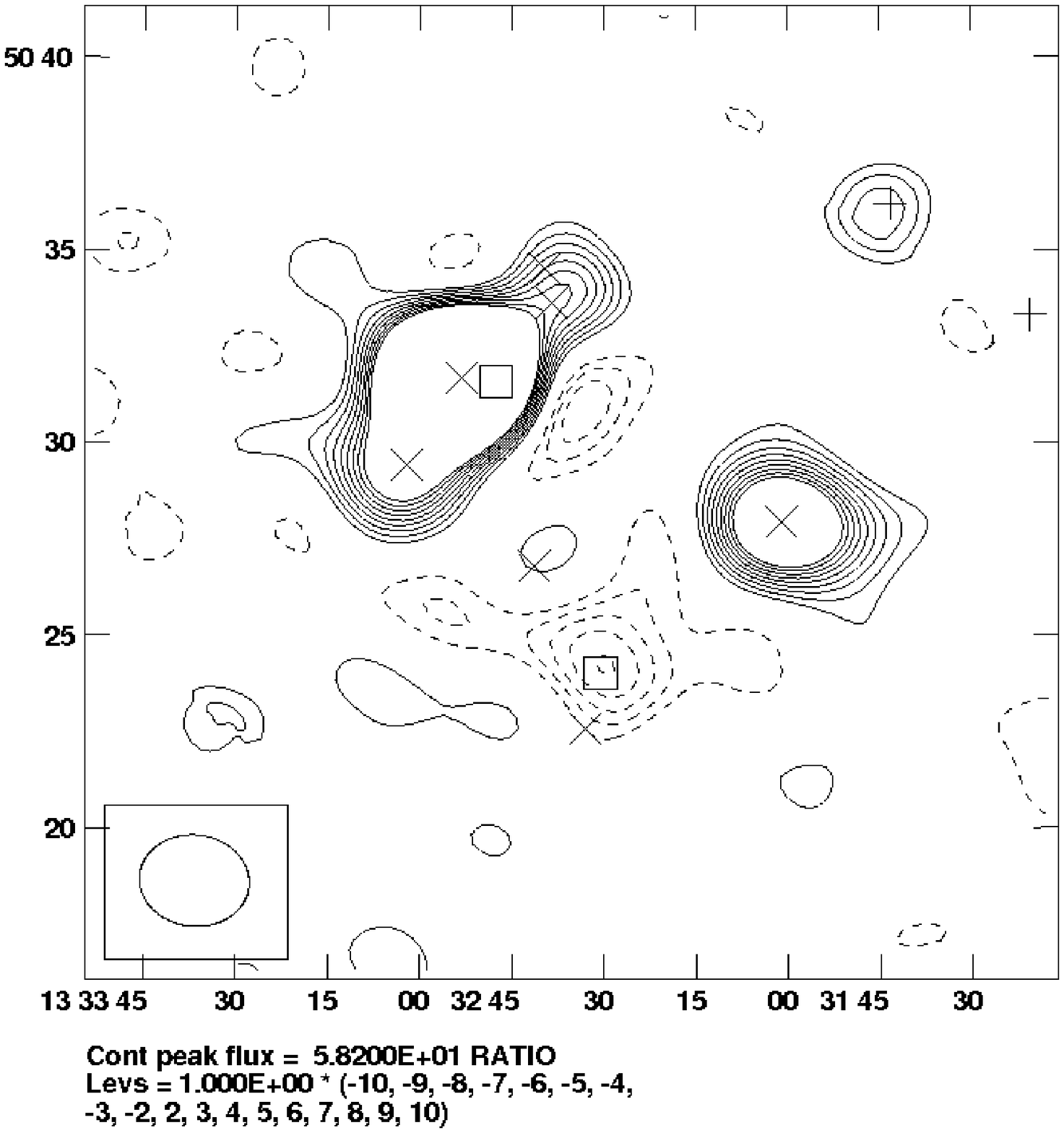}\qquad\includegraphics[width=7.5cm,height= 7.5cm,clip=,angle=0.]{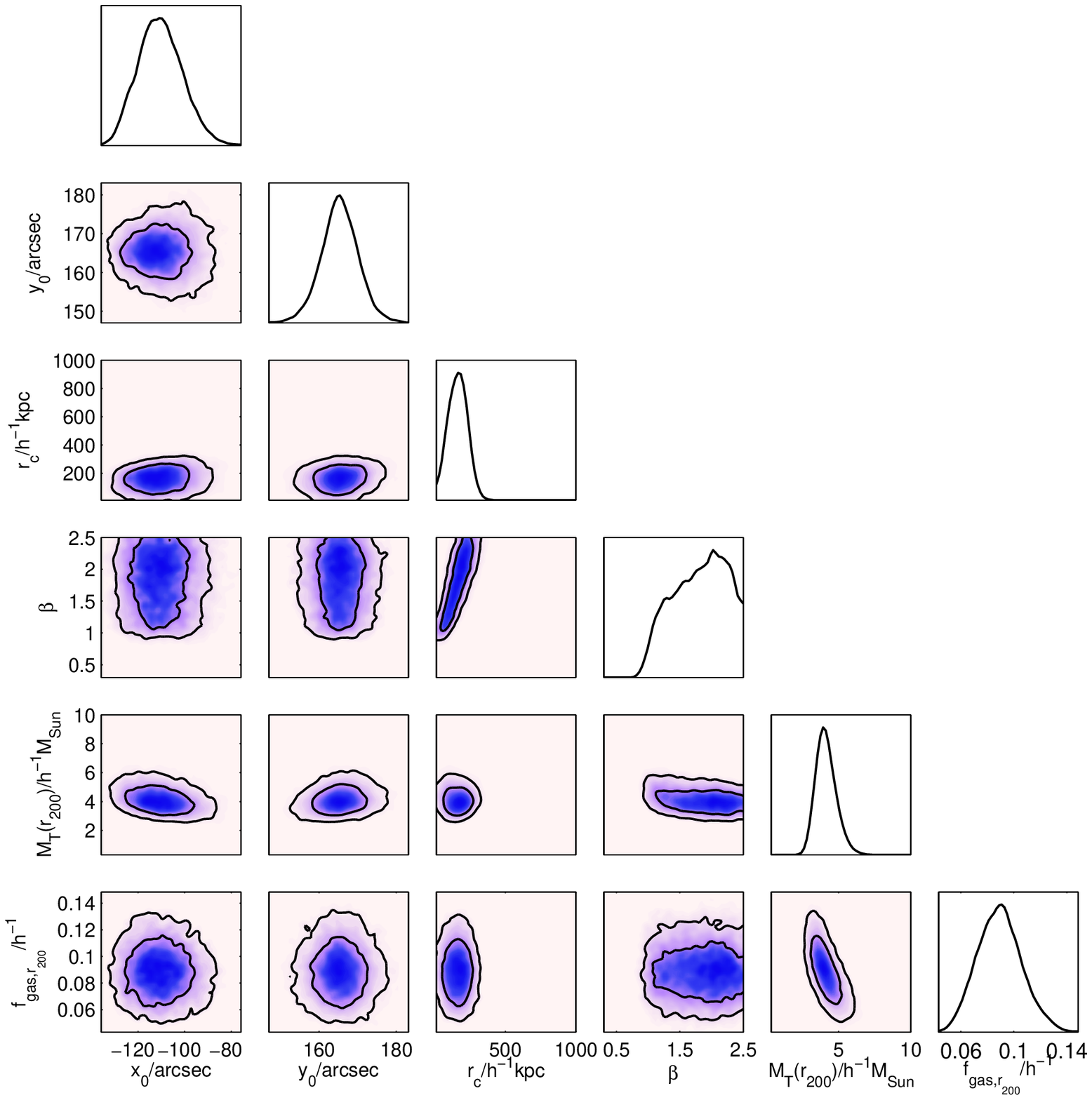}}
 \centerline{\includegraphics[width=7.5cm,height= 7.5cm,clip=,angle=0.]{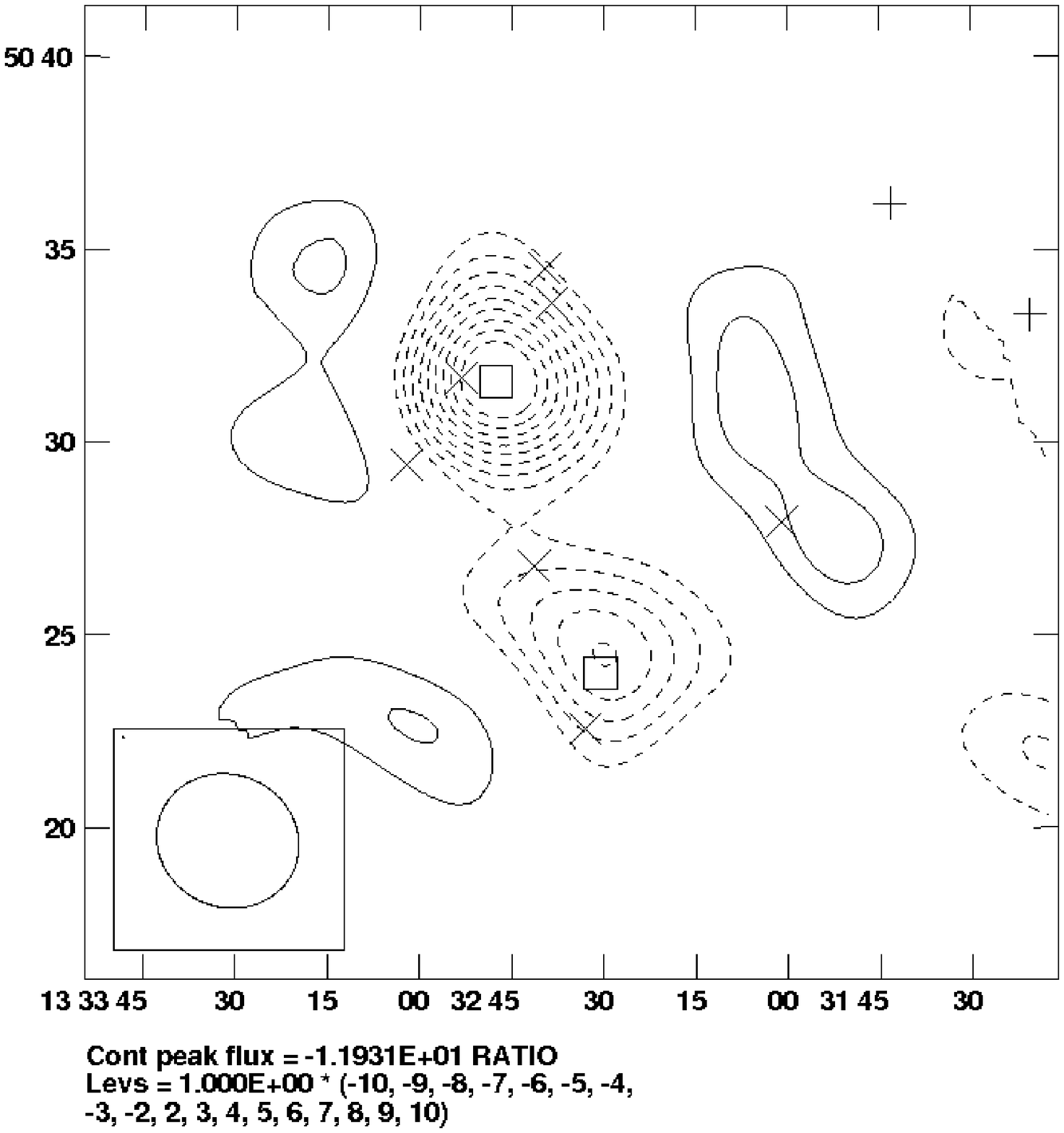}\qquad\includegraphics[width=7.5cm,height= 7.5cm,clip=,angle=0.]{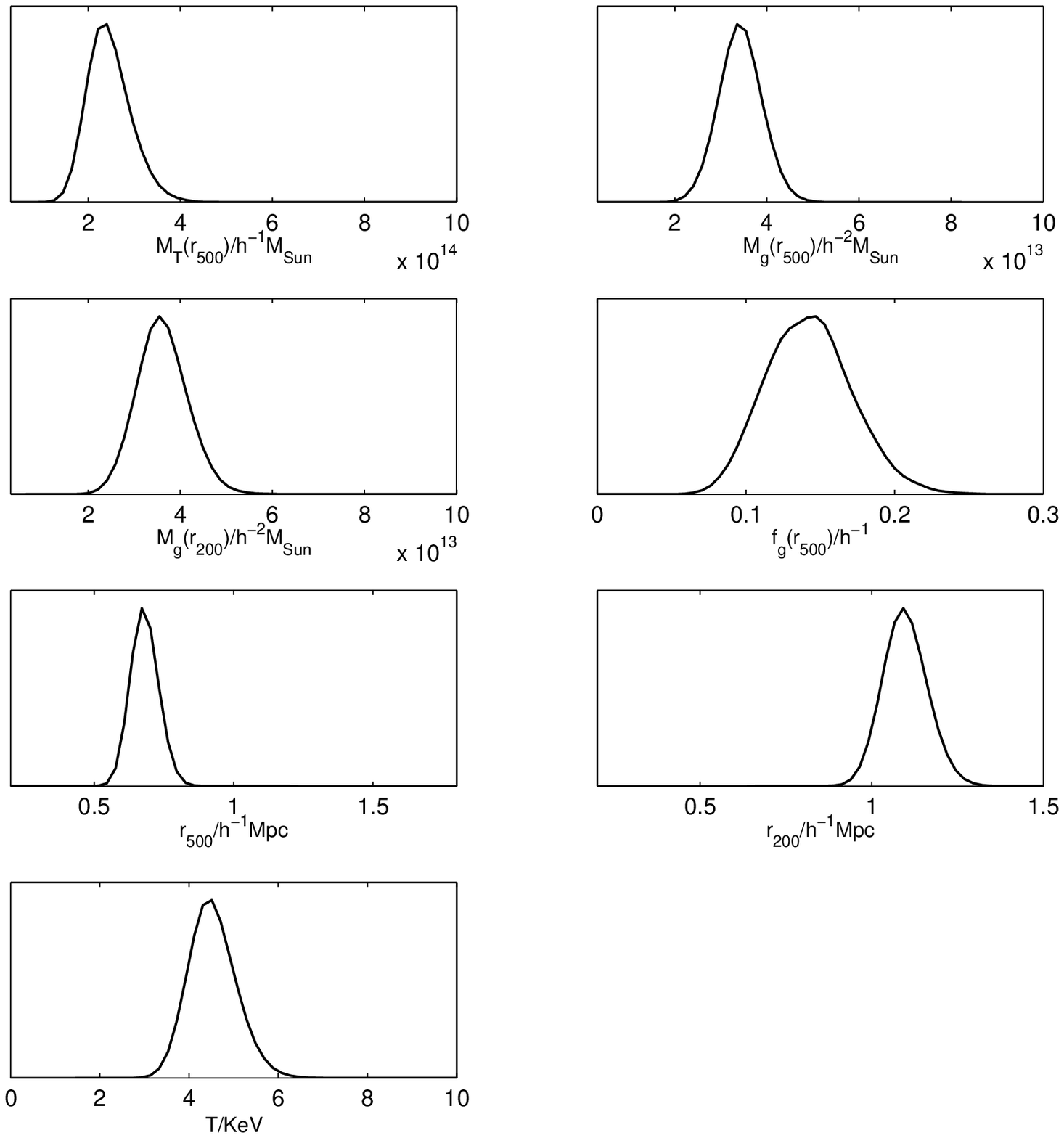}}
 \centerline{\includegraphics[width=7.5cm,height= 6.5cm,clip=,angle=0.]{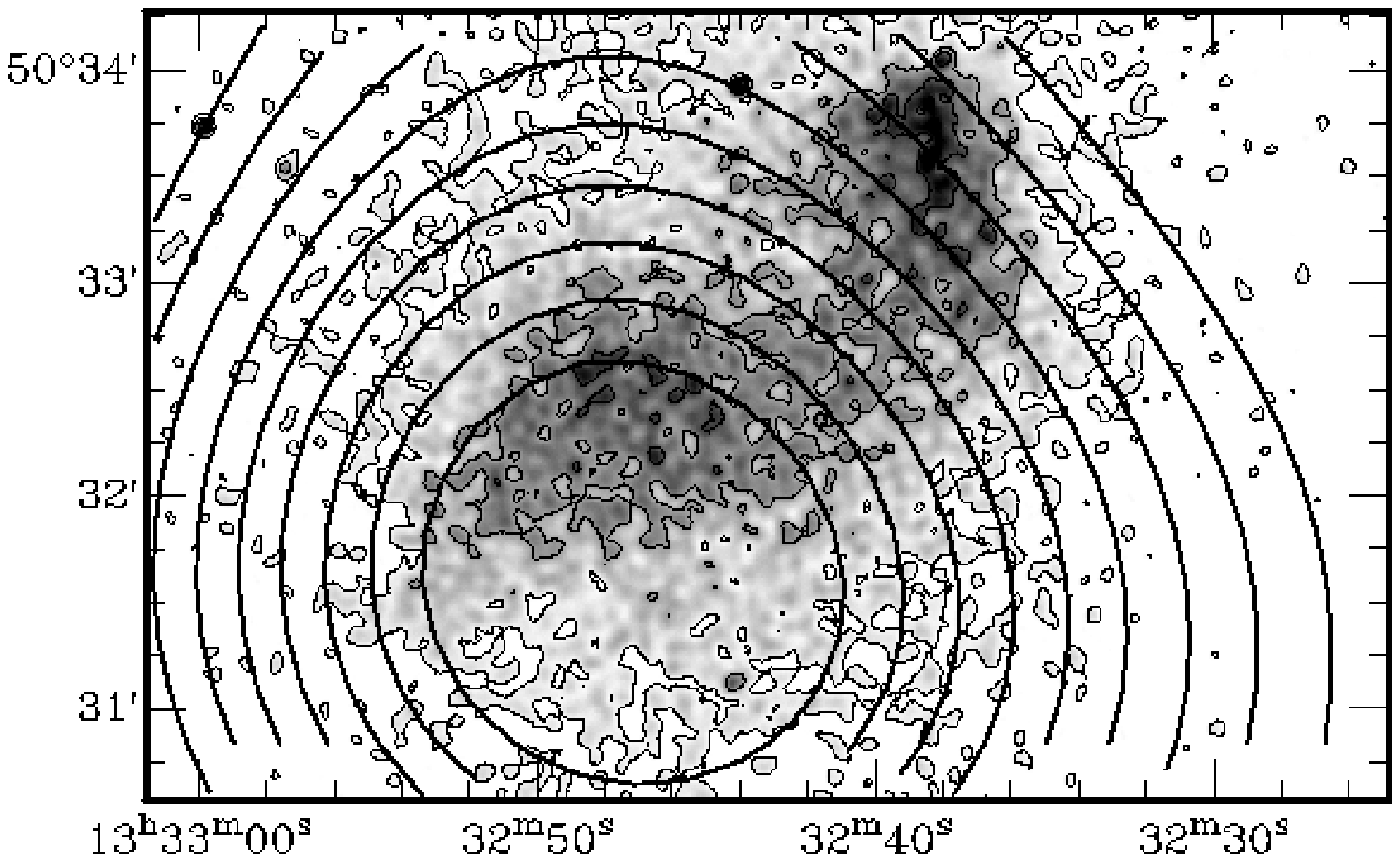}\qquad\includegraphics[width=7.5cm,height= 6.5cm,clip=,angle=0.]{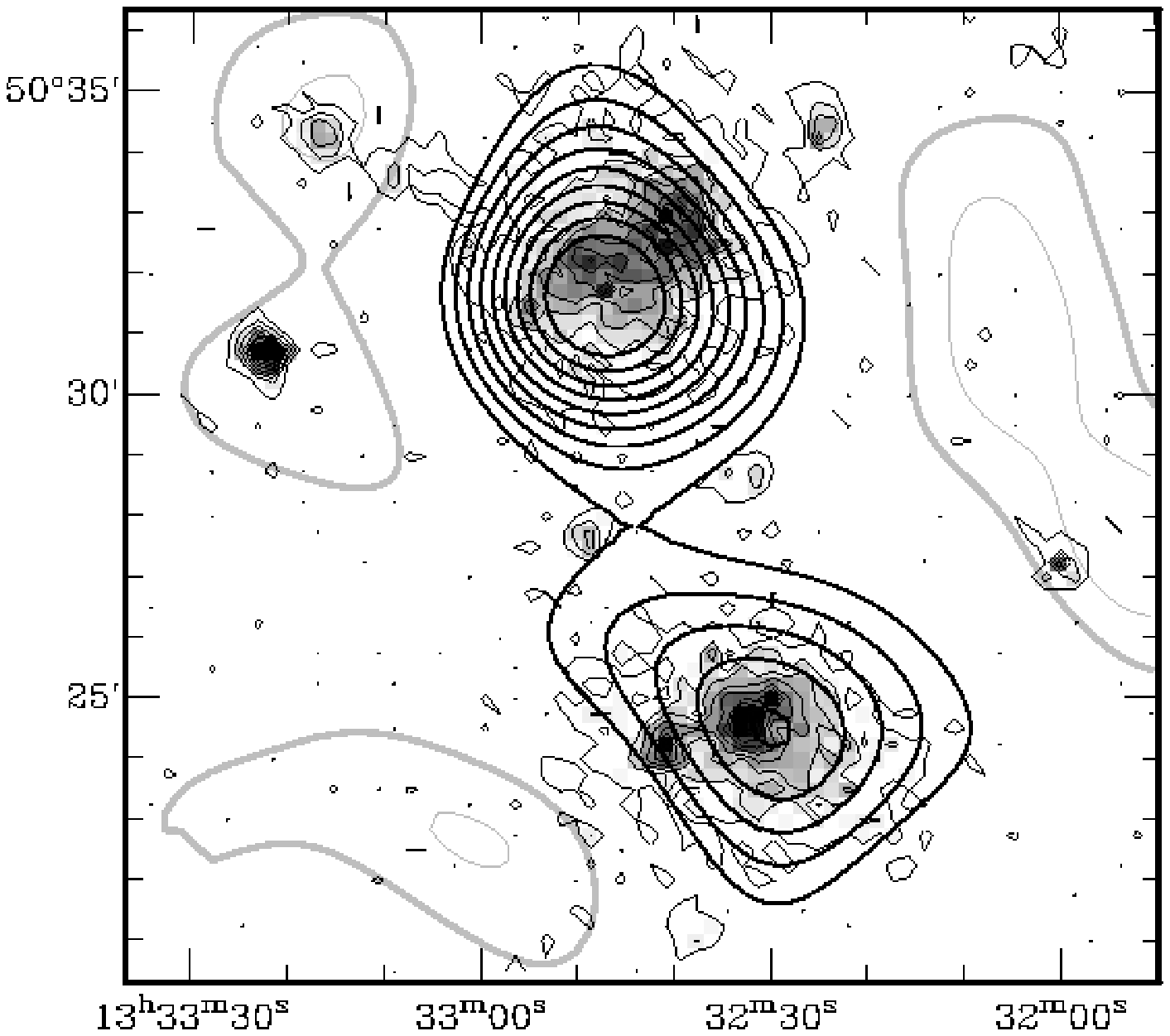}}
\caption{The top left image shows the SA map before subtraction, the map in the middle left has had the sources removed. The maps shown here are primary beam corrected signal-to-noise maps cut off at 0.3 of the primary beam. The noise level is $\approx$ 115$\mu$Jy towards the upper cluster (Abell 1758A) and $\approx$ 130$\mu$Jy towards the lower cluster (Abell 1758B). The source subtracted \textsc{uv} tapered map at the top right has a noise level $\approx$ 20$\%$ higher. The top right panel shows the cluster parameters that we sample from in our Bayesian analysis and the middle right plot presents several cluster parameters derived from our sampling parameters. The image at the right bottom shows the \emph{ROSAT} PSPC X-ray map overlayed with SA contours, whilst the bottom left shows a \emph{Chandra} image with SA countours.}
\label{fig:A1758A}
\end{figure*}

\begin{figure*}
\centerline{Abell 1758B}
 \centerline{\includegraphics[width=7.5cm,height= 7.5cm,clip=,angle=0.]{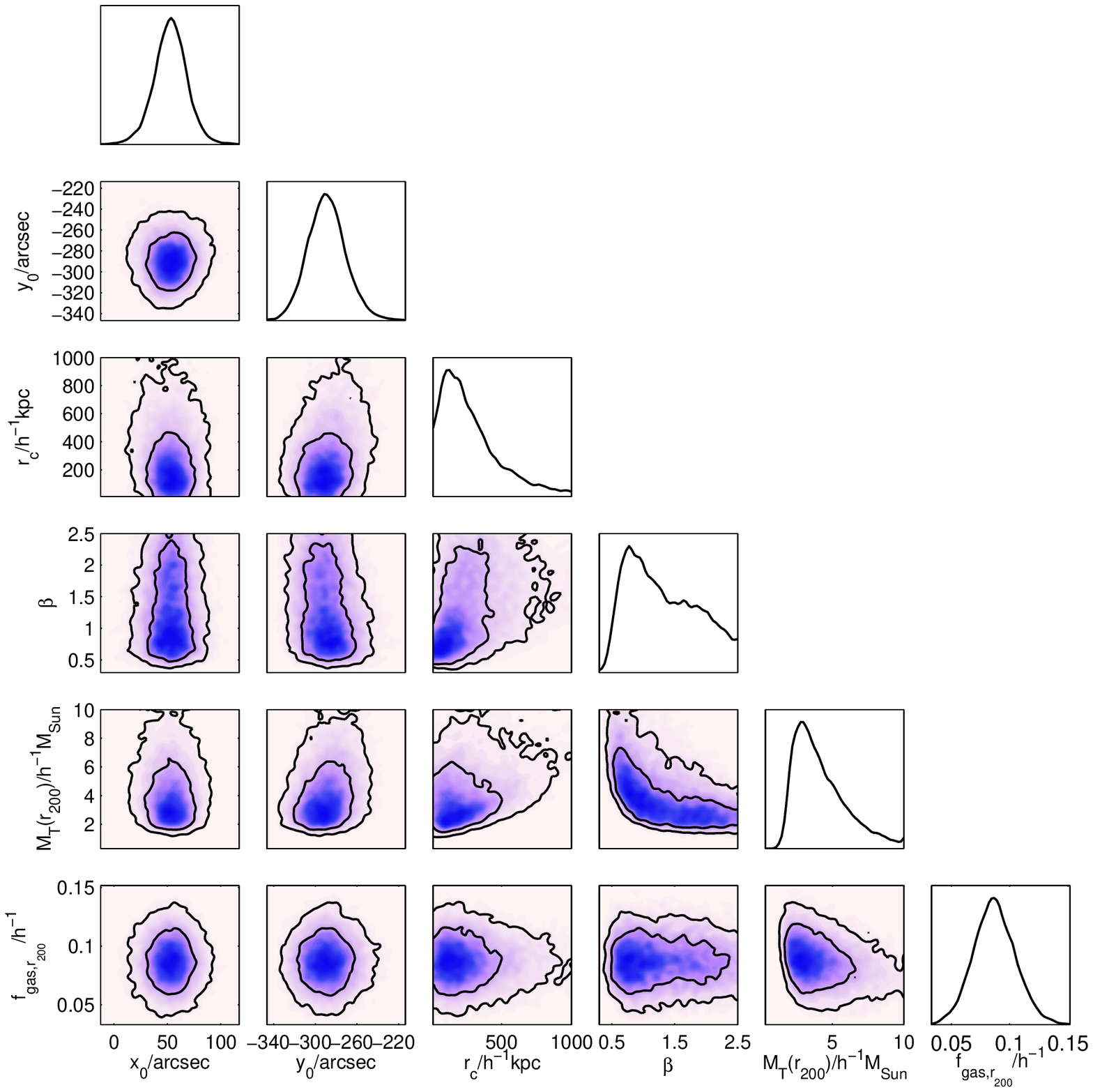}\qquad\includegraphics[width=7.5cm,height= 7.5cm,clip=,angle=0.]{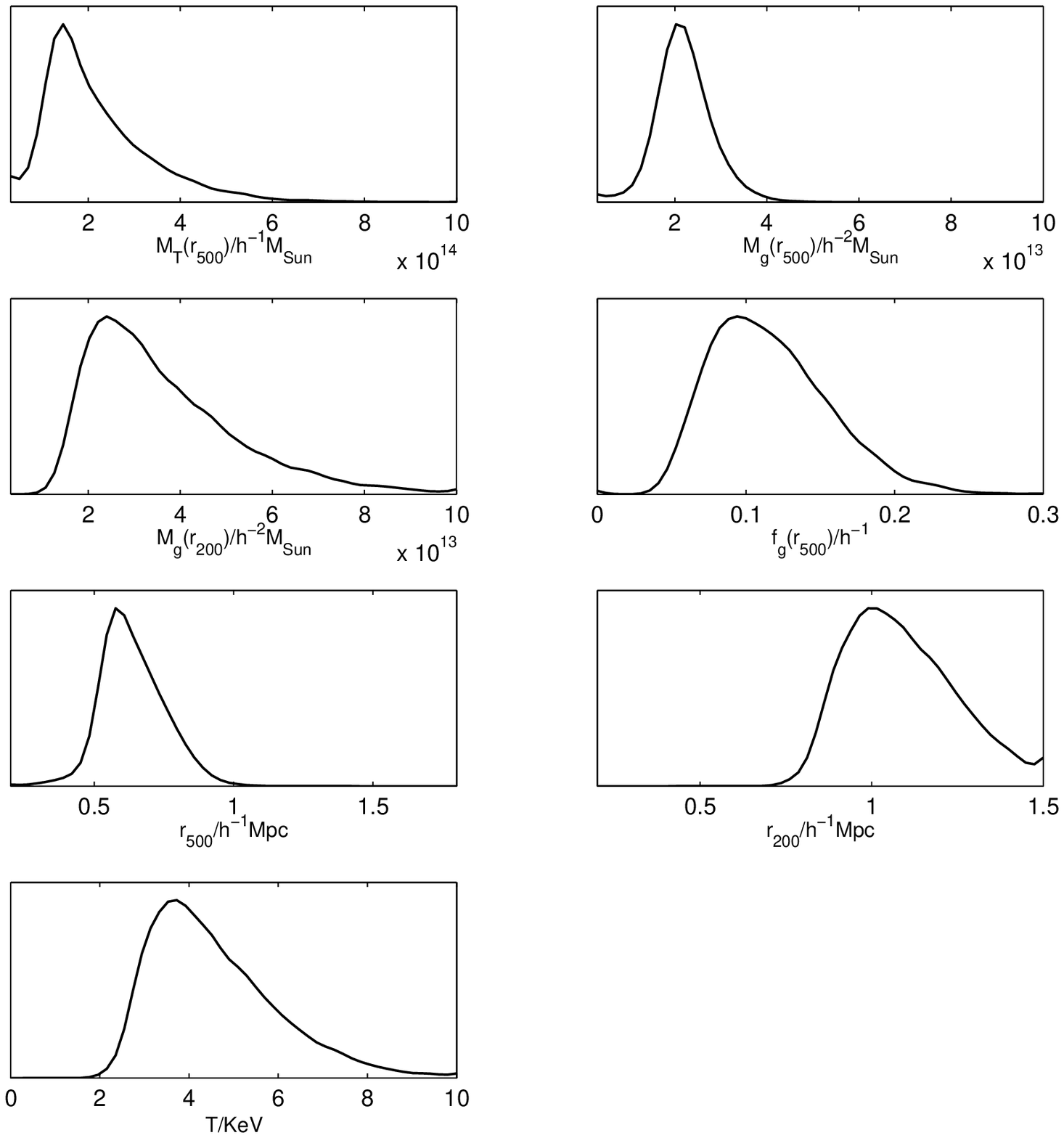}}
\caption{On the left we show the cluster parameters that we sample from and on the right we present some cluster parameters derived from our sampling parameters.}
\label{fig:A1758B}
\end{figure*}

An analysis of \emph{ROSAT} images clearly shows that this system consists of two interacting clusters, Abell~1758A and Abell~1758B, separated by 8$'$ (\citealt{ROSAT_A1758}). 
In Figure \ref{fig:A1758A} we present a single map that contains combined data from observations towards both clusters and the derived parameters for cluster Abell~1758A. We present the derived parameters for cluster Abell~1758A and X-ray maps from both the \emph{Chandra} data archive and \emph{ROSAT}%
\footnote{We acknowledge the use of NASA's SkyView facility
     (http://skyview.gsfc.nasa.gov) located at NASA Goddard
     Space Flight Center.}%
. In Figure \ref{fig:A1758B} we show the derived parameters for cluster Abell~1758B.

A detailed analysis of \emph{XMM-Newton} and \emph{Chandra} by \cite{2004ApJ...613..831D} indicates that the clusters Abell~1758A and Abell~1758B are likely to be in an early stage of merging and that both of these clusters are also undergoing major mergers with other smaller systems. A recent analysis of Spitzer/MIPS 24$\mu$m data by \cite{LoCuSS_A1758} classifies Abell 1758 as the most active system they have observed at that wavelength. They also identify numerous smaller mass peaks and filamentary structures, which are likely to indicate the presence of infalling galaxy groups, in support of the David \& Kempner observations. Zhang et al. use \emph{XMM-Newton} to study Abell~1758A and found $M_{500}$ = 1.1 $\pm$0.3 $\times 10^{15}\rm{M_{\odot}}$, where $r_{500}$ = 1.43Mpc. They assume isothermality, spherical symmetry and $h_{70}=1$.

\subsection{Zw1454.8+2233}

\begin{figure*}
\centerline{Zw1454.8+2233}
\centerline{\includegraphics[width=7.5cm,height= 7.5cm,clip=,angle=0.]{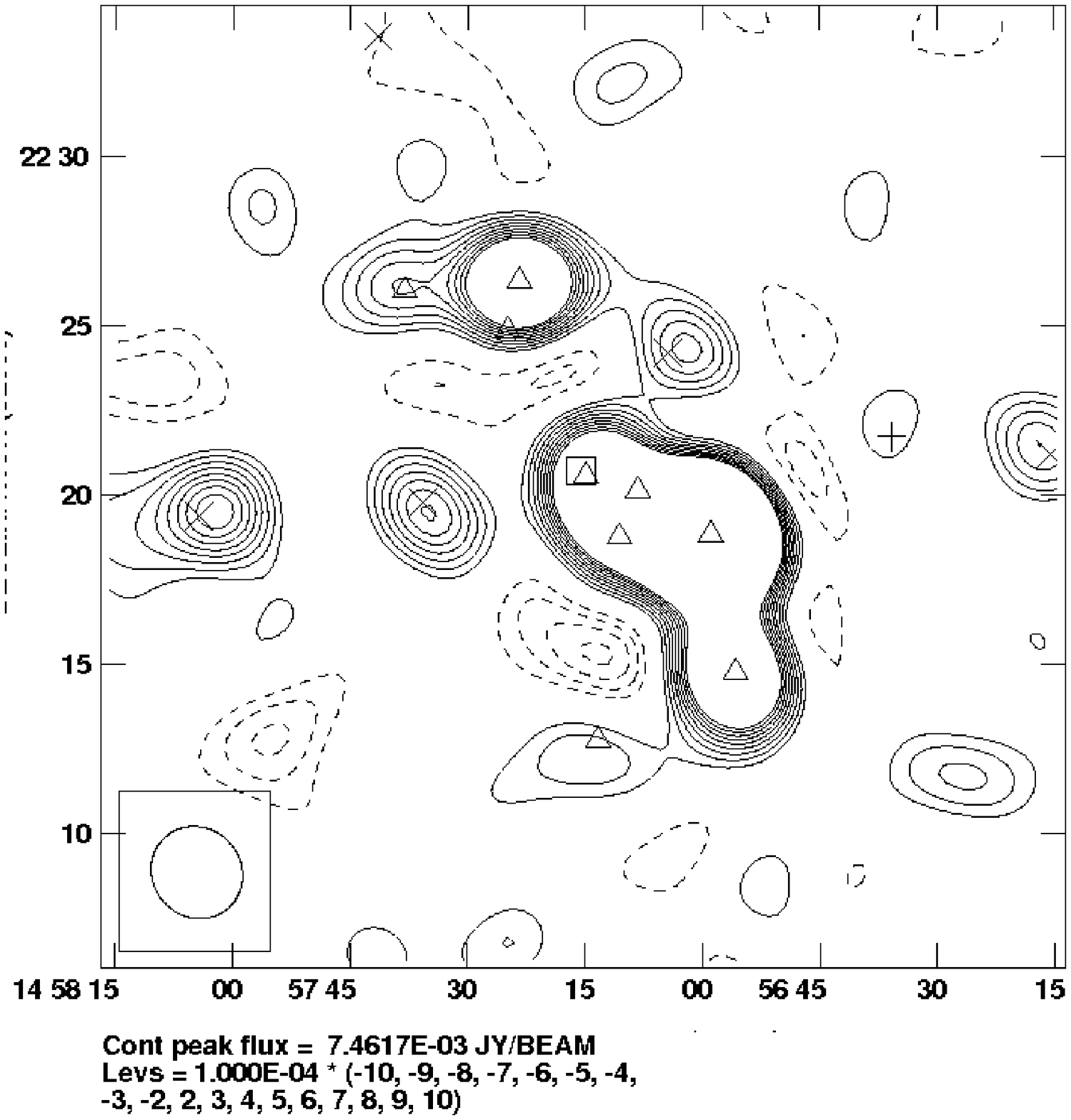}\qquad\includegraphics[width=7.5cm,height= 7.5cm,clip=,angle=0.]{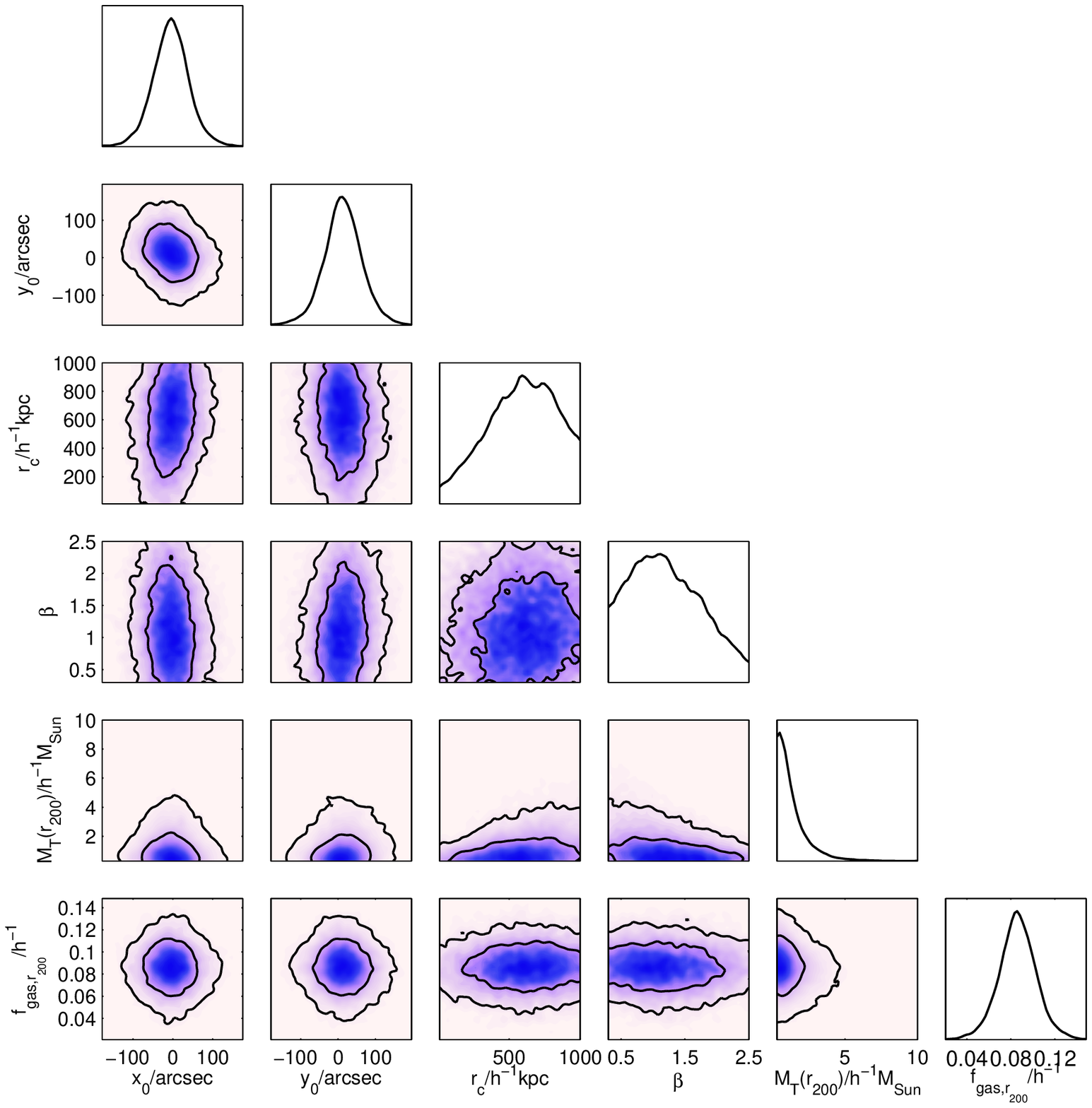}}
 \centerline{\includegraphics[width=7.5cm,height= 7.5cm,clip=,angle=0.]{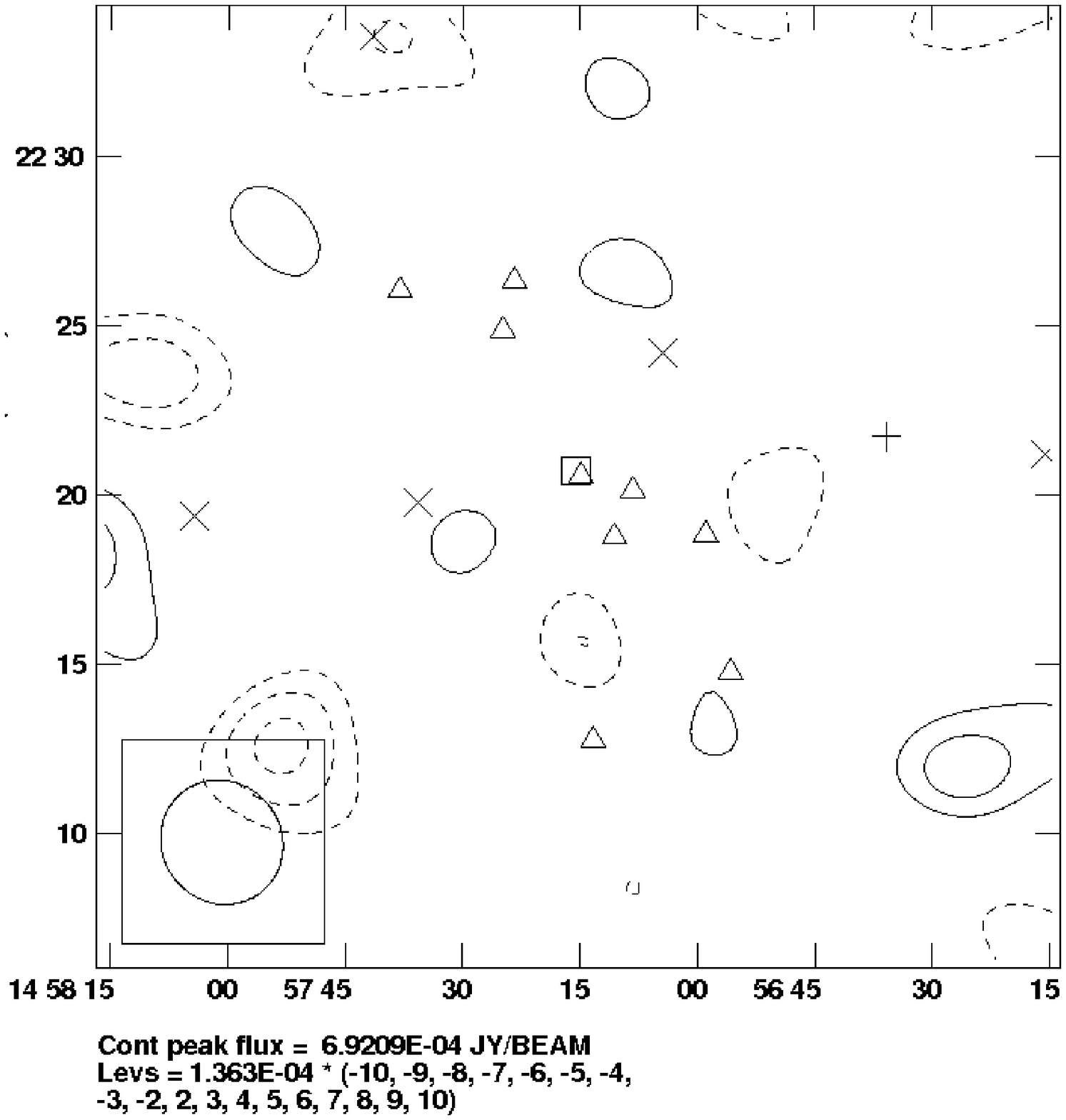}\qquad\includegraphics[width=7.5cm,height= 7.5cm,clip=,angle=0.]{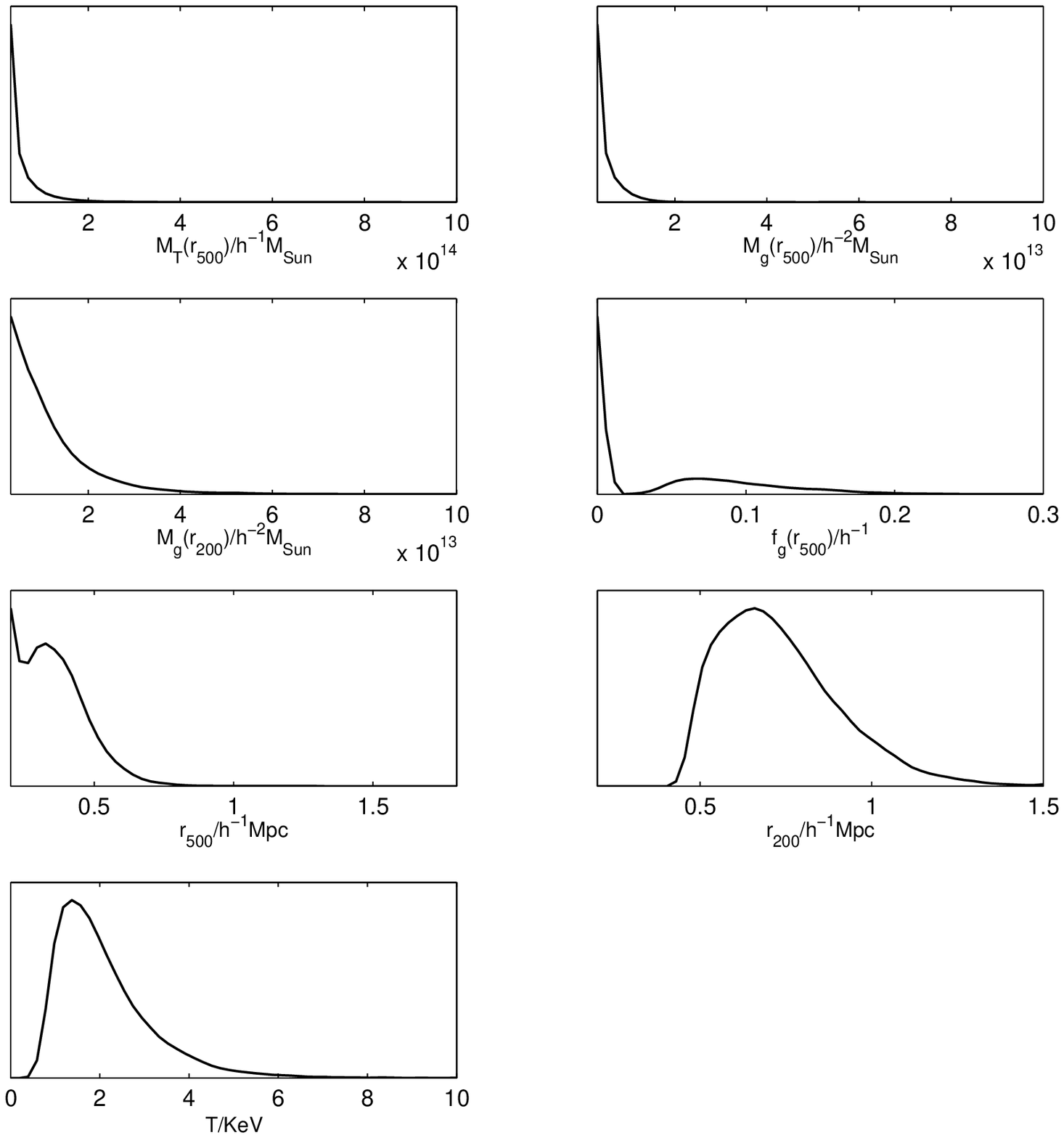}}
 \centerline{\includegraphics[width=7.5cm,height= 6.5cm,clip=,angle=0.]{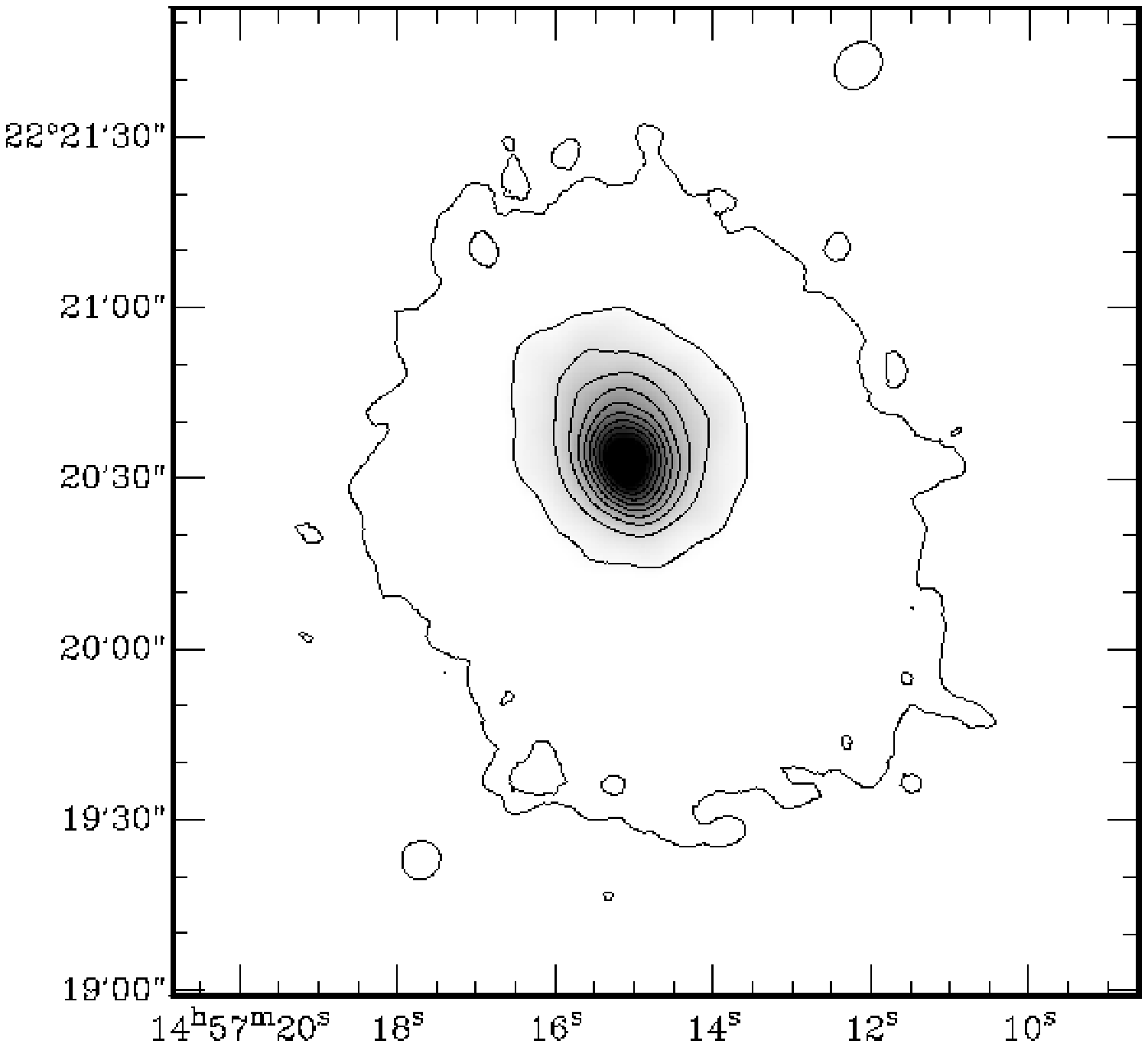}}
\caption{The \emph{null detection} of Zw1454.8+2233 in SZ. The top left image is the SA map before subtraction, showing the challenging source environment. the map in the middle left has had the sources removed, however, no decrement is visible. The middle panel on the left shows the sampling parameters and on the right we show the derived cluster parameters, these parameters are what we would expect from a null detection, they indicate mass with a high probability of being 0.0. The image at the bottom shows the \emph{Chandra} X-ray map overlayed with SA contours.}
\label{fig:ZW7160}
\end{figure*}

We detect several sources close to the cluster centre -- a point source with a flux density of 7.97~mJy at 14:56:59.11 +22:18:55.97 and a 4.67~mJy source at 14:57:25.38 +22:37:33.03. We do not detect an SZ effect from this cluster even though we would expect to, given the low noise levels of our SA maps. 
The SA maps and derived parameters are shown in Figure \ref{fig:ZW7160}. The derived parameters for this non-detection are as expected: we find that $M_{\rm{T},r200}$ approaches our lower prior limit ($0.32$ $\times$ $10^{14}M_{\odot}h_{100}^{-1}$) and that $M_g$ shows simliar behaviour; both $r_{200}$ and $T_{SZ,MT}$ are well constrained because both these parameters are derived from $M_{\rm{T},r200}$ which itself is well constrained at the value of its lower prior limit.

Zhang et al. used XMM Newton to study Zw1454.8+2233 and found $M_{500}$ = 2.4 $\pm$0.7 $\times 10^{14}\rm{M_{\odot}}$, where $r_{500}$ = 0.87Mpc. They assumed isothermality, spherical symmetry and $h_{70}=1$. \cite{Zw1454.8+2233_HALO} observed the cluster with the GMRT at 610MHz and found that the cluster contains a core-halo source. This is in agreement with the value obtained from \emph{Subaru} weak lensing observations by Okabe et al. The \emph{Chandra} X-ray observations (\citealt{Zw1454.8+2233_COOLING}) also reveal that Zw1454.8+2233 is a cooling core cluster; these are often associated with core-halos.

\subsection{RXJ1720.1+2638}

\begin{figure*}
\centerline{RXJ1720.1+2638}
\centerline{\includegraphics[width=7.5cm,height= 7.5cm,clip=,angle=0.]{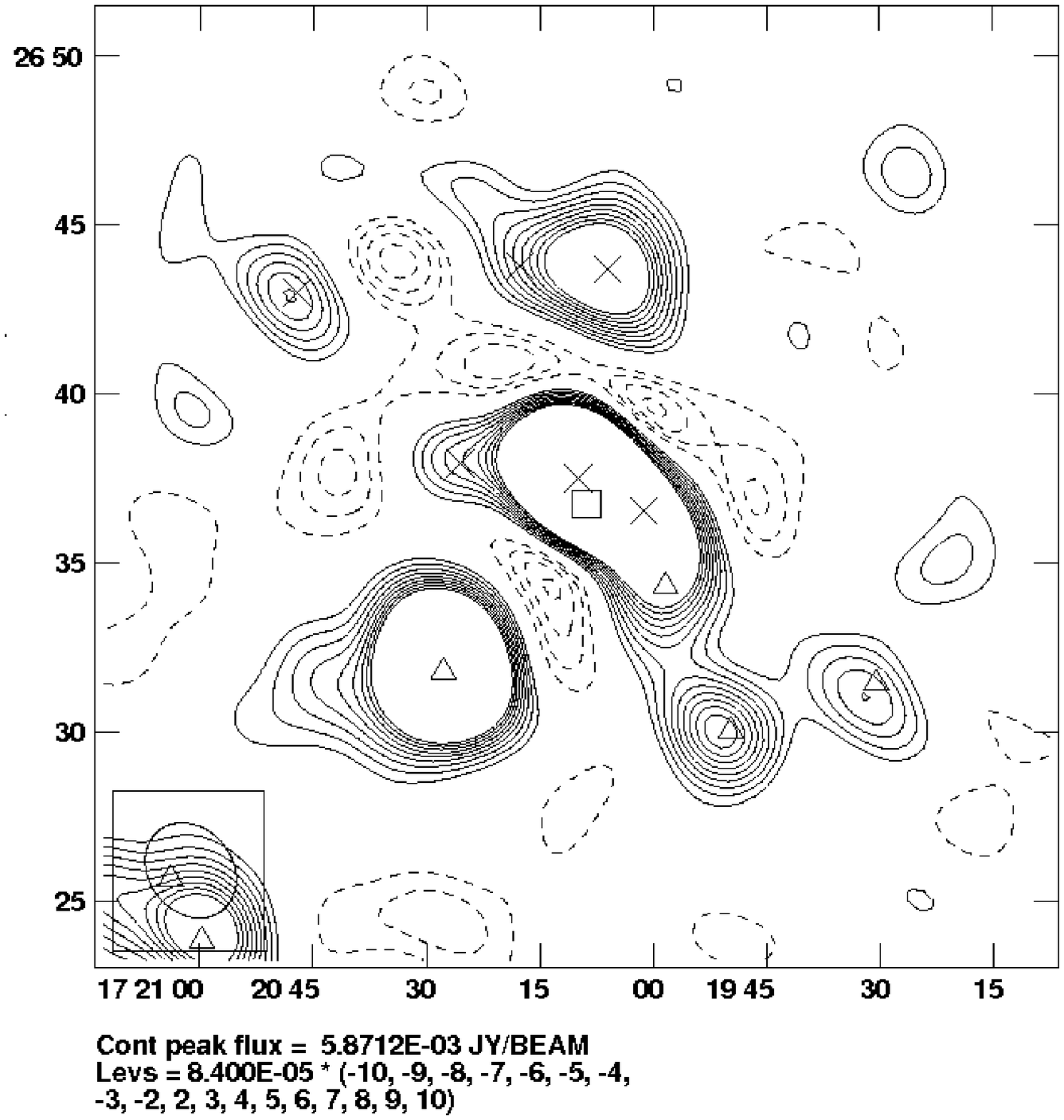}\qquad\includegraphics[width=7.5cm,height= 7.5cm,clip=,angle=0.]{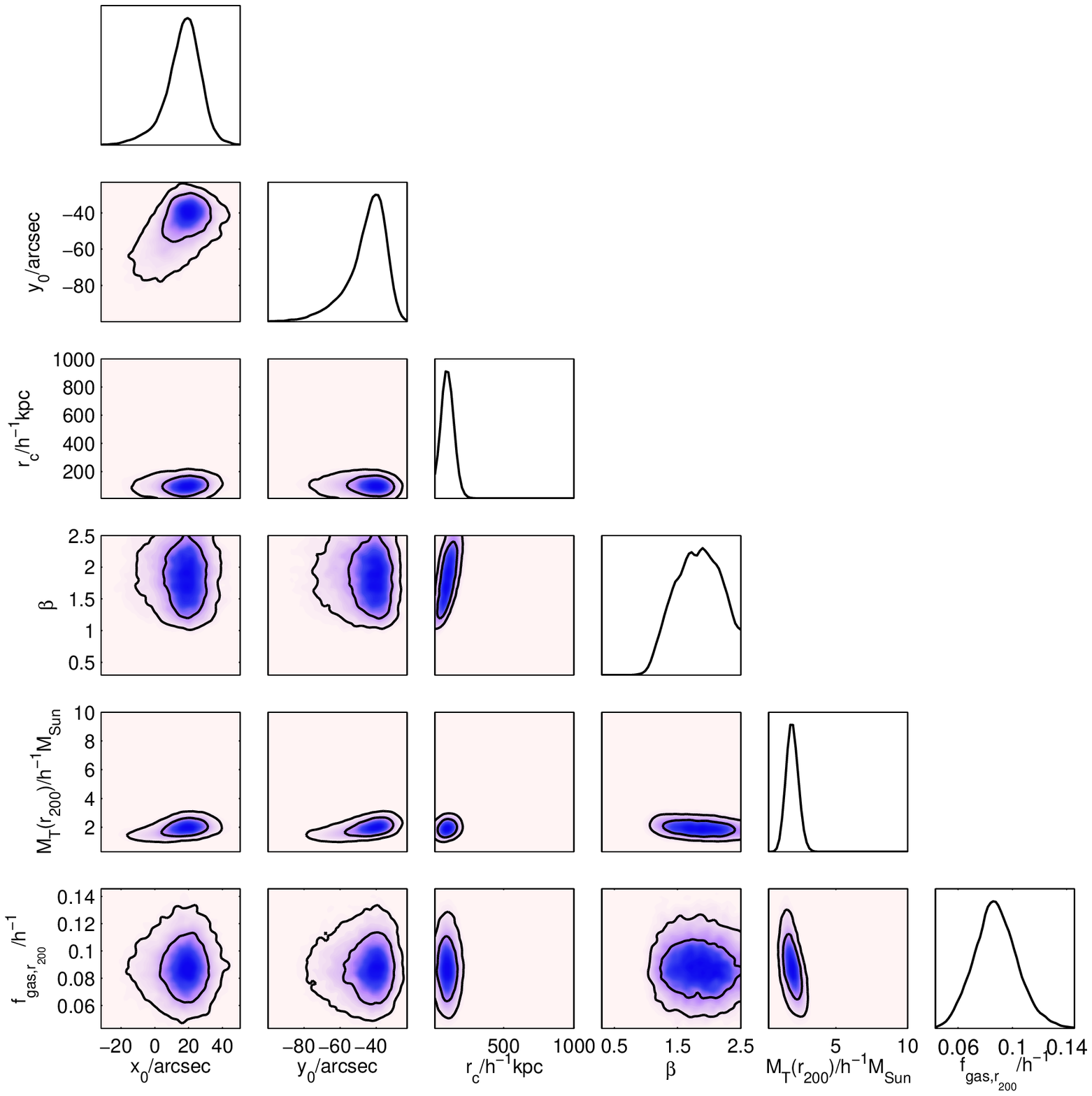}}
 \centerline{\includegraphics[width=7.5cm,height= 7.5cm,clip=,angle=0.]{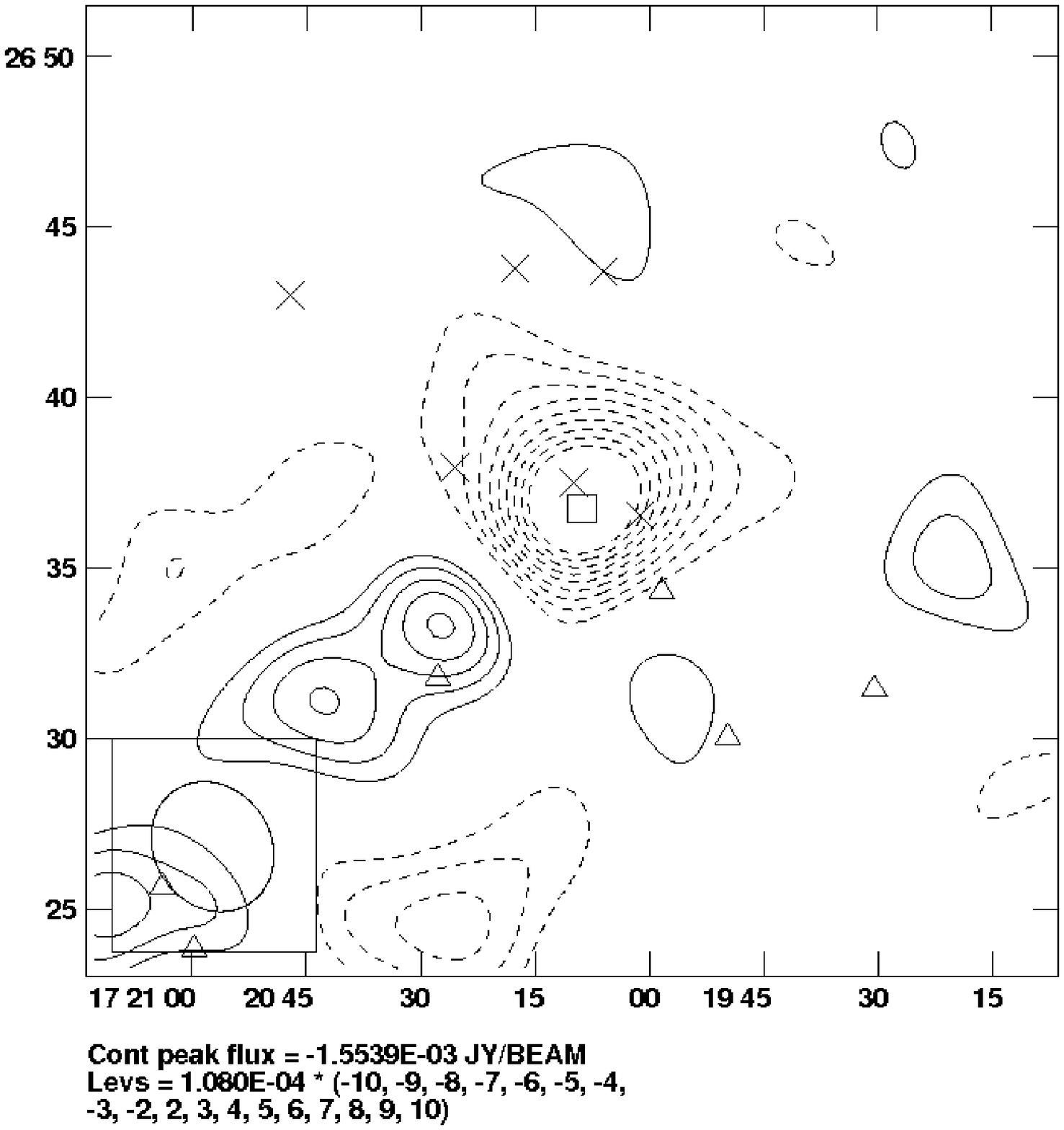}\qquad\includegraphics[width=7.5cm,height= 7.5cm,clip=,angle=0.]{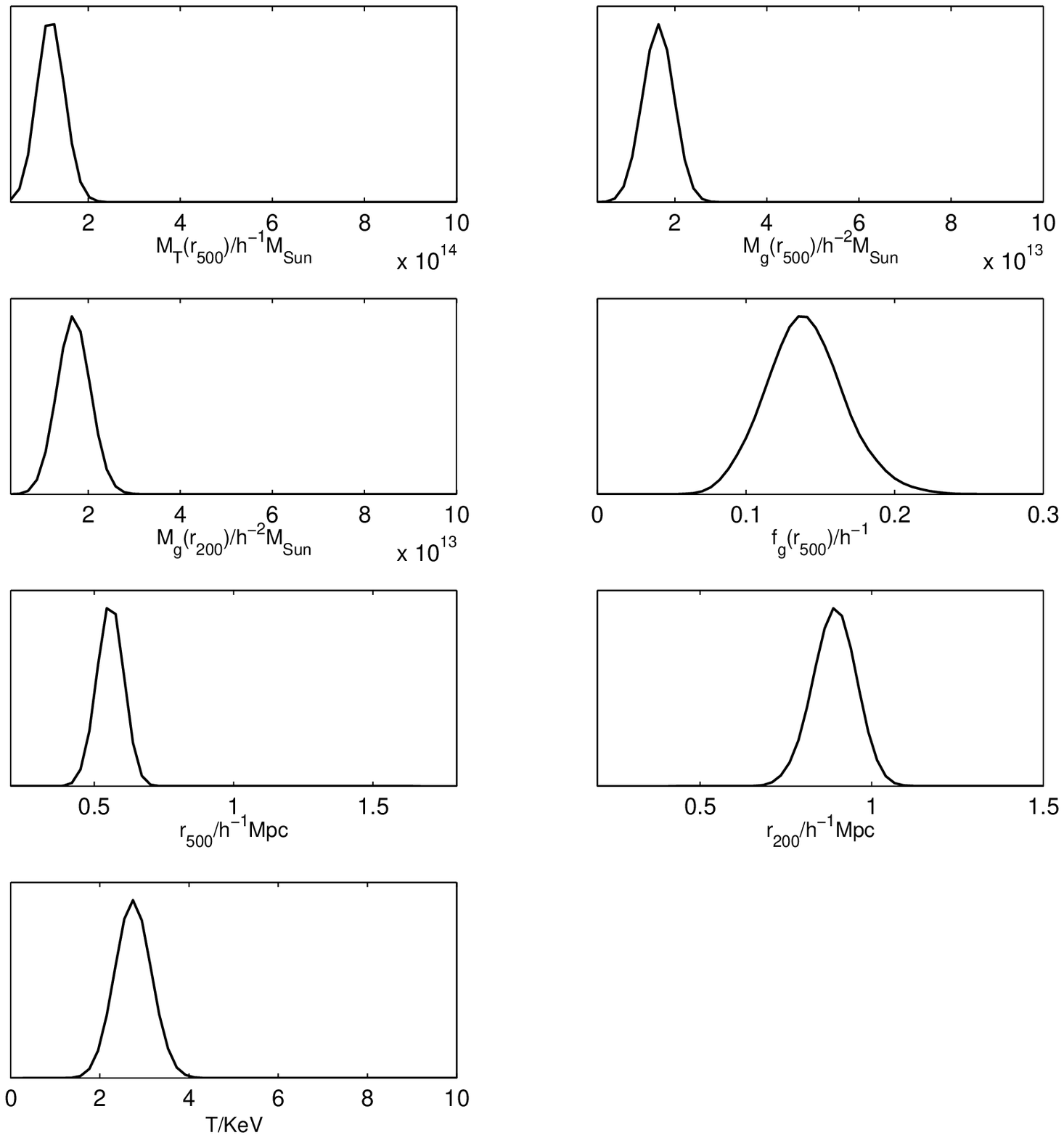}}
 \centerline{
\includegraphics[width=7.5cm,height= 6.5cm,clip=,angle=0.]{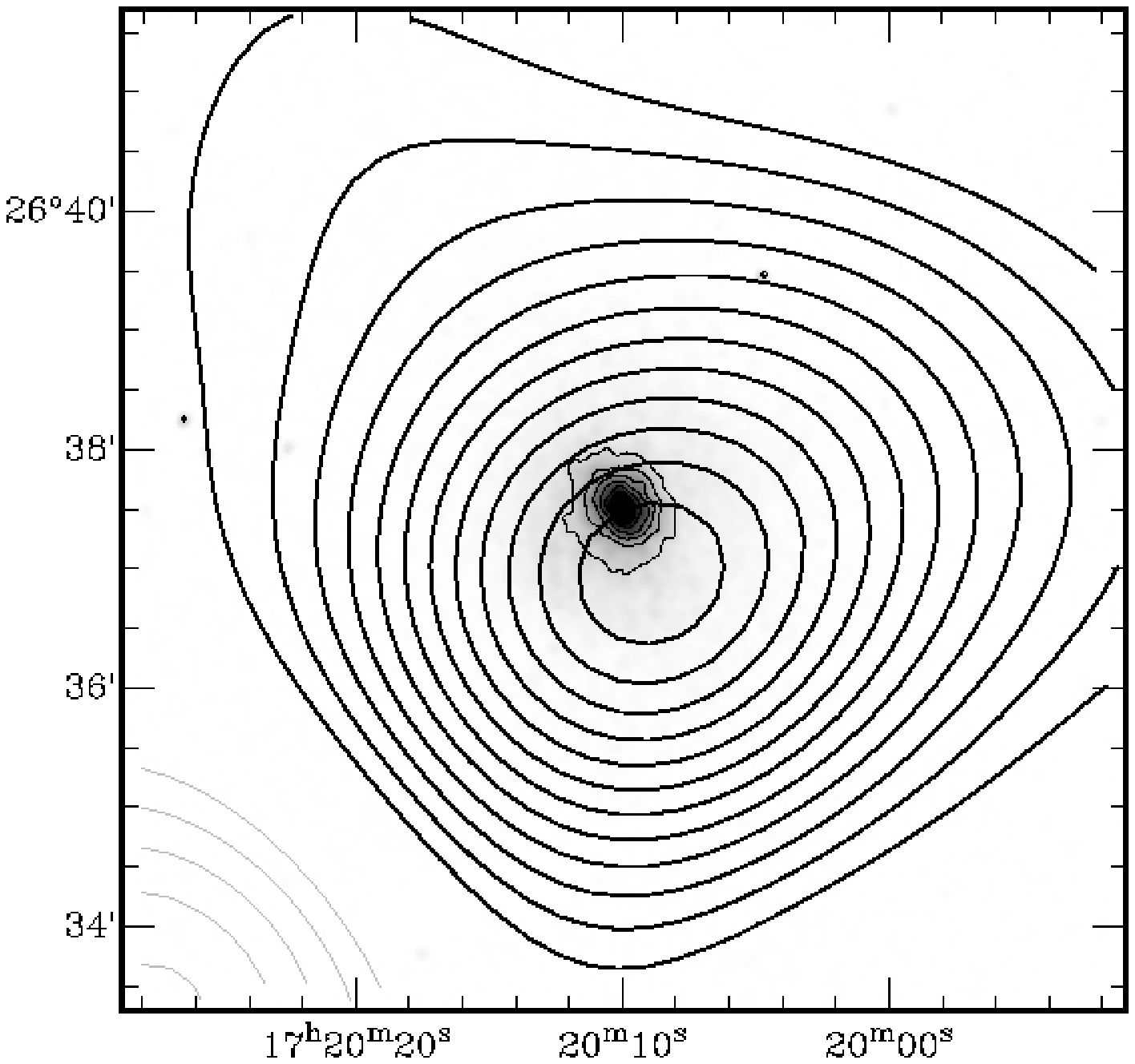}}
\caption{The top left image shows the SA map before subtraction, the map in the middle left has had the sources removed, the top right panel shows the cluster parameters that we sample from in our Bayesian analysis and the middle right plot presents several cluster parameters derived from our sampling parameters. The image at the bottom shows the \emph{Chandra} X-ray map overlayed with SA contours.}
\label{fig:RXJ1720.1+2638}
\end{figure*}

At 16~GHz the source environment around the cluster is challenging: in our LA data we detect a 3.9~mJy source at the same position as the cluster, and several other sources with comparable flux densitites $<500 \arcsec$ from the cluster centre. However, using our Bayesian analysis we are able to accurately model the positions, flux densities and spectral indices of these sources such that, after they are subtracted from our SA maps, we see a significant decrement. The AMI SA maps before and after source subtraction are shown in Figure \ref{fig:RXJ1720.1+2638}, as are the derived cluster parameters and a \emph{Chandra} image of the cluster. Our SZ-effect map shows that the cluster may have an irregular shape; we see low signal-to-noise emission to the SE and NW of the cluster X-ray centroid; however, the centre of the SZ emission is coincident with the X-ray centroid.

\emph{Chandra} observations  (\citealt{RXJ1720_CHANDRA}) indicate that, although the cluster does not have an irregular shape or elongation, it has discontinuities in its density profile; this may indicate it is in the latter stages of merging. The largest discontinuity is observed in the SE sector of the cluster and is noted to have a structure similar to a cold front observed in other merging systems such as Abell~2142 and Abell~3667. Mazzotta et al. determined the mass profile for the cluster assuming hydrostatic equilibrium using two distinct regions (SE and NW) to model the cluster density profile: each region was separately analysed and used to calculate $M_{1000\rm{kpc}}$ = 4 $\pm$ 10 $\times 10^{14}h^{-1}_{50}\rm{M_{\odot}}$. 

\subsection{Cluster Temperatures}\label{sec:TEMP_COMP}

\begin{figure*}
\centerline{\includegraphics[width=13.2cm,height= 18.7cm,clip=,angle=-90.0]{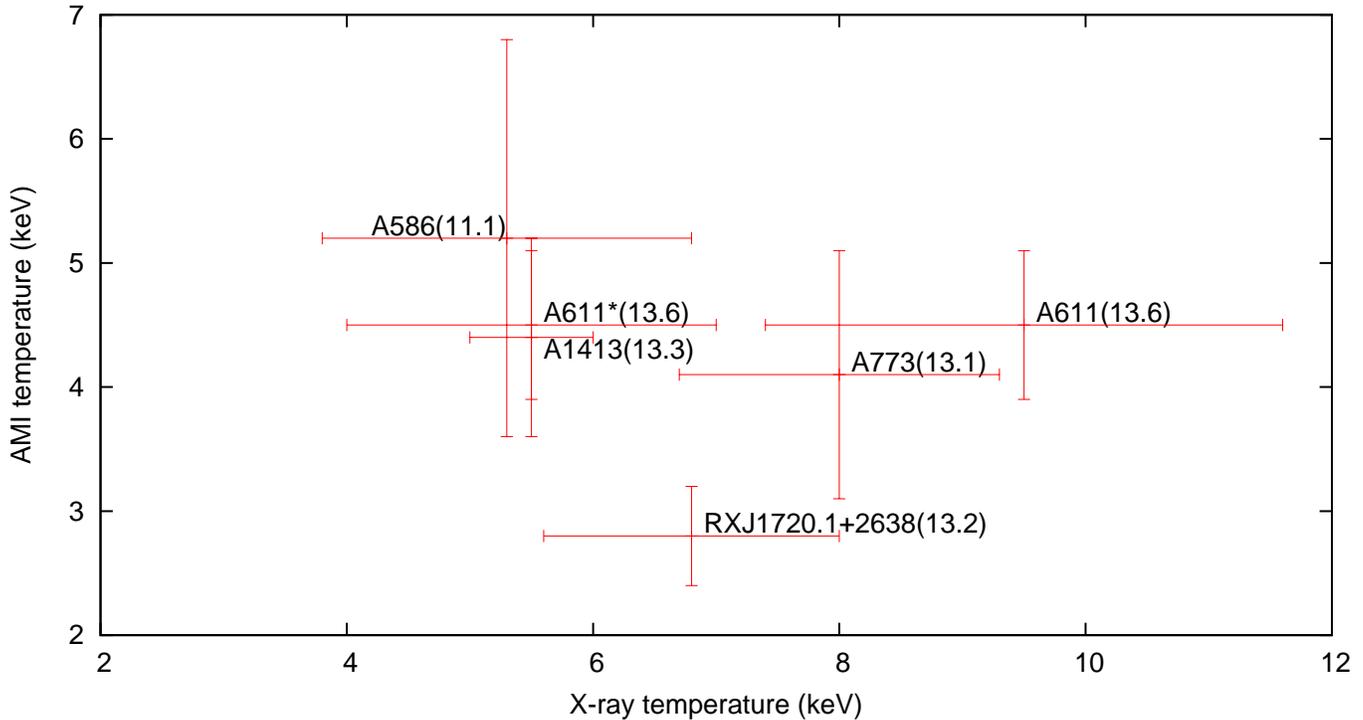}}
\caption{The AMI mean temperature within $r_{200}$ versus the X-ray temperature, each point is labelled with the cluster name and X-ray luminosity. Most of the X-ray measurements are large-radius temperatures from the ACCEPT archive (\citealt{ACCEPT}) with 90\% confidence bars. The radius of the measurements taken from the ACCEPT archive are 400-600kpc for Abell 586,  300-700kpc for Abell 611, 300-600kpc for Abell 773 and for RXJ1720.1+2638 r = 550-700kpc. The A1413 X-ray temperature is estimated from the 700-1200kpc measurements made with the Suzaku satellite (\citealt{Suzaka_A1413}), this value is consistant with \citealt{Chandra-A1413} and \citealt{XMM_snowden}. The Abell 611* temperature is the 450-750kpc value with $\sigma$ error bars (\citealt{Xray_lense_Abell 611_2}). The ACCEPT archive temperature for Abell~1758A is 16$\pm$7keV at r= 475-550kpc with SZ temperature 4.5$\pm$0.5, for clarity this has not been included on the plot.}
\label{TEMP_COMP1}
\end{figure*}

Following on from Rodr{\'i}guez-Gonz{\'a}lvez et al. (2011), in Figure \ref{TEMP_COMP1} we compare the AMI SA observed cluster temperatures within $r_{200}$ ($T_{SZ,MT}$) with large-radius X-ray values ($T_{X}$) from \emph{Chandra} or \emph{Suzaku} that we have been able to find in the literature. We use large radius X-ray temperature values as these ignore the complexities of the cluster core and are representative of the average cluster temperature within $\approx$1Mpc which is measured by AMI. Before comment on these we deal with two technical points. First, for Abell~611, we have plotted two X-ray values (from \emph{Chandra} data); one from the ACCEPT archive (\citealt{ACCEPT}) which is higher than our AMI SA measurement, while the second X-ray measurement from \emph{Chandra} (Donnarumma et al.) is consistent with our measurement. Secondly, the ACCEPT archive r= 475-550kpc temperature for Abell~1758A is  16$\pm$7keV and for clarity is not included on the plot. 

\begin{figure*}
\centerline{\includegraphics[width=13.2cm,height= 18.7cm,clip=,angle=-90.0]{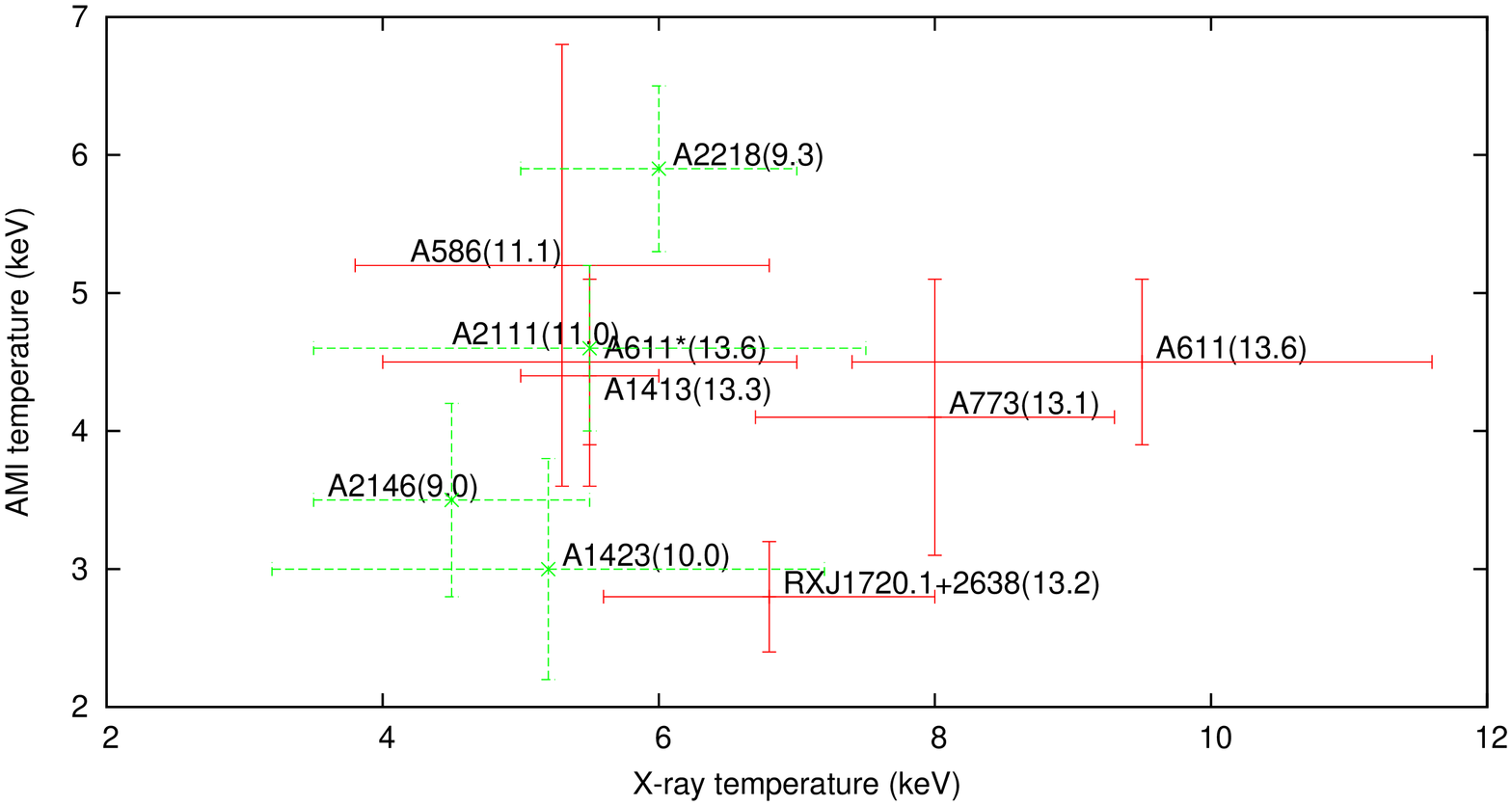}}
\caption{The AMI mean temperature within $r_{200}$ versus the X-ray temperature including values from Rodr{\'i}guez-Gonz{\'a}lvez et al. (2011). Again Abell~1758A is not shown.}
\label{TEMP_COMP2}
\end{figure*}

Evidently Abell~586, Abell~611 (with the Donnarumma et al. X-ray temperature) and Abell~1413 have corresponding SZ and X-ray temperatures while  Abell~773, Abell~1758A and RXJ1720.1+2638 have X-ray temperatures significantly higher than their SZ temperature. The position is made clearer by combining the values with those in Rodr{\'i}guez-Gonz{\'a}lvez et al. (2011). The combined data are shown in Figure \ref{TEMP_COMP2}, in which there is reasonable correspondence between SZ and X-ray temperatures at lower X-ray luminosity, with excess (over SZ) X-ray temperatures at higher X-ray luminosity. An exception to this is Abell~1413 which despite its high X-ray luminosity is in good agreement with our SZ value, but for this cluster we have been able to use \emph{Suzuka} measurements over r= 700-1200kpc. It is noteworthy that Abell~773, Abell~1758A and RXJ1720.1+2638  are major mergers, and we emphasize the $n_{e}^2$ weighting of X-ray temperature measurements.




\section{Conclusion}\label{sec:CONCLUSION}
 
i) We have good SZ detections for eight clusters and a non-detection for Zw1454.8+2233.  \\
ii) For the seven detected clusters with $L_{X}>11$$\times 10^{37}$W ($h_{50}=1$), we fit $\beta$ profiles to the cluster signals and find $M_{g,r200}$ values of 1.7-4.3$\times 10^{13} h_{100}^{-1}M_{\odot}$ and values $M_{T,r200}$ of 2.0-5.1$\times 10^{14} h_{100}^{-1}M_{\odot}$.  \\
iii) For Abell~611 and Abell~773 our values of $M_{g,r200}$ and $M_{T,r200}$ are lower than those in Zwart et al. (2010) which they thought to be biased high, because they use a high value for $T_{SZ,MT}$ (estimated from a low-radius X-ray measurement) and assume this value to be constant throughout the cluster. \\
iv) For the six clusters in the work of this paper for which we have found large-radius $r \geq 500$kpc X-ray spectroscopic temperatures in the literature, we find that $T_{X}$ and $T_{SZ,MT}$ values correspond reasonably well for Abell~586, Abell~611 (with the Donnarumma et al. X-ray temperature rather than the ACCEPT archive value) and Abell~1413, but that correspondence falls away for Abell~773, Abell~1758A and RXJ1720.1+2638 which have a high $T_{X}$, for these, $T_{SZ,MT}$ is less than  $T_{X}$. \\
v) The picture seems to become clearer -- although all of this work involves only very small numbers -- when we add in the data of Rodr{\'i}guez-Gonz{\'a}lvez et al. (2011). We find that there is reasonable $T_{X}$:$T_{SZ,MT}$ correspondence for the six clusters at lower $T_{X}$ but that the correspondence breaks down at high  $T_{X}$. However, two points are evident. The more general one is that
the breakdown of the $T_{X}$:$T_{SZ,MT}$ correspondence tends to be associated with high $L_{X}$ and with major mergers.
The more specific one is that Abell~1413 has values of $T_{X}$ and $T_{SZ,MT}$ that correspond yet has high $L_{X}$: but we have used $T_{X}$ measured by the \emph{Suzaku} at very high radius. \\
vi) We suspect this points to agreement between large-radius SZ estimates and larger-radius spectroscopic temperature measurements, but that substantial mergers bias $T_{X}$ measurements more than $T_{SZ,MT}$; however we stress again that our sample from that and our companion paper is very small.


\section{Acknowledgements}

We thank PPARC/STFC for support of AMI and its operations. We are grateful to
the staff of the Cavendish Laboratory and the Mullard Radio Astronomy
Observatory for the maintenance and operation of AMI. CRG, MO, MPS
and TWS acknowledge PPARC/STFC studentships. This work was performed using the
Darwin Supercomputer of the University of Cambridge High Performance Computing
Service (http://www.hpc.cam.ac.uk/), provided by Dell Inc. using Strategic
Research Infrastructure Funding from the Higher Education Funding Council for
England, and the Altix 3700 supercomputer at DAMTP, University of Cambridge
supported by HEFCE and STFC. We are grateful to Stuart Rankin and Andrey
Kaliazin for their computing assistance.

\bsp
\label{lastpage}

\end{document}